\documentclass[prl,twocolumn,superscriptaddress,nofootinbib,floatfix]{revtex4-2}

\usepackage{amsmath,amssymb,amsfonts,amsbsy}
\usepackage{bm}
\usepackage{slashed}
\usepackage{graphicx}
\usepackage{mathrsfs}
\usepackage{braket}
\usepackage{subcaption}
\usepackage{xcolor}
\usepackage[colorlinks=true,allcolors=blue]{hyperref}

\newcommand{\bbar}{\bar{b}}
\newcommand{\mubar}{\bar{\mu}}
\newcommand{\dmu}{\delta\mu}
\newcommand{\Nz}{N_0}
\newcommand{\cs}{c_s^2}
\newcommand{\DOmega}{\Delta\Omega}

\begin{document}

\title{Dynamically Generated Fermi Surface Mismatch and Relativistic Superfluidity\texorpdfstring{\\}{ }in a Two-Component Massless Fermionic Theory}
\author{Susobhan Mandal}
\email{susobhan@labs.iisertirupati.ac.in}
\affiliation{Department of Physics, Indian Institute of Science Education and Research (IISER)
Tirupati, Tirupati - 517507, Andhra Pradesh, India}

\author{Sambuddha Sanyal}
\email{sambuddha.sanyal@iisertirupati.ac.in}
\affiliation{Department of Physics, Indian Institute of Science Education and Research (IISER)
Tirupati, Tirupati - 517507, Andhra Pradesh, India}

\date{\today}

\begin{abstract}
When fermions pair across mismatched Fermi surfaces, the mismatch reflects a built-in inequivalence between the species. We show it can instead arise dynamically by spontaneous symmetry breaking. In a massless two-component Dirac theory with exact SU(2) flavor symmetry, a self-interacting vector boson condenses, splitting the Fermi surfaces while preserving time reversal. Pairing then yields a stable relativistic superfluid, promoting the Chandrasekhar–Clogston line to a surface in coupling space, the mismatch fixed self-consistently by the symmetry-breaking coupling.
\end{abstract}

\maketitle

\section{Introduction}

Cooper pairing in the standard Bardeen–Cooper–Schrieffer (BCS) theory of superconductivity rests on a degeneracy: the fermions that pair are related by a symmetry, their Fermi surfaces coincide, and Cooper pairs condense at zero total momentum. How the pairing survives, or transforms, when the degeneracy is lifted, is a question posed across widely separated areas of physics. A mismatch between the Fermi surfaces frustrates the BCS kinematics and forces a competition between distinct ground states. Depending on the underlying system properties, this frustration is resolved via the gapless Sarma (breached-pair) state~\cite{Sarma1963,LiuWilczek2003,Gubankova2005}, the crystalline Fulde--Ferrell--Larkin--Ovchinnikov (FFLO) phase~\cite{FuldeFerrell1964,LarkinOvchinnikov1965,Matsuda2007,Kinnunen2018,Zheng2014}, or the macroscopic phase separation, with the homogeneous paired state surviving only below the Chandrasekhar--Clogston limit~\cite{Chandrasekhar1962,Clogston1962}. Nature poses this competition in the cores of compact stars, where two-flavor color-superconducting quark matter pairs $u$ and $d$ quarks whose Fermi surfaces are split by the strange-quark mass and by electric and color neutrality constraints~\cite{Gubankova2003,Reddy2005,Gubankova2006}. In ultracold quantum gas experiments the imbalance between two hyperfine populations is tuned directly~\cite{Zwierlein2006,Partridge2006,Schunck2007}; and it appears again in Dirac and Weyl materials, where an interaction-induced valley or orbital polarization plays the same role~\cite{Wehling2014,Nandkishore2016,Bednik2015,Yang2014,Meng2012,Bai2025}. The thermodynamics of mismatched pairing has accordingly been mapped in detail, non-relativistically~\cite{Gubbels2012} and relativistically~\cite{Boettcher2015,He2006,He2006b,He2009,Liao2003,Huang2007}.

When the Fermi-surface splitting originates from an underlying inequivalence between the species, the degeneracy is lifted explicitly. A Zeeman field or a prepared population imbalance imposes the mismatch; in neutral quark matter it descends from the quark masses and electric charges, its magnitude set self-consistently by the neutrality conditions together with the gap equation~\cite{ShovkovyHuang2003}. The question posed in the literature has therefore remained the same throughout: given such a mismatch, imposed externally or fixed by constraints, does pairing persist? Here we pose the opposite question: can a Fermi-surface mismatch instead be an emergent, dynamical feature of a stable superconducting ground state, arising from the spontaneous breaking of an exact internal symmetry that treats the two fermion species identically?

In this letter, we study two massless Dirac fermions that form an exact $\mathrm{SU}(2)$ flavor doublet, coupled to a flavor-triplet vector boson with a quartic self-interaction of strength $\zeta$~\cite{Schmitt2010}. We find that above a critical value of $\zeta$ the temporal component of the vector field condenses along a preferred flavor direction, spontaneously breaking $\mathrm{SU}(2)\to\mathrm{U}(1)$. Crucially, this condensate acts as an effective flavor-dependent chemical potential that self-consistently splits the two Fermi surfaces by mismatch $\dmu$. Because time reversal acts within each flavor rather than exchanging the two, the flavor charge,  and hence the condensate that couples to it, is even under $TR$. The mismatched ground state therefore preserves $TR$, unlike the time-reversal-odd polarization of an itinerant (Stoner) ferromagnet or a valley-polarized Dirac material, where the splitting, though spontaneous, breaks $TR$. To the best of our knowledge, this is the first instance in which a Fermi-surface mismatch between the pairing species is generated by the spontaneous breaking of an exact internal symmetry, while preserving time reversal, rather than being seeded by an explicit or environmental asymmetry.

The dynamical origin restructures the paired state. Introducing a scalar-mediated BCS interaction on the split surfaces and computing the free energy within the two-particle-irreducible (2PI) effective action framework~\cite{CJT1974,Schmitt2015}, we obtain the gap equation, the quasiparticle spectrum, and an analytic condition for the favorability of the mismatched superfluid in which $\dmu$ is no longer an independent control parameter, but is dynamically locked to the pairing gap $\Delta$ by the symmetry-breaking coupling $\zeta$. Consequently, the Chandrasekhar--Clogston boundary, traditionally a line in the $(\delta\mu, \Delta)$ plane, is reorganized into a stability surface spanned by $\zeta$, the pairing coupling $G$, and the net chemical potential $\mu$. Finally, a Gaussian fluctuation analysis yields the expected Goldstone mode of broken $\mathrm{U}(1)$ symmetry, establishing this mismatched superfluid as a locally stable minimum of the free energy.

\section{Model and Spontaneous Fermi Surface Mismatch}

We consider two massless fermion species in a Dirac doublet $\psi = (\psi_1, \psi_2)^T$ coupled to an $\mathrm{SU}(2)$ triplet vector field $\bm{b}_\mu$ with field strength $b_{\mu\nu} = \partial_\mu b_\nu - \partial_\nu b_\mu$. The action~\cite{Schmitt2010} is given by,
\begin{align}
S = \int d^4 x \Big[ \bar{\psi}\, i\slashed{\partial}\, \psi &- \tfrac{1}{4} \bm{b}_{\mu\nu} \cdot \bm{b}^{\mu\nu} + \tfrac{1}{2} m_b^2\, \bm{b}_\mu \cdot \bm{b}^\mu \nonumber \\
&- \tfrac{g}{2} \bar{\psi} \gamma^\mu \bm{\tau} \cdot \bm{b}_\mu \psi - \tfrac{\zeta g^4}{24} (\bm{b}_\mu \cdot \bm{b}^\mu)^2 \Big],
\label{eq:action}
\end{align}
with $\bm{\tau}$ the Pauli matrices in flavor space, $g$ the vector coupling, and $m_b$ the boson mass~\cite{Walecka1974,Bodmer1991,MuellerSerot1996}. The action is invariant under the global vector symmetry $\mathrm{SU}(2)_V$ ($\psi \to U\psi$, $\bm{\tau}\cdot\bm{b}_\mu \to U\,\bm{\tau}\cdot\bm{b}_\mu\,U^\dagger$) and under the $\mathrm{U}(1)$ phase rotation $\psi \to e^{i\theta}\psi$.

Under the relativistic mean-field ansatz $\langle b_\mu^a \rangle = \bbar\, \delta_\mu^0\, \delta^{a3}$, the vector field acquires a uniform expectation value along the temporal Lorentz component and the third flavor direction. This background field couples to the fermions as a flavor-dependent chemical potential shift, yielding effective chemical potentials $\mu_\pm = \mu \pm \frac{g}{2}\bar{b}$ and massless linear dispersions $\varepsilon_\pm(\bm{k}) = |\bm{k}| \pm \frac{g}{2}\bar{b}$, with $\mu$ the chemical potential of the conserved $\mathrm{U}(1)$ charge. Evaluated in the degenerate regime $\beta\mu \gg 1$ ($\beta \equiv 1/T$, with $T$ the temperature), the partition function can be computed in closed form. Expressing the resulting free-energy density directly via the Fermi-surface mismatch parameter $\delta\mu \equiv \frac{g}{2}\bar{b}$, we obtain: 
\begin{equation}
f = \frac{g^4}{24}\Big(\zeta - \frac{1}{4\pi^2}\Big)\bbar^4 - \bbar^2 \Big(\frac{m_b^2}{2} + \frac{g^2 \mu^2}{4\pi^2}\Big) - \frac{\mu^4}{6\pi^2}.
\label{eq:free_energy}
\end{equation}

The quartic coefficient changes sign at a critical self-interaction threshold $\zeta_c = 1/(4\pi^2)$. Below this threshold ($\zeta < \zeta_c$), the unique minimum sits at $\delta\mu = 0$: both fermion species share a common Fermi surface, and the $\mathrm{SU}(2)_V$ symmetry remains unbroken. Above the threshold ($\zeta > \zeta_c$), the system minimizes its free energy by developing a nonzero expectation value:
\begin{equation}
|\delta\mu| = \sqrt{\frac{3}{2}\, \frac{m_b^2/g^2 + \mu^2 / (2\pi^2)}{\zeta - 1/(4\pi^2)}},
\label{eq:bbar}
\end{equation}
spontaneously breaking $\mathrm{SU}(2)_V \to \mathrm{U}(1)$ and splitting the system into two distinct Fermi surfaces with radii $k_F^\pm = \mu \mp |\delta\mu|$. Both Fermi spheres remain physical ($k_F^- > 0$) provided $\mu^2 > 3 m_b^2 / [2g^2(\zeta - 1/\pi^2)]$ for $\zeta > 1/\pi^2$; within the intermediate coupling window $1/(4\pi^2) < \zeta < 1/\pi^2$, a single polarized Fermi sphere persists.

The Fermi surface asymmetry is driven by the self-interaction dynamics rather than being an artifact of the variational ansatz; we verify this by examining the theory at $\zeta = 0$. Here, the stationarity condition reduces to $m_b^2 \bar{b} = (g/2)(n_1 - n_2)$, where $n_{1,2} = \mu_\mp^3 / (3\pi^2)$ represent the respective flavor number densities. Algebraically, this coupled system yields no stable asymmetric solutions for $\bar{b}\neq0$; at $\bar{b}=0$, the unpolarized state $n_1 = n_2$ is unique, demonstrating that a stable Fermi-surface mismatch cannot occur without the stabilizing feedback of the quartic self-interaction ($\zeta > \zeta_c$), which is the isovector counterpart of the $\omega$ self-coupling that governs the high-density equation of state in mean-field models~\cite{Bodmer1991,MuellerSerot1996}.

\begin{figure}[t]
 \includegraphics[height = 5cm, width=\columnwidth]{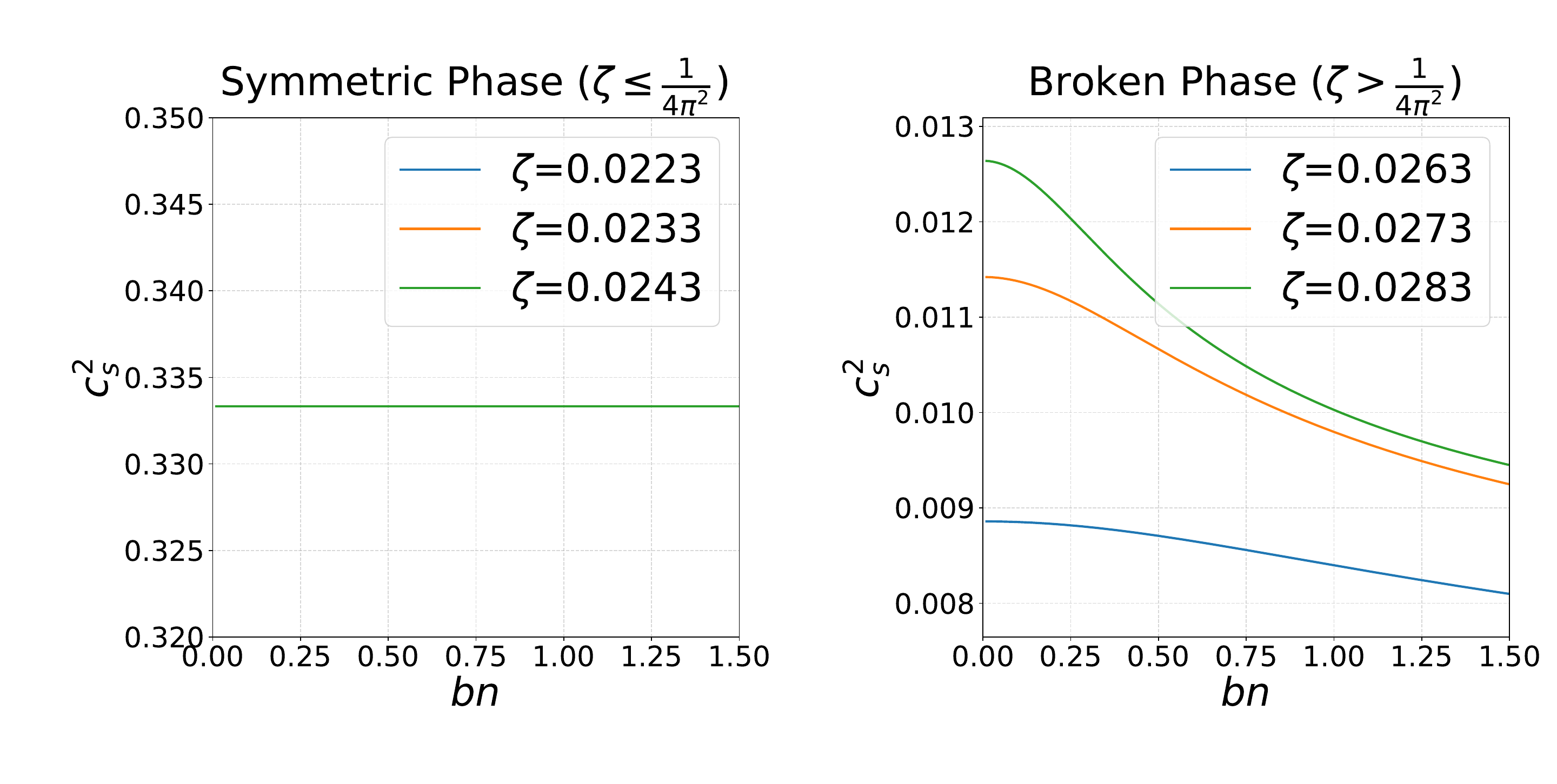}
\centering
\caption{Speed of sound $\cs$ as a function of dimensionless number density $bn$ for representative values of $\zeta$.}
\label{fig:speed_of_sound}
\end{figure}

This restructuring of the Fermi surfaces imprints a distinct thermodynamic signature on the speed of sound, $c_s^2 = dP/d\rho$. In the symmetry-preserving phase, the system retains the conformal value $c_s^2 = 1/3$ characteristic of massless fermions at a common chemical potential. Upon crossing $\zeta_c$, the onset of the Fermi surface mismatch drives $c_s^2$ along a nontrivial curve as a function of the dimensionless number density $bn$, with $b \equiv 3\pi^2/m_b^3$ (shown in Fig.~\ref{fig:speed_of_sound}). Because the vector condensate $\bar{b}$ is even under time reversal, $TR$ is preserved across the transition, unlike the field-induced imbalances that break it.

\section{Cooper Pairing with Dynamical Mismatch}
We introduce a Cooper-pairing channel on the split Fermi surfaces by coupling the fermions to a real scalar field $\varphi$ of mass $M$ via $\mathcal{L}_{\mathrm{int}} = -g_\varphi \bar{\psi}\psi\,\varphi$~\cite{Schmitt2015,Ohsaku2004}, which in the heavy-boson limit ($M \to \infty$) reduces to a four-fermion effective contact interaction with coupling $G = g_\varphi^2/(2M^2)$. We select an even-parity spin-singlet pairing channel between the two flavors, characterized by a pairing gap $\Delta$ and a gap matrix $\Phi^\pm = \pm \Delta\, \tau_1 \gamma_5$. This pairing explicitly breaks the global $\mathrm{U}(1)$ phase symmetry ($\Phi^\pm \to e^{\pm 2i\alpha}\Phi^\pm$). Because the broken symmetry is global rather than gauged, the condensate describes a relativistic superfluid, whose phase mode is a physical, massless Goldstone boson.

Diagonalizing the inverse fermion propagator in the combined Dirac $\otimes$ Nambu--Gorkov $\otimes$ flavor space yields a quasiparticle spectrum featuring eight poles at
\begin{equation}
E_{\bm{k}}^{e,\pm} = \varepsilon_{\bm{k}}^e \pm \delta\mu, \quad \varepsilon_{\bm{k}}^e = \sqrt{(\mu - ek)^2 + \Delta^2},
\label{eq:poles}
\end{equation}
for $e = \pm 1$ (particle and antiparticle branches) and their negatives. The mismatch $\delta\mu$ rigidly splits each branch. For $\delta\mu > \Delta$, the lower quasiparticle excitation branch crosses zero and becomes gapless, defining the breached-pair (Sarma) regime~\cite{Sarma1963,LiuWilczek2003}. In the weak-coupling limit ($\Delta \ll \Lambda \ll \mu$), with $\Lambda$ the pairing cutoff around the Fermi surface, the gap equation reduces to the BCS form $\Delta_0 \simeq 2\Lambda\, e^{-2\pi^2/(G\mu^2)}$~\cite{BCS1957}.

The total free-energy density of the system is computed within the Cornwall--Jackiw--Tomboulis (2PI) effective action framework~\cite{CJT1974,Schmitt2015}, which self-consistently incorporates Gaussian-level fluctuations at one-loop order:
\begin{equation}
\Omega = -\frac{1}{2}\frac{T}{V}\sum_K \mathrm{Tr}\log\frac{\mathcal{G}^{-1}}{T} + \frac{1}{4}\frac{T}{V}\sum_K \mathrm{Tr}[1 - \mathcal{G}_0^{-1}\mathcal{G}],
\label{eq:CJT}
\end{equation}
here $\mathcal{G}$ and $\mathcal{G}_0$ represent the full and bare fermion propagators in the combined space, respectively. 
At zero temperature, the pairing-sector contribution to the free-energy difference relative to the normal state is isolated as $\Delta\Omega_{\text{pair}} \equiv \Omega_{\text{superfluid}} - \Omega_{\text{normal}}\big|_{\bar{b}=0}$, which evaluates to (Supplemental Material~\cite{SM} Sec.~S4):
\begin{equation}
\Delta\Omega_{\text{pair}} \simeq \frac{\mu^2 \delta\mu^2}{\pi^2} - \frac{\mu^2 \Delta^2}{2\pi^2},
\label{eq:delta_omega}
\end{equation}
for $0 < \delta\mu < \Delta$. The first term in Eq.~\eqref{eq:delta_omega} represents the kinetic energy cost of maintaining the Fermi-surface splitting, while the second accounts for the pairing condensation energy. Their competition recovers the Chandrasekhar--Clogston thermodynamic stability bound $\delta\mu < \Delta/\sqrt{2}$~\cite{Chandrasekhar1962,Clogston1962}, beyond which a uniform superfluid becomes unstable.

The structure of the mechanism appears once this pairing energy is combined with the symmetry-breaking vector sector. We split the normal-state free energy of Eq.~\eqref{eq:free_energy} into mismatch-independent and mismatch-dependent parts, the latter defining the vector contribution $\Delta\Omega_{\text{vec}} \equiv \frac{2}{3}(\zeta - \frac{1}{4\pi^2})\delta\mu^4 - \frac{2m_b^2}{g^2}\delta\mu^2 - \frac{\mu^2\delta\mu^2}{\pi^2}$ (Supplemental Material~\cite{SM} Sec.~S5). Forming the complete free-energy shift relative to the unpolarized normal state, $\Delta\Omega \equiv \Delta\Omega_{\text{vec}} + \Delta\Omega_{\text{pair}}$, the cross-terms proportional to $\mu^2\delta\mu^2$ cancel identically, yielding:
\begin{equation}
\Delta\Omega = \frac{2}{3}\Big(\zeta - \frac{1}{4\pi^2}\Big)\delta\mu^4 - \frac{2m_b^2}{g^2}\delta\mu^2 - \frac{\mu^2\Delta^2}{2\pi^2}.
\label{eq:delta_omega_1}
\end{equation}
Because the mismatch $\delta\mu$ is constrained by the underlying coupling parameters via the classical field equations [Eq.~\eqref{eq:bbar}], the mismatched superfluid is energetically preferred over the unpolarized state only when
\begin{equation}
\bigg|\frac{\delta\mu}{\mu}\bigg| < \sqrt{\frac{3}{2\pi^2(\zeta - 1/(4\pi^2))}}.
\label{eq:favorability}
\end{equation}
This condition relates the self-interaction coupling $\zeta$ that generates the mismatch to the thermodynamic stability of the paired state.

\begin{figure}[t]
\includegraphics[width=\columnwidth]{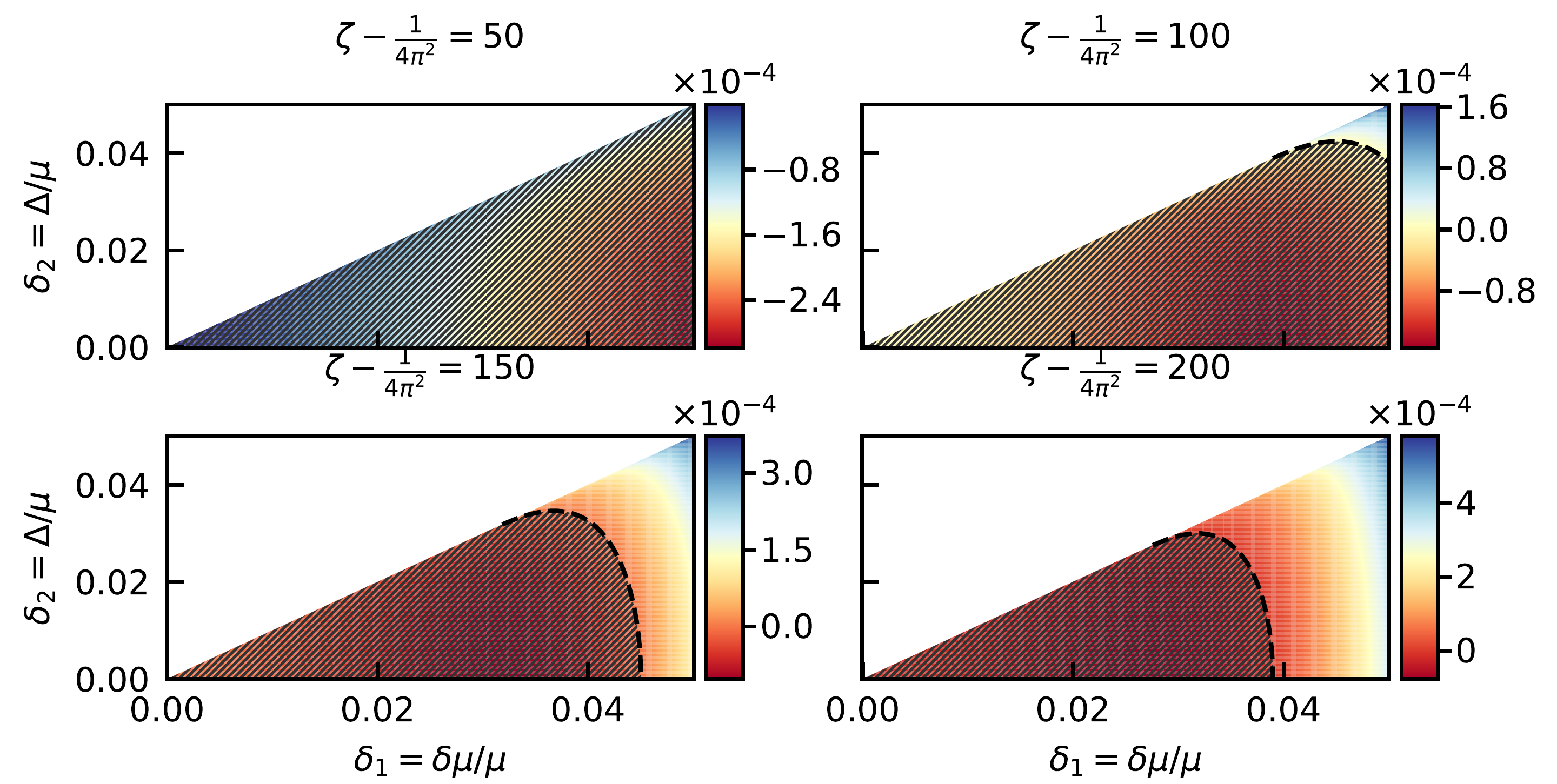}
\centering
\caption{The mismatch-dependent free-energy density correction $\Delta\Omega/\mu^4$ evaluated for representative self-interaction strengths $\zeta$. The solid red curve maps the self-consistent boundary enclosing the thermodynamically stable mismatched superfluid phase.}
\label{fig:color_plot}
\end{figure}

Combined with the Chandrasekhar--Clogston bound, Eq.~\eqref{eq:favorability} maps the phase-stability region of the mismatched superfluid in the three-dimensional parameter space $(\zeta, G, \mu)$ (Fig.~\ref{fig:color_plot}). As $\zeta$ increases beyond $1/\pi^2$, the stable region contracts because a stronger self-interaction drives a wider splitting that eventually overwhelms the pairing condensation energy. The promotion of the static Chandrasekhar--Clogston boundary from a simple line in the $(\delta\mu, \Delta)$ plane to a dynamical surface spanned by the Lagrangian's fundamental parameters $(\zeta, G, \mu)$ is the central result of this Letter. Within this stable region, the homogeneous superfluid remains fully gapped: the Chandrasekhar--Clogston bound enforces $\delta\mu < \Delta/\sqrt{2}$, i.e., the gapless Sarma window ($\delta\mu > \Delta$) opens only beyond this boundary. Past this thermodynamic threshold, the uniform phase yields to inhomogeneous configurations, phase separation, or breached-pair instabilities~\cite{LiuWilczek2003,Forbes2005}; the local dynamical stability of the uniform phase up to this boundary is verified via the collective mode analysis detailed below.

\section{Fluctuations and Collective Modes}
The mean-field analysis identifies the mismatched superfluid as a stationary point of the effective action. Its stability against order-parameter fluctuations, alongside the spectrum of collective modes it supports, is determined by a Gaussian expansion within the 2PI framework. Allowing the pairing gap to acquire spatial and temporal variations, we parameterize the complex fluctuation field as
\begin{equation}
\Delta(X) = \Delta\, e^{2i\bm{q}\cdot\bm{x}} + \eta(X),
\label{eq:gap_fluct}
\end{equation}
where $\Delta$ denotes the uniform mean-field expectation value, $\bm{q}$ represents a superflow momentum necessary for describing superfluid hydrodynamics, and $\eta(X)\in\mathbb{C}$ captures the space-time fluctuations around the condensate. A field redefinition of the fermion doublet $\psi'(X) = e^{i\bm{q}\cdot\bm{x}}\psi(X)$ and the fluctuation field $\eta'(X) = e^{-2i\bm{q}\cdot\bm{x}}\eta(X)$ eliminates explicit coordinate dependence from the kinetic terms, yielding the fluctuation Lagrangian $\mathcal{L}_{\mathrm{MF}+\mathrm{fl}} = \bar{\psi}'(\mathcal{G}^{-1} + h)\psi' - \Delta^2/G - (\Delta/G)(\eta'+\eta'^*) - |\eta'|^2/G$. Here, the fluctuation vertex $h$ is linear in $\eta'$ and $\eta'^*$, while the background inverse propagator $\mathcal{G}^{-1}$ carries the superflow via diagonal $\pm\bm{\gamma}\cdot\bm{q}$ Doppler shifts.
Integrating out the bilinear fermion fields and expanding the functional trace $\mathrm{Tr}\log(\mathcal{G}^{-1}+h)$ to quadratic order decomposes the effective action as $S_E = S^{(0)} + S^{(1)} + S^{(2)}$. The zeroth-order term $S^{(0)}$ recovers the fermionic mean-field action, the linear fluctuation term $S^{(1)}$ vanishes by the gap equation, and the quadratic contribution $S^{(2)}$, expressed in terms of the real and imaginary components $\eta' = (\eta_1',\eta_2')$, reads:
\begin{equation}
S^{(2)} = \frac{1}{2}\sum_K \eta'(K)\,\frac{D^{-1}(K)}{T^2}\,\eta'^{\dagger}(K),
\label{eq:S2}
\end{equation}
where $K \equiv (\omega_n, \bm{k})$ is the bosonic 4-momentum. The inverse bosonic propagator matrix is given by:
\begin{equation}
D^{-1}(K) = \begin{pmatrix} \tfrac{1}{G} - \bar{\Pi}(K) + \Sigma(K) & i\,\delta\Pi(K) \\[3pt] -i\,\delta\Pi(K) & \tfrac{1}{G} - \bar{\Pi}(K) - \Sigma(K) \end{pmatrix}.
\label{eq:Dinv}
\end{equation}
The symmetric ($\bar{\Pi}$) and antisymmetric ($\delta\Pi$) polarization functions are constructed from the normal ($G^\pm$) and anomalous ($F^\pm$) fermion propagators evaluated within the split background as $\bar{\Pi}(K) \equiv \tfrac{1}{2}[\Pi(K)+\Pi(-K)]$ and $\delta\Pi(K) \equiv \tfrac{1}{2}[\Pi(K)-\Pi(-K)]$, with $\Sigma(K)$ isolating the purely anomalous loop contribution. Their explicit momentum-space loop integrals are compiled in the End Matter, and the analytic Matsubara summations are detailed in the Supplemental Material~\cite{SM} Sec.~S6.

\begin{figure}[t]
\includegraphics[height = 5cm, width=\columnwidth]{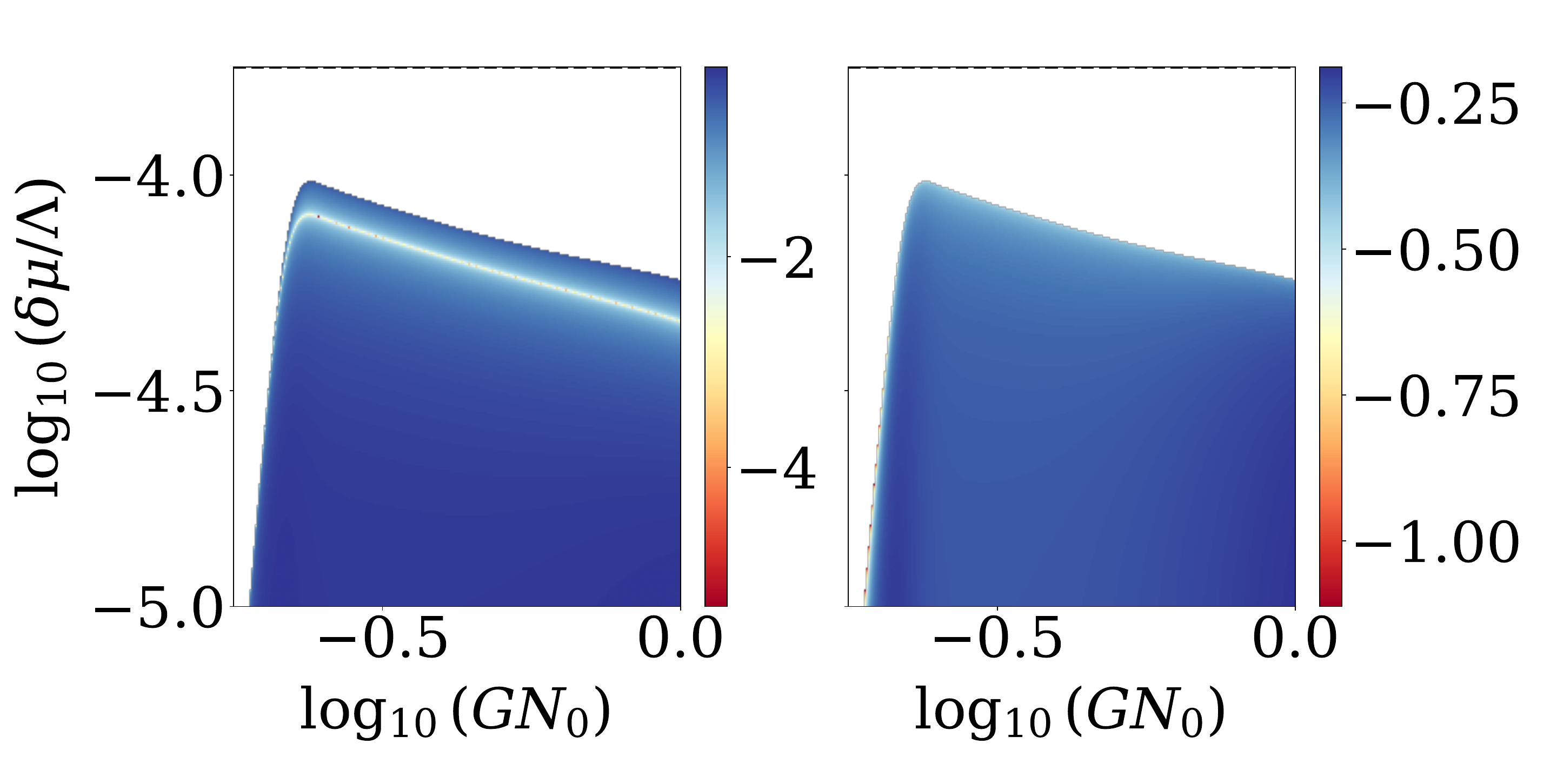}
\centering
\caption{Goldstone mode velocities $u_+$ (left) and $u_-$ (right) in $\log_{10}$ scale as functions of 
$G\Nz$ and $\dmu/\Lambda$.}
\label{fig:goldstone}
\end{figure}

The collective modes follow from $\det[D^{-1}(K)] = 0$ in the long-wavelength limit. The spontaneously broken global $\mathrm{U}(1)$ symmetry yields a gapless Goldstone mode with asymmetric relativistic dispersion $\omega = u_\pm|\bm{q}|$, whose velocities depend on the dimensionless coupling $G\Nz$ and on $\dmu/\Lambda$ (Fig.~\ref{fig:goldstone}), where $\Nz = \mubar^2/\pi^2$ is the density of states at the average Fermi surface and $\Lambda$ the ultraviolet cutoff around it. Concurrently, the amplitude (Higgs-like) fluctuation mode remains gapped at a threshold of $\omega \sim 2\Delta$. The eigenvalues of $D^{-1}(K)$ remain non-negative in the long-wavelength limit throughout the parameter domain bounded by the favorability condition [Eq.~\eqref{eq:favorability}]. At the quadratic order, this establishes the uniform, split superfluid as a stable local minimum of the free energy: the interaction-driven Fermi-surface asymmetry is stable against collective instabilities, so a paired state can coexist with a spontaneously broken flavor symmetry. Its relation to the gapless-state instabilities of mismatched pairing (the Sarma, negative-superfluid-density, and chromomagnetic instabilities) is discussed in the End Matter.

\section{Discussion}
In this framework, the Fermi-surface imbalance is an emergent output of the underlying dynamics. Given the vector coupling $g$, the self-interaction strength $\zeta$, the scalar pairing coupling $G$, and the total chemical potential $\mu$, the ground state develops an internal flavor splitting $\delta\mu$ and a pairing gap $\Delta$ simultaneously, their ratio fixed by the self-consistency of the field equations [Eqs.~\eqref{eq:bbar} and~\eqref{eq:favorability}]. From this follow two structural consequences. First, the stability of the paired state is governed by Eq.~\eqref{eq:favorability}, which links the symmetry-breaking coupling to the pairing sector. The traditional Chandrasekhar--Clogston boundary, typically a static line in the $(\delta\mu, \Delta)$ plane for an imposed imbalance~\cite{Gubankova2006,Gubbels2012,Boettcher2015}, becomes a surface in the $(\zeta, G, \mu)$ parameter space whose geometry is shaped by the interaction channels. Second, because the order parameter is even under time reversal, the mismatched superfluid preserves $TR$ and remains free of the $TR$-odd response coefficients that accompany field-induced population imbalances.

This mechanism relies on a generic low-energy structure: two Dirac species, an attractive pairing channel, and a flavor symmetry whose spontaneous breaking drives the asymmetry via a Stoner-type instability. A short-range interaction in the flavor (spin, valley, or isospin) channel polarizes two otherwise degenerate species once a critical coupling threshold is crossed. This configuration represents the relativistic, paired counterpart to itinerant ferromagnetism in two-component repulsive Fermi gases~\cite{Jo2009}, spontaneous valley polarization in Dirac materials~\cite{ValleyPol2015,TBG2021}, and isovector mean-field condensation in dense nuclear matter. Each physical domain provides a clear realization. In two-flavor color-superconducting quark matter, the $u$--$d$ Fermi-surface splitting controlling the competition between fully gapped, gapless~\cite{ShovkovyHuang2003}, and crystalline phases is conventionally fixed by the strange-quark mass and neutrality constraints~\cite{Gubankova2003,Reddy2005,Gubankova2006}. An intrinsic, interaction-driven splitting tied to the pairing dynamics could shift the phase boundaries in compact-star interiors. In ultracold gases, this scenario is directly testable: a two-component Fermi gas trapped in a Dirac-like (e.g., honeycomb) optical lattice, featuring a Stoner-type on-site repulsion alongside a Feshbach-tuned attraction, would manifest the mismatch as a spontaneous transition out of an unpolarized state. This would be signaled by the simultaneous, abrupt onset of population imbalance and superfluid coherence rather than a continuously tuned external asymmetry. In Dirac and Weyl materials, an interaction-induced valley or orbital polarization acts as the flavor mismatch, where the same favorability surface and collective-mode structure dictate the superconductivity.
One structural distinction is worth noting: when the two species form a time-reversal-conjugate pair, as the valleys of a Dirac material do, the spontaneous polarization breaks $TR$, whereas for an internal flavor doublet the condensate is $TR$-even. The same mechanism describes both: $TR$ is broken when the polarized degree of freedom is itself $TR$-odd, and preserved when it is not. The dynamically generated mismatch places the system in proximity to finite-momentum (FFLO) instabilities~\cite{Zheng2014,Kinnunen2018,Koponen2007}; whether the self-consistent feedback between $\bbar$ and $\Delta$ suppresses or promotes inhomogeneous pairing calls for a beyond-Gaussian treatment: an FRG analysis extending Ref.~\cite{Boettcher2015} to dynamical imbalance, or a Ginzburg–Landau expansion at finite pairing momentum. A finite-temperature signature of the mechanism is given in the End Matter.

\begin{acknowledgments}
\paragraph*{\textbf{Acknowledgments}}
S.~S. acknowledges support from the Anusandhan National Research Foundation (Department of Science and Technology) Govt. of India, under grant no. CRG/2023/007457 and an internal grant from the Indian Institute of Science Education and Research, Tirupati. \\

Note added.—While completing this work we became aware of a contemporaneous study~\cite{Yang2026} in which an intrinsic electron–hole Fermi-surface mismatch drives spontaneous rotational symmetry breaking and nematic superconductivity. There the mismatch is the input and the spontaneously broken symmetry is spatial, a mirror image of the internal-symmetry, time-reversal-preserving mechanism studied here.
\end{acknowledgments}

\onecolumngrid
\bigskip
\begin{center}
\textbf{\large End Matter}
\end{center}
\twocolumngrid

\section{Explicit propagators and polarization functions}
\label{em:propagators}

This appendix records the explicit momentum-space quantities entering the inverse bosonic propagator $D^{-1}(K)$ of Eq.~\eqref{eq:Dinv}; the remaining Matsubara summations and the small-momentum extraction of the mode velocities are given in the Supplemental Material~\cite{SM} Sec.~S6. Setting the superflow to zero, $\bm{q} = 0$, the normal and anomalous fermion propagators in the mismatched background are
\begin{widetext}
\begin{equation}
{\scriptstyle
G^{\pm}(K) = \sum_{e}\gamma^{0}\Lambda_{k}^{\mp e}
\begin{bmatrix}
\dfrac{k_{0} \mp (\mu_{2} - ek)}{(k_{0} \pm \dmu)^{2} - (\varepsilon_{k}^{e})^{2}} & 0\\[10pt]
0 & \dfrac{k_{0} \mp (\mu_{1} - ek)}{(k_{0} \mp \dmu)^{2} - (\varepsilon_{k}^{e})^{2}}
\end{bmatrix},
\qquad
F^{\pm}(K) = \pm\sum_{e}\Delta\gamma_{5}\Lambda_{k}^{\mp e}
\begin{bmatrix}
0 & \dfrac{1}{(k_{0} \mp \dmu)^{2} - (\varepsilon_{k}^{e})^{2}}\\[10pt]
\dfrac{1}{(k_{0} \pm \dmu)^{2} - (\varepsilon_{k}^{e})^{2}} & 0
\end{bmatrix},
}
\label{eq:GF_propagators}
\end{equation}
\end{widetext}
where $\Lambda_k^{\pm e}$ are the energy projectors, $e=\pm$ labels particles and antiparticles, and $\varepsilon_k^e = \sqrt{(\xi_k^e)^2 + \Delta^2}$ with $\xi_k^e = k - e\mubar$; $\mu_{1,2} = \mu \mp \tfrac{g}{2}\bbar$ are the two flavor chemical potentials. The gap parameter satisfies
\begin{equation}
\frac{\Delta}{G} = \sum_{e=\pm}\int\frac{d^3p}{(2\pi)^3}\,\frac{\Delta}{2\varepsilon_p^e}\big[1 - 2f(\varepsilon_p^e)\big],
\label{eq:gap_eqn}
\end{equation}
with $f$ the Fermi--Dirac distribution. The polarization functions are defined by $\Pi^\pm(K) = \tfrac{1}{2}\tfrac{T}{V}\sum_P \mathrm{Tr}[G^\pm(P)\gamma_5\tau_1 G^\mp(P+K)\gamma_5\tau_1]$ and $\Sigma^\pm(K) = \tfrac{1}{2}\tfrac{T}{V}\sum_P \mathrm{Tr}[F^\pm(P)\gamma_5\tau_1 F^\pm(P+K)\gamma_5\tau_1]$. Writing $Q \equiv P+K$, $\varepsilon_1 \equiv \varepsilon_p^{e_1}$, $\varepsilon_2 \equiv \varepsilon_q^{e_2}$, $\xi_1 \equiv p - e_1\mubar$, $\xi_2 \equiv q - e_2\mubar$, and using the Dirac trace $-\mathrm{Tr}[\gamma^0\Lambda_p^{-e_1}\gamma_5\gamma^0\Lambda_q^{e_2}\gamma_5] = 1 + e_1 e_2\,\hat{\bm{p}}\cdot\hat{\bm{q}}$, the normal polarization evaluates to
\begin{widetext}
\begin{equation}
\Pi(K) = -\frac{1}{2}\frac{T}{V}\sum_P\sum_{e_1,e_2}(1 + e_1 e_2\,\hat{\bm{p}}\cdot\hat{\bm{q}})
\left[
\frac{(p_0 + \dmu) + e_1\xi_1}{(p_0+\dmu)^2 - \varepsilon_1^2}\cdot\frac{(q_0 - \dmu) - e_2\xi_2}{(q_0-\dmu)^2 - \varepsilon_2^2}
+ \frac{(p_0 - \dmu) + e_1\xi_1}{(p_0-\dmu)^2 - \varepsilon_1^2}\cdot\frac{(q_0 + \dmu) - e_2\xi_2}{(q_0+\dmu)^2 - \varepsilon_2^2}
\right].
\label{eq:Pi_explicit}
\end{equation}
\end{widetext}
Using $G^+(-K) = -G^-(K)$ and $F^+(-K) = -F^-(K)$, which follow directly from Eq.~\eqref{eq:GF_propagators}, one finds $\Pi^+(K) = \Pi^-(-K)$ and $\Sigma^+(K) = \Sigma^-(K)$, so that all entries of $D^{-1}(K)$ are expressed through the combinations $\bar{\Pi}(K) = \tfrac{1}{2}[\Pi(K)+\Pi(-K)]$, $\delta\Pi(K) = \tfrac{1}{2}[\Pi(K)-\Pi(-K)]$, and $\Sigma(K)$, as in Eq.~\eqref{eq:Dinv}. Integrating out the fluctuations, the free-energy density acquires the corresponding Gaussian correction
\begin{equation}
\Omega = \frac{\Delta^2}{G} - \frac{1}{2}\frac{T}{V}\sum_K \log\frac{\mathcal{G}^{-1}}{T} + \frac{1}{2}\frac{T}{V}\sum_K \mathrm{Tr}\log\frac{D^{-1}(K)}{T^2},
\label{eq:Omega_fluct}
\end{equation}
whose first two terms reproduce the mean-field result and whose last term carries the collective-mode contribution. The poles of $D(K) = [D^{-1}(K)]^{-1}$ at small $|\bm{q}|$ give the Goldstone velocities of Fig.~\ref{fig:goldstone}, and the non-negativity of the eigenvalues of $D^{-1}(K)$ throughout the region of Eq.~\eqref{eq:favorability} establishes the stability quoted in the main text.

\section{Stability and the gapless instabilities}
The instabilities that historically afflict mismatched pairing are properties of the gapless regime $\dmu > \Delta$: the thermodynamic Sarma instability~\cite{Sarma1963,LiuWilczek2003}, the negative superfluid density of the interior-gap state~\cite{WuYip2003}, and the chromomagnetic instability (an imaginary Meissner mass) of gapless color superconductors~\cite{HuangShovkovy2004}. In each, the gapless branch destabilizes the homogeneous state toward inhomogeneous (FFLO) order~\cite{Forbes2005}. The present uniform phase avoids this regime: the Chandrasekhar--Clogston bound confines the stable region to $\dmu < \Delta/\sqrt{2}$, where the spectrum is fully gapped. Because the broken $\mathrm{U}(1)$ is global rather than gauged, the relevant diagnostic is not a Meissner mass but the superfluid stiffness, encoded in the $q^2$ coefficient of the Goldstone dispersion $\omega = u_\pm|\bm{q}|$. The reality of $u_\pm$ throughout the favorability region [Eq.~\eqref{eq:favorability}] is therefore a positive superfluid density, and the uniform phase is stable there.

\section{Heat capacity in the Sarma regime}
\label{em:heat_capacity}

\begin{figure}[t]
\includegraphics[width=\columnwidth]{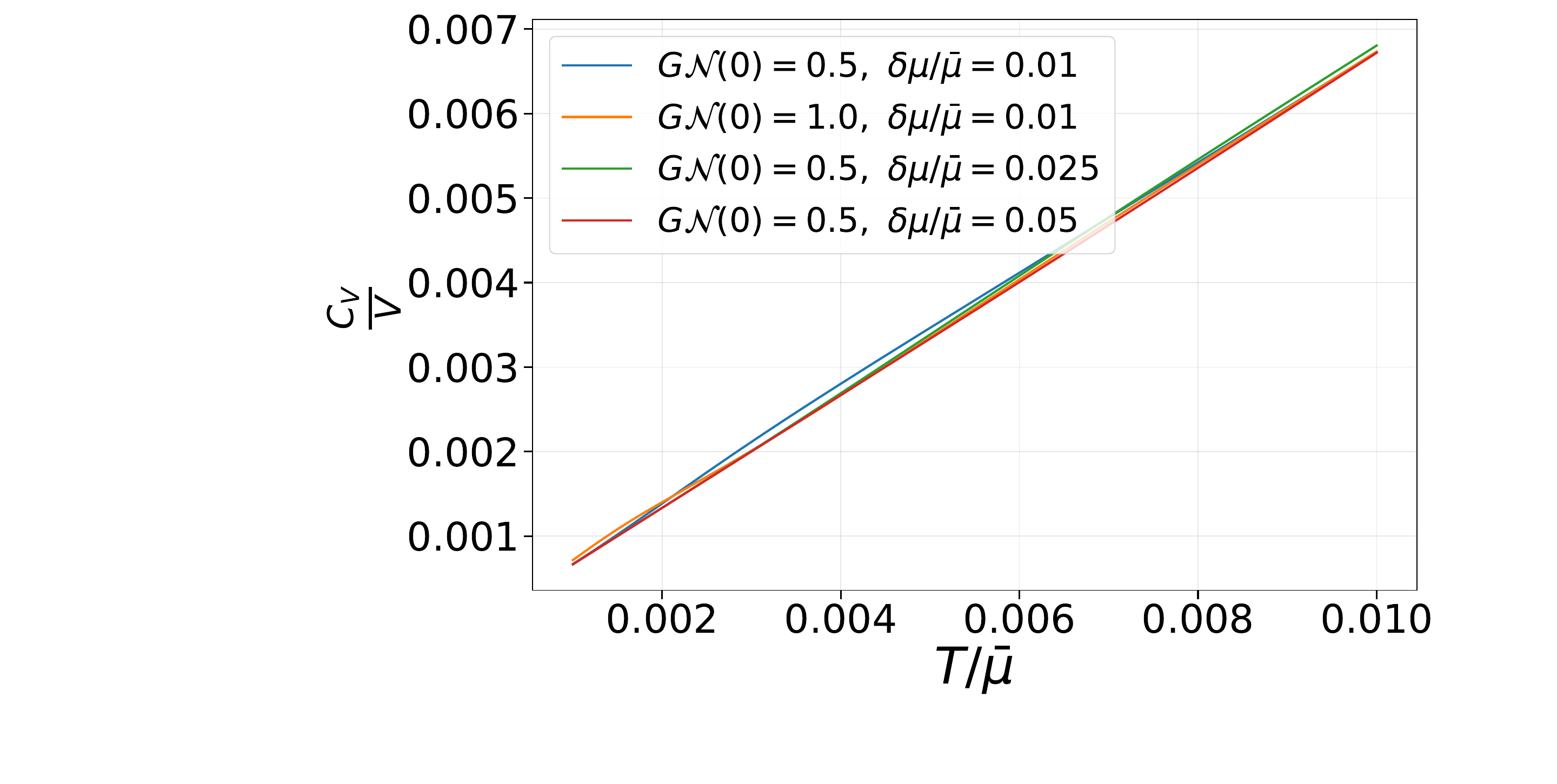}
\centering
\caption{Heat capacity $C_V$ as a function of temperature for representative values of $G\Nz$ and $\dmu/\mubar$ in the low-temperature regime.}
\label{fig:heat_cap}
\end{figure}

At finite temperature the partially gapless spectrum of the breached-pair (Sarma) regime, $\dmu\gtrsim\Delta$, dominates the low-temperature thermodynamics. With the two quasiparticle branches $E_{\bm{k}}^{\pm} = E_{\bm{k}}\pm\dmu$, $E_{\bm{k}} = \sqrt{(k-\mubar)^2+\Delta^2}$, the grand potential is
\begin{equation}
\Omega = \frac{\Delta^2}{G} - 2\sum_{\bm{k}}\Big[E_{\bm{k}} + T\log\big(1+e^{-E_{\bm{k}}^+/T}\big) + T\log\big(1+e^{-E_{\bm{k}}^-/T}\big)\Big],
\label{eq:Omega_T}
\end{equation}
and the heat capacity $C_V = -T\,\partial^2\Omega/\partial T^2$, evaluated with $\Delta \simeq 2\Lambda\,e^{-2\pi^2/(G\mubar^2)}$ and $\Nz = \mubar^2/\pi^2$ (Supplemental Material~\cite{SM} Sec.~S7), gives at low $T$ a linear-in-temperature behavior $C_V\propto T$, in contrast to the exponentially suppressed $C_V\sim e^{-\Delta/T}$ of a fully gapped BCS state (Fig.~\ref{fig:heat_cap}). This linear-$T$ behavior is a direct thermodynamic signature of the residual gapless fermionic excitations in the Sarma regime, and provides an observable diagnostic for systems realizing the dynamical-mismatch mechanism, whether in dense quark matter, where it would affect transport and cooling, or in ultracold Fermi gases, where the heat capacity is directly accessible.

\clearpage
\onecolumngrid
\setcounter{secnumdepth}{3}
\setcounter{section}{0}
\setcounter{equation}{0}
\setcounter{figure}{0}
\setcounter{table}{0}
\renewcommand{\thesection}{S\arabic{section}}
\renewcommand{\thesubsection}{S\arabic{section}.\arabic{subsection}}
\renewcommand{\theequation}{S\arabic{equation}}
\renewcommand{\thefigure}{S\arabic{figure}}
\renewcommand{\thetable}{S\arabic{table}}
 
\begin{center}
{\large\bfseries Supplemental Material:\\[2pt]
Dynamically Generated Fermi Surface Mismatch and Relativistic Superfluidity\\
in a Two-Component Massless Fermionic Theory}\\[6pt]
Susobhan Mandal and Sambuddha Sanyal
\end{center}
\vspace{1em}

\section{Detailed Model and Mean-Field Analysis}
\label{sec:S1}

The purpose of this Supplemental Material is to provide a comprehensive and technically detailed derivation of the results presented in the main article, entitled \emph{``Dynamically Generated Fermi Surface Mismatch and Relativistic Superfluidity in a Two-Component Massless Fermionic Theory''}. In the main text, several intermediate steps and algebraic developments have been omitted for the sake of clarity and conciseness; here, we systematically restore these details to ensure full transparency and reproducibility of the analysis. In particular, we present an explicit construction of the two-component massless fermionic framework, followed by a careful derivation of the effective low-energy action incorporating interaction effects. We further elaborate on the procedure leading to the dynamically generated Fermi surface mismatch, including all intermediate field redefinitions, Fourier-space manipulations, and regularization conventions employed in the computation.

Moreover, we provide a detailed account of the mean-field decomposition of the interacting theory, together with a rigorous derivation of the associated gap equations and their self-consistent solutions. Special attention is devoted to the stability analysis of the emergent relativistic superfluid phase, where all fluctuation corrections and consistency conditions are explicitly derived. This Supplemental Material is intended to serve as a self-contained reference, ensuring that all results quoted in the main text can be independently verified and that the underlying physical and mathematical structure of the theory is made fully explicit.

\subsection{Action and symmetries}

Here we consider a system consisting of two massless fermion species denoted by Dirac fields $\psi_{1}$ and $\psi_{2}$. The model that we start is described by the following action
\begin{equation}\label{action 1}
\begin{split}
S & = \int d^{4}x \Big[\bar{\psi}i\slashed{\partial}\psi - \frac{1}{4}\mathbf{b}_{\mu\nu}.\mathbf{b}^
{\mu\nu} + \frac{1}{2}m_{b}^{2}\mathbf{b}_{\mu}.\mathbf{b}^{\mu} - \frac{g}{2}\bar{\psi}\gamma^{\mu}
\boldsymbol{\tau}.\mathbf{b}_{\mu}\psi - \frac{\zeta g^{4}}{24}(\mathbf{b}_{\mu}.\mathbf{b}^{\mu})^{2}\Big],
\end{split} 
\end{equation} 
where
\begin{equation}\label{two component Dirac field}
\psi = \begin{bmatrix}
\psi_{1}\\
\psi_{2}
\end{bmatrix}, \ \bar{\psi} = \begin{bmatrix}
\bar{\psi}_{1} & \bar{\psi}_{2}
\end{bmatrix}
\end{equation}
and $\mathbf{b}_{\mu} = (b_{\mu}^{1}, b_{\mu}^{2}, b_{\mu}^{3})$ is a triplet vector field with the field strength tensor being
\begin{equation}\label{field strength tensor}
\mathbf{b}_{\mu\nu} = \partial_{\mu}\mathbf{b}_{\nu} - \partial_{\nu}\mathbf{b}_{\mu}.
\end{equation}
The above action (\ref{action 1}) is invariant under $SU(2)_{V}$ group defined by
\begin{equation}\label{symmetry}
\psi \rightarrow U\psi, \ \bar{\psi} \rightarrow \bar{\psi}U^{\dagger}, \ \boldsymbol{\tau}.\mathbf{b}_{\mu} \rightarrow U\boldsymbol{\tau}.\mathbf{b}_{\mu}U^{\dagger},
\end{equation}
where $U\in SU(2)$. In the action (\ref{action 1}), $\boldsymbol{\tau} = (\tau^{1}, \tau^{2}, \tau^{3})$ are the Pauli matrices. The action (\ref{action 1}) is invariant under $SU(2)_{V} \times U(1)$.

\subsection{Mean-field reduction}

Under the relativistic mean field approximation, we assume $\langle b_{\mu}^{a}\rangle = \bar{b}\delta_{\mu}^{0}\delta^{a3}$ which reduces the action (\ref{action 1}) to the following mean-field approximation
\begin{equation}\label{RMF action 1}
S = \int d^{4}x \Big[\bar{\psi}i\slashed{\partial}\psi + \frac{1}{2}m_{b}^{2}\bar{b}^{2} - \frac{g}{2}\bar{\psi}\gamma^{0}\tau^{3}\bar{b}\psi - \frac{\zeta g^{4}}{24}\bar{b}^{4}\Big].
\end{equation}
At finite temperature and chemical potential, the Euclidean action is given by
\begin{equation}\label{Euclidean action 1}
\begin{split}
S^{\beta} & = \int_{0}^{\beta}d\tau\int d^{3}x \ \Big[\bar{\psi}\left(\gamma^{0}(\partial_{\tau} - \mu) + i\gamma^{k}\partial_{k} + \frac{g}{2}\gamma^{0}\tau^{3}\bar{b}\right)\psi
 - \frac{1}{2}m_{b}^{2}\bar{b}^{2} + \frac{\zeta g^{4}}{24}\bar{b}^{4}\Big],
\end{split}
\end{equation}
and as a result, the Euclidean Lagrangian density for the two-component spinor field can be expressed as
\begin{equation}\label{Euclidean Lagrangian 1}
\mathcal{L}_{\psi}^{\beta} = \bar{\psi}\mathcal{M}\psi,
\end{equation}
where
\begin{equation}\label{expression of the quadration form}
\mathcal{M} = \begin{bmatrix}
\gamma^{0}(\partial_{\tau} - \mu_{-}) + i\gamma^{k}\partial_{k} & 0\\
0 & \gamma^{0}(\partial_{\tau} - \mu_{+}) + i\gamma^{k}\partial_{k}
\end{bmatrix},
\end{equation}
where $\mu_{\pm} = \mu \pm \frac{g}{2}\bar{b}$.

\subsection{Partition function in the degeneracy limit}

Under the degeneracy limit (or small temperature limit) such that $\beta\mu \gg 1$, the partition function defined by
\begin{equation}\label{partition function}
\mathcal{Z} = \int\mathcal{D}\bar{\psi}\mathcal{D}\psi \ e^{-S^{\beta}},
\end{equation}
can be computed exactly and expressed as
\begin{equation}\label{logarithm of partition function}
\begin{split}
\frac{1}{\beta V}\log\mathcal{Z} & = \frac{\mu_{+}^{4}}{12\pi^{2}} + \frac{\mu_{-}^{4}}{12\pi^{2}} 
 + \frac{1}{2}m_{b}^{2}\bar{b}^{2} - \frac{\zeta g^{4}}{24}\bar{b}^{4}\\
 & = \frac{\mu^{4}}{6\pi^{2}} + \bar{b}^{2}\left(\frac{m_{b}^{2}}{2} + \frac{g^{2}\mu^{2}}{4\pi^{2}}
 \right) + \left(\frac{g^{4}}{96\pi^{2}} - \frac{\zeta g^{4}}{24}\right)\bar{b}^{4}.
\end{split}
\end{equation}
The associated free-energy density can be expressed as
\begin{equation}\label{free-energy density}
f = - \frac{1}{\beta V}\log\mathcal{Z} = \frac{g^{4}}{24}\left(\zeta - \frac{1}{4\pi^{2}}\right)
\bar{b}^{4} - \bar{b}^{2}\left(\frac{m_{b}^{2}}{2} + \frac{g^{2}\mu^{2}}{4\pi^{2}}
 \right) - \frac{\mu^{4}}{6\pi^{2}}.
\end{equation}

\subsection{Free-energy density and extremization}

If we consider for time-being $\zeta = 0$, then the from the field equation, we get
\begin{equation}\label{zeta = 0 equation of motion}
m_{b}^{2}\bar{b} = \frac{g}{2}\langle\bar{\psi}\gamma^{0}\tau^{3}\psi\rangle = \frac{g}{2}
(n_{1} - n_{2}).
\end{equation}
On the other hand, in that case, the total number density can be expressed as
\begin{equation}\label{number density}
\begin{split}
n & = \frac{\partial}{\partial\mu}\left(\frac{1}{\beta V}\log\mathcal{Z}\right) = 
\frac{\partial}{\partial\mu_{+}}\left(\frac{1}{\beta V}\log\mathcal{Z}\right) + 
\frac{\partial}{\partial\mu_{-}}\left(\frac{1}{\beta V}\log\mathcal{Z}\right)\\
 & = \frac{\mu_{+}^{3}}{3\pi^{2}} + \frac{\mu_{-}^{3}}{3\pi^{2}}. 
\end{split}
\end{equation}
The above expression can also be expressed as
\begin{equation}\label{number density relations}
n = n_{1} + n_{2}, \ n_{1} = \frac{\mu_{-}^{3}}{3\pi^{2}}, \ n_{2} = \frac{\mu_{+}^{3}}{3\pi^{2}}.
\end{equation} 
From the above expression, we also find the following relation
\begin{equation}\label{relation 1}
- g\bar{b} = (3\pi^{2}n_{1})^{1/3} - (3\pi^{2}n_{2})^{1/3},
\end{equation}
which combines with (\ref{zeta = 0 equation of motion}) leads to the following relation
\begin{equation}\label{relation 2}
\begin{split}
(bn_{1})^{1/3} - (bn_{2})^{1/3} & = - \frac{g^{2}}{6\pi^{2}}(bn_{1} - bn_{2})\\
\implies (bn_{1})^{2/3} + (bn_{1})^{1/3}(bn_{2})^{1/3} & + (bn_{2})^{2/3} = - \frac{6\pi^{2}}{g^{2}},
\end{split}
\end{equation}
where $b = \frac{3\pi^{2}}{m_{b}^{3}}$. Defining $S = (bn_{1})^{1/3} + (bn_{2})^{1/3}$ and $P = (bn_{1})^{1/3}(bn_{2})^{1/3}$, we will find the following relations
\begin{equation}
S^{2} - P = - \frac{6\pi^{2}}{g^{2}}, \ S^{3} - 3PS = bn,
\end{equation}
and combining the above set of equations, we get the following relation
\begin{equation}
2S^{3} + \frac{18\pi^{2}}{g^{2}}S + bn = 0,
\end{equation}
whose real root is given by
\begin{equation}
\begin{split}
S & = - \Bigg[\left(\frac{bn}{4} + \sqrt{\left(\frac{bn}{4}\right)^{2} + \left(\frac{3\pi^{2}}{g^{2}}\right)^{3}}\right)^{1/3}
 - \left( - \frac{bn}{4} + \sqrt{\left(\frac{bn}{4}\right)^{2} + \left(\frac{3\pi^{2}}{g^{2}}\right)^{3}}\right)^{1/3}\Bigg],
\end{split}
\end{equation}
which shows $S < 0$, however, $S$ by definition should be positive definite. Therefore, the only solution for $\zeta = 0$ case is $n_{1} = n_{2}$.

On the other hand, for $\zeta > \frac{1}{4\pi^{2}}$, $\bar{b}$ chooses a non-zero expectation value which minimizes the free-energy density, and its expectation value is given by
\begin{equation}\label{SU(2) breaking order parameter}
|\bar{b}| = \sqrt{\frac{6}{g^{4}}\frac{\left(m_{b}^{2} + \frac{g^{2}\mu^{2}}{2\pi^{2}}\right)}{\left(\zeta - \frac{1}{4\pi^{2}}\right)}}.
\end{equation}

\subsection{Fermi radii and split condition}

Note that the dispersion relation for the fermions can be obtained vanishing of the following determinant in momentum space
\begin{equation}
\begin{split}
\mathbf{M} & = \begin{bmatrix}
\slashed{p} - \frac{g}{2}\bar{b}\gamma^{0} & 0\\
0 & \slashed{p} + \frac{g}{2}\bar{b}\gamma^{0}
\end{bmatrix}\\
\text{det}[\mathbf{M}] & = \Big[\left(\varepsilon - \frac{g}{2}\bar{b}\right)^{2} + \mathbf{k}^{2}\Big]
\Big[\left(\varepsilon + \frac{g}{2}\bar{b}\right)^{2} + \mathbf{k}^{2}\Big],
\end{split}
\end{equation}
which shows the energy dispersion relations are
\begin{equation}
\varepsilon_{\pm}(\mathbf{k}) = |\mathbf{k}| \pm \frac{g}{2}\bar{b}.
\end{equation}
This shows that for a given chemical potential $\mu$, the Fermi radius of two particle species become different 
\begin{equation}
k_{F}^{\pm} = \mu_{\mp} = \mu \mp \frac{g}{2}\bar{b} = \mu \mp \sqrt{\frac{3}{2g^{2}}\frac{\left(m_{b}^{2} + \frac{g^{2}\mu^{2}}{2\pi^{2}}\right)}{\left(\zeta - \frac{1}{4\pi^{2}}\right)}}.
\end{equation}
In order for $k_{F}^{+} > 0$, the following condition must be satisfied
\begin{equation}
\begin{split}
\frac{2g^{2}}{3}\mu^{2} & > \frac{\left(m_{b}^{2} + \frac{g^{2}\mu^{2}}{2\pi^{2}}\right)}{\left(\zeta - \frac{1}{4\pi^{2}}\right)}\\
\implies \mu^{2} &  > \frac{3m_{b}^{2}}{2g^{2}\left(\zeta - \frac{1}{\pi^{2}}\right)}.
\end{split}
\end{equation}
The above equation shows that for $\zeta > \frac{1}{\pi^{2}}$, if the chemical potential satisfies the above inequality, two different Fermi surfaces of different radius form. On the other hand, for $\frac{1}{\pi^{2}} > \zeta > \frac{1}{4\pi^{2}}$, there will be a single Fermi sphere whereas for $0 \leq \zeta < \frac{1}{4\pi^{2}}$, $\bar{b}$ satisfy the following equation of motion
\begin{equation}
\begin{split}
m_{b}^{2}\bar{b} & = \frac{\zeta g^{4}}{6}\bar{b}^{3} + \frac{g}{2}(n_{1} - n_{2}) = \frac{\zeta g^{4}}{6}\bar{b}^{3} - \frac{g}{6\pi^{2}}(\mu_{+}^{3} - \mu_{-}^{3})\\
\implies m_{b}^{2}\bar{b} & = \frac{\zeta g^{4}}{6}\bar{b}^{3} - \frac{\mu^{2}}{2\pi^{2}}g^{2}\bar{b}
- \frac{g^{4}}{24\pi^{2}}\bar{b}^{3}\\
\implies \bar{b} & \Big[\frac{g^{4}}{6}\left(\zeta - \frac{1}{4\pi^{2}}\right)\bar{b}^{2} - \left(m_{b}^{2} + \frac{g^{2}\mu^{2}}{2\pi^{2}}\right)\Big] = 0,
\end{split}
\end{equation}
which shows that $\bar{b} = 0$ for $0 \leq \zeta < \frac{1}{4\pi^{2}}$.

\section{Equation of State and Thermodynamic Quantities}
\label{sec:S2}

\subsection{Pressure, energy density, and number density}

The total pressure exerted by the system can be expressed in terms of chemical potential as
\begin{equation}
P = \frac{\mu^{4}}{6\pi^{2}} + \bar{b}^{2}\left(\frac{m_{b}^{2}}{2} + \frac{g^{2}\mu^{2}}{4\pi^{2}}
 \right) + \left(\frac{g^{4}}{96\pi^{2}} - \frac{\zeta g^{4}}{24}\right)\bar{b}^{4},
\end{equation}
where $\mu$ can be expressed in terms of total number density as
\begin{equation}
n = \frac{\partial P}{\partial\mu} = \frac{2\mu^{3}}{3\pi^{2}} + \frac{g^{2}\mu}{2\pi^{2}}\bar{b}^{2}.
\end{equation}
Here the expression of $\bar{b}$ can be obtained through the minimization of free-energy density
which only becomes non-zero when $\zeta > \frac{1}{4\pi^{2}}$ and it is given as in (\ref{SU(2) breaking order parameter}). As a result, the above equation becomes
\begin{equation}
n = \frac{2\mu^{3}}{3\pi^{2}} + \frac{3\mu}{g^{2}\pi^{2}}\frac{\left(m_{b}^{2} + \frac{g^{2}\mu^{2}}{2\pi^{2}}\right)}{\left(\zeta - \frac{1}{4\pi^{2}}\right)}.
\end{equation}
We must invert this solution in order to obtain $\mu$ as a function of number density $n$. The above expression can also be expressed in terms of dimensionless number density as
\begin{equation}
bn = 2\left(\frac{\mu}{m_{b}}\right)^{3} + \frac{9\mu}{m_{b}g^{2}}\frac{\left(1 + \frac{g^{2}\mu^{2}}{2\pi^{2}m_{b}^{2}}\right)}{\left(\zeta - \frac{1}{4\pi^{2}}\right)},
\end{equation}
and in similar manner pressure can be expressed as
\begin{equation}
P = m_{b}^{4}\Bigg[\frac{\mu^{4}}{6\pi^{2}m_{b}^{4}} + \frac{\bar{b}^{2}}{m_{b}^{2}}\left(\frac{1}{2} 
+ \frac{g^{2}\mu^{2}}{4\pi^{2}m_{b}^{2}}\right) + \left(\frac{g^{4}}{96\pi^{2}} - \frac{\zeta g^{4}}{24}\right)\frac{\bar{b}^{4}}{m_{b}^{4}}\Bigg],
\end{equation}
where
\begin{equation}
\frac{|\bar{b}|}{m_{b}} = \sqrt{\frac{6}{g^{4}}\frac{\left(1 + \frac{g^{2}\mu^{2}}{2\pi^{2}m_{b}^{2}}\right)}{\left(\zeta - \frac{1}{4\pi^{2}}\right)}}.
\end{equation}
On the other hand, the energy density can be expressed as
\begin{equation}
\rho = \mu n - P_{0} + P_{1},
\end{equation}
where
\begin{equation}
\begin{split}
P_{0} & =  m_{b}^{4}\Bigg[\frac{\mu^{4}}{6\pi^{2}m_{b}^{4}} + \frac{\bar{b}^{2}}{m_{b}^{2}}\left(\frac{g^{2}\mu^{2}}{4\pi^{2}m_{b}^{2}}\right) + \left(\frac{g^{4}}{96\pi^{2}}\right)\frac{\bar{b}^{4}}{m_{b}^{4}}\Bigg]\\
P_{1} & = m_{b}^{4}\Big[\left(\frac{\zeta g^{4}}{8}\frac{\bar{b}^{4}}{m_{b}^{4}}\right) 
+ \frac{\bar{b}^{2}}{2m_{b}^{2}}\Big].
\end{split}
\end{equation}
\begin{figure}[b]
\includegraphics[height = 10cm, width = 15cm]{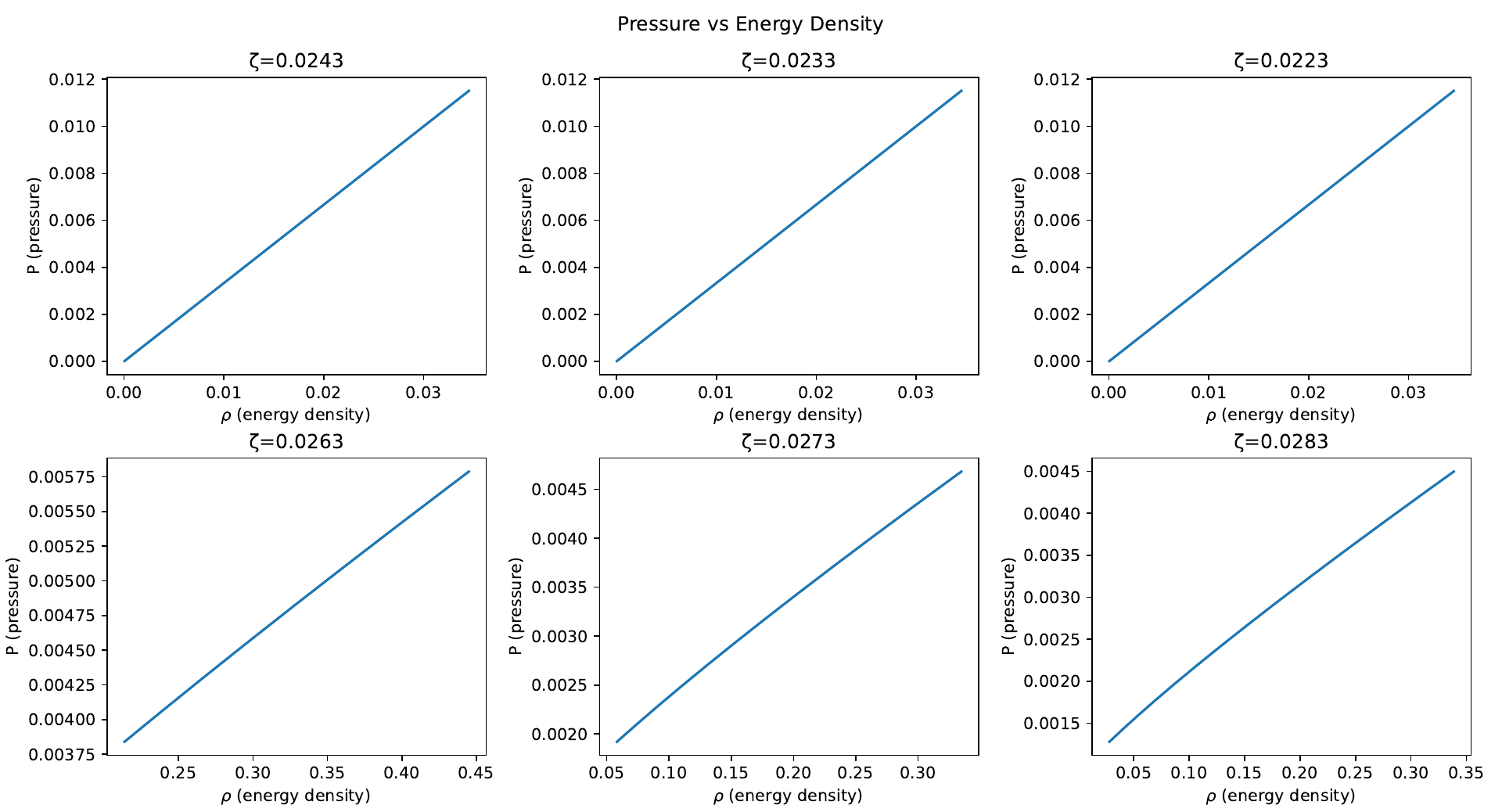}
\includegraphics[height = 10cm, width = 15cm]{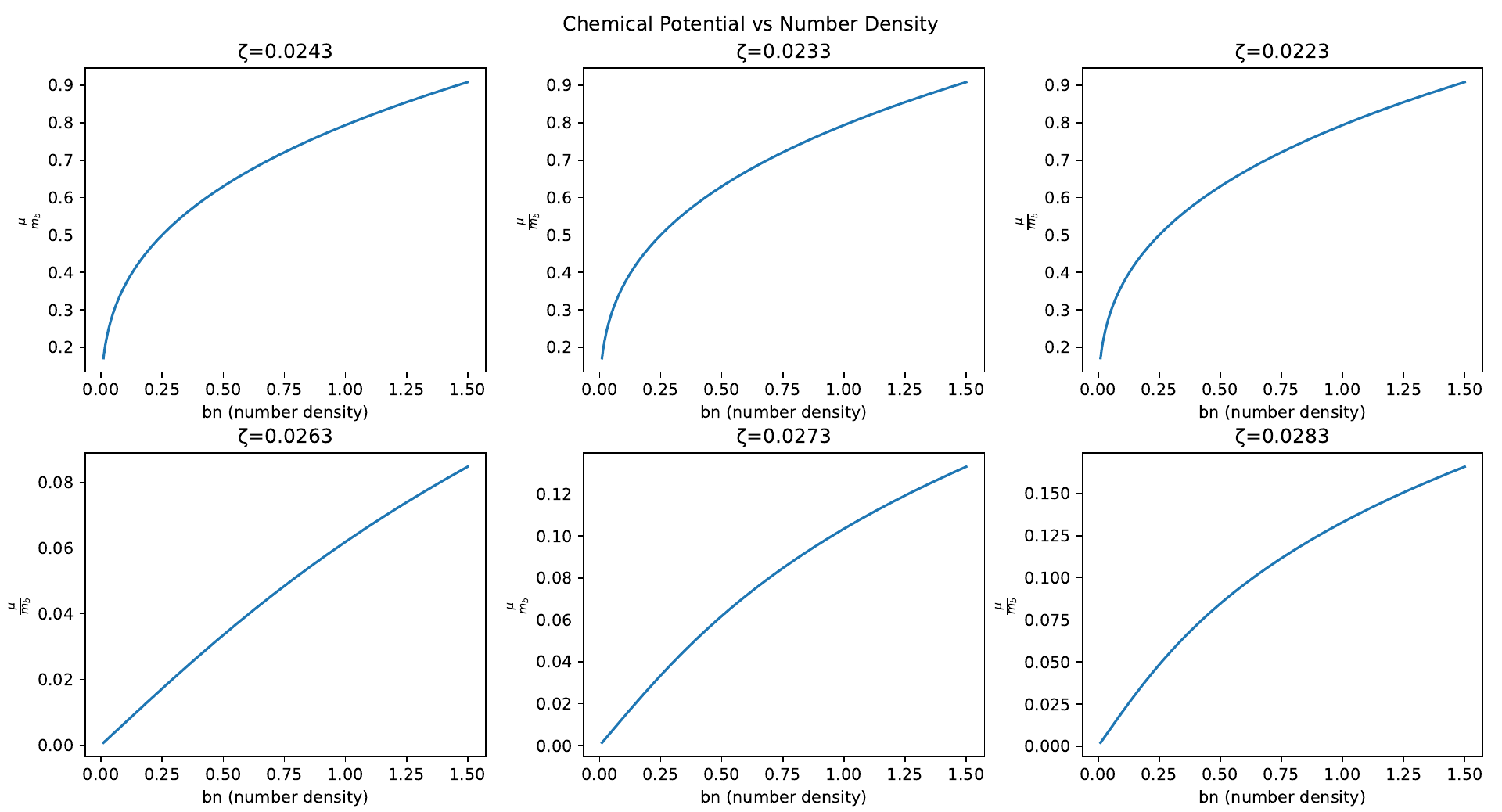}
\caption{Plot of $\frac{P}{m_{b}^{4}}$ as a function of $\frac{P}{m_{b}^{4}}$, and chemical potential potential $\frac{\mu}{m_{b}}$ as a function of number density $bn$.}
\label{Figure EOS}
\end{figure}
\begin{figure}[b]
\includegraphics[height = 10cm, width = 15cm]{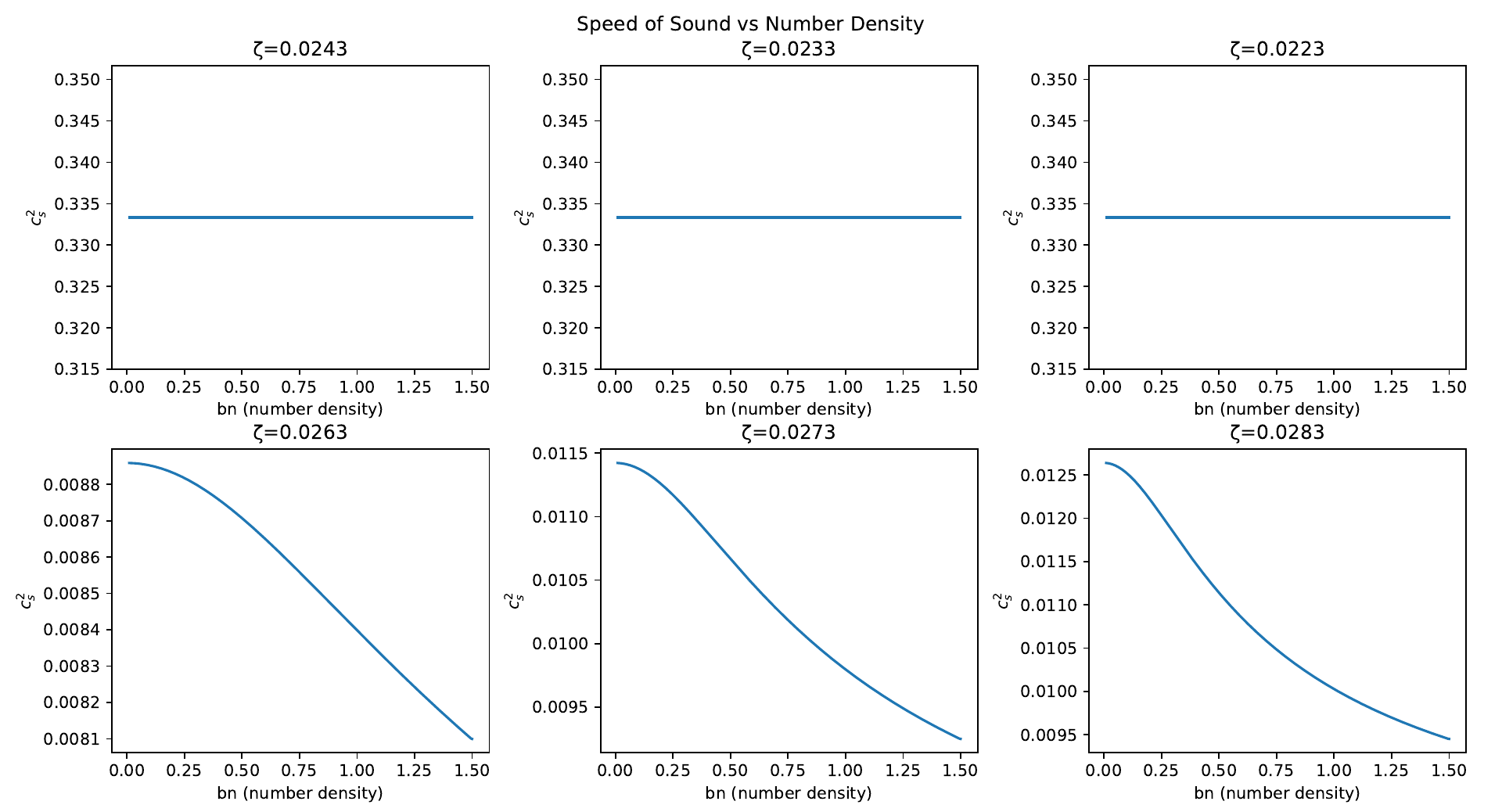}
\caption{Plot of speed of sound as a function of number density $bn$.}
\label{Figure SoS}
\end{figure}
In figure \ref{Figure EOS}, we have plotted the equation of state of matter and the relation between number density and chemical potential, whereas in figure \ref{Figure SoS}, we have plotted the speed of sound as a function of number density. In figure \ref{Figure Bulk Compressibility}, we plotted bulk compressibility $K = \rho\frac{dP}{d\rho}$ as a function of $bn$ for six different values of $\zeta$.
\begin{figure}[b]
\includegraphics[height = 10cm, width = 15cm]{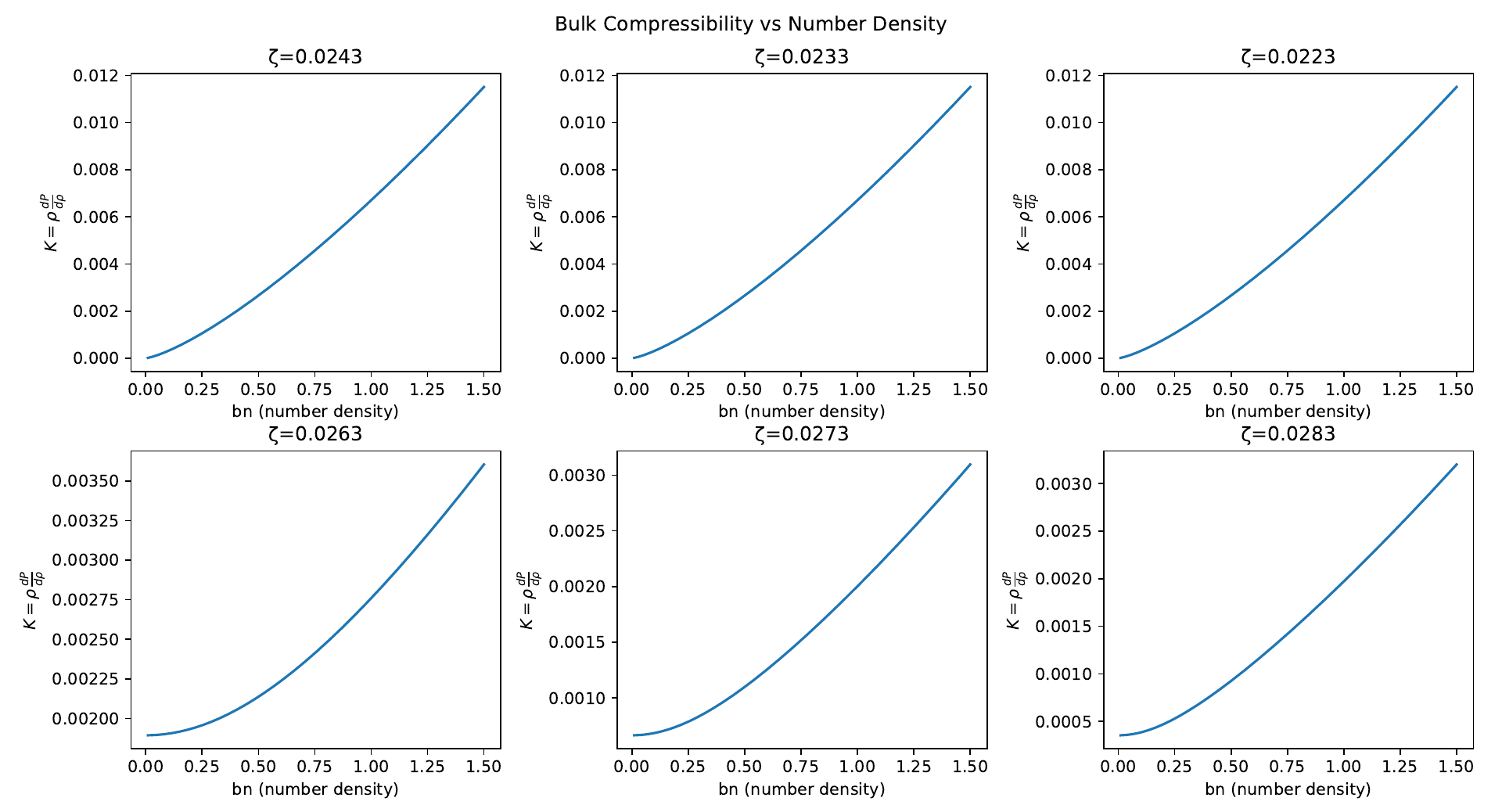}
\caption{Plot of bulk compressibility as a function of number density $bn$.}
\label{Figure Bulk Compressibility}
\end{figure}
%

\section{Nambu--Gorkov Formalism for One-Component Dirac Field}
\label{sec:S3}

Now we consider a real scalar field interaction with the two massless fermions which is described by adding the following Lagrangian densities to the action that we have considered in last section
\begin{equation}\label{added Lagrangian density}
\mathcal{L}_{\text{bosons}} = \frac{1}{2}\partial_{\mu}\varphi\partial^{\mu}\varphi
 - \frac{1}{2}M^{2}\varphi^{2}, \ \mathcal{L}_{\text{interactions}} = - g\bar{\psi}\psi\varphi,
\end{equation}
where $g > 0$ is the coupling constant.

The partition function action can now be expressed as
\begin{equation}
\mathcal{Z} = \int\mathcal{D}\bar{\psi}\mathcal{D}\psi\mathcal{D}\varphi \ 
e^{ - S_{\text{tot}}^{(E)}},
\end{equation}
where $S_{\text{tot}}^{(E)}$ is total Euclidean Lagrangian density of which we are interested now in the following part 
\begin{equation}
\begin{split}
S_{1}^{(E)} & = - \Big[\int_{x, y}\bar{\psi}(x)G_{0}^{-1}(x,y)\psi(y) - \int_{x, y}\frac{1}{2}\varphi(x)D^{-1}(x, y)\varphi(y) - g\int_{x}\bar{\psi}(x)\psi(x)\varphi(x)\Big],
\end{split}
\end{equation}
where we have abbreviated the spacetime integral as
\begin{equation}
\int_{x} = \int_{0}^{\beta}d\tau\int d^{3}x,
\end{equation}
and where 
\begin{equation}
G_{0}^{-1}(x, y) = \delta^{(4)}(x - y)(i\slashed{\partial} + \gamma^{0}\mu),
\end{equation}
is the inverse fermionic tree level propagator, and $D^{-1}(x, y)$ is the inverse bosonic propagator
whose specific form is not important for the present discussion.

\subsection{Mean-field decomposition and Cooper pair condensate}

In this section, we find an approximation for the four-fermion interaction term so that we can write the product of two fermion spinors as its expectation value plus fluctuations around that value. The expectation value corresponds to a condenstate of fermion pair. For a di-fermionic condensate, we consider the fermion-fermion pair which describes Cooper pairs in a superfluid or a superconductor. This can be done by considering charge conjugate spinor $\psi_{C}$ such that a Cooper pair of fermions can be described by $\psi\bar{\psi}_{C}$ and a Cooper of anti-fermions would be described by $\psi_{C}\bar{\psi}$. Introducing the charge-conjugation matrix as $C = i\gamma^{2}\gamma^{0}$, we define
\begin{equation}
\psi_{C} \equiv C\bar{\psi}^{T},
\end{equation}
which also implies $\bar{\psi}_{C} = \psi^{T}C, \ \psi = C\bar{\psi}_{C}^{T}, \ \bar{\psi} = \psi_{C}^{T}C$. Here $\bar{\psi}_{C}$ is defined by first charge-conjugating, the taking Hermitian conjugate and multiplying by $\gamma^{0}$ from right. Since
\begin{equation}
\bar{\psi}_{C}\psi_{C} = \psi^{T}CC\bar{\psi}^{T} = - \psi^{T}\bar{\psi}^{T} = (\bar{\psi}\psi)^{T} = \bar{\psi}\psi,
\end{equation}
where $C = - C^{-1}$ and the Grassmann property of spinors have been used. With this, we may now write the following relation
\begin{equation}
\begin{split}
\bar{\psi}(x)\psi(x)\bar{\psi}(y)\psi(y) & = \frac{1}{2}[\bar{\psi}_{C}(x)\psi_{C}(x)\bar{\psi}(y)
\psi(y) + \bar{\psi}(x)\psi(x)\bar{\psi}_{C}(y)\psi_{C}(y)]\\
 & = - \frac{1}{2}\text{Tr}[\psi_{C}(x)\bar{\psi}(y)\psi(y)\bar{\psi}_{C}(x) + \psi(x)\bar{\psi}_{C}(y)\psi_{C}(y)\bar{\psi}(x)], 
\end{split}
\end{equation}
where the trace is taken over Dirac space, and the minus sign arises as the fermion field is a Grassmann variable. Now we consider the following decomposition
\begin{equation}
\begin{split}
\psi_{C}(x)\bar{\psi}(y) & = \langle\psi_{C}(x)\bar{\psi}(y)\rangle + [\psi_{C}(x)\bar{\psi}(y)
 - \langle\psi_{C}(x)\bar{\psi}(y)\rangle]\\
\psi(y)\bar{\psi}_{C}(x) & = \langle\psi(y)\bar{\psi}_{C}(x)\rangle + [\psi(y)\bar{\psi}_{C}(x)
 - \langle\psi(y)\bar{\psi}_{C}(x)\rangle],
\end{split}
\end{equation}
where we consider the terms insider the square brackets to be fluctuations. Neglecting terms quadratic in fluctuations, we arrive at the following relation
\begin{equation}
\begin{split}
\int_{x, y} & D(x, y)\bar{\psi}(x)\psi(x)\bar{\psi}(y)\psi(y) = \int_{x, y}D(x, y) \text{Tr}
[\langle\psi_{C}(x)\bar{\psi}(y)\rangle\langle\psi(y)\bar{\psi}_{C}(y)\rangle]\\
 & - \int_{x, y}D(x, y) \text{Tr}[\langle\psi_{C}(x)\bar{\psi}(y)\rangle\psi(y)\bar{\psi}_{C}(x)
 + \langle\psi(x)\bar{\psi}_{C}(y)\rangle\psi_{C}(y)\bar{\psi}(x)], 
\end{split}
\end{equation}
where the boson propagator being symmetric in position space $D(x, y) = D(y, x)$ is assumed. Since the first term does not depend on the fermion fields, we can pull it out out of the functional integral,
\begin{equation}
\mathcal{Z} = \mathcal{Z}_{\text{bosons}}\mathcal{Z}_{0}\int\mathcal{D}\bar{\psi}\mathcal{D}\psi \ 
e^{- S''},
\end{equation}
where
\begin{equation}
\mathcal{Z}_{0} = e^{\frac{g^{2}}{2}\int_{x, y}D(x, y)\text{Tr}[\langle\psi_{C}(x)\bar{\psi}(y)\rangle 
\langle\psi(y)\bar{\psi}_{C}(y)\rangle]}.
\end{equation}
In the upcoming derivation of the gap equation, we are going to show that $\mathcal{Z}_{0}$ will play no role although this contribution is important for the thermodynamic potential. It is also possible to derive the same gap equation by minimizing the thermodynamic potential \textit{w.r.t} the gap parameter and in that case $\mathcal{Z}_{0}$ must be considered.

The new action abbreviated action $S''$ is essential expressed as
\begin{equation}
\begin{split}
S'' & = - \int_{x, y}\Big[\bar{\psi}(x)G_{0}^{-1}(x,y)\psi(y) + \frac{1}{2}[\bar{\psi}_{C}(x)
\Phi^{+}(x, y)\psi(y) + \bar{\psi}(x)\Phi^{-}(x, y)\psi_{C}(y)]\Big],
\end{split}
\end{equation}
where we have introduced the following notations
\begin{equation}\label{Gap parameters}
\begin{split}
\Phi^{+}(x, y) & \equiv g^{2} D(x, y)\langle\psi_{C}(x)\bar{\psi}(y)\rangle\\
\Phi^{-}(x, y) & \equiv g^{2} D(x, y)\langle\psi(x)\bar{\psi}_{C}(y)\rangle.
\end{split}
\end{equation}
From the above definitions, the following relation can be verified
\begin{equation}
\Phi^{-}(y, x) = \gamma^{0}[\Phi^{+}(x, y)]^{\dagger}\gamma^{0}.
\end{equation}
It is important to note that so far we have not considered the interaction of fermions with the vector field to which we will come soon.

\subsection{Nambu--Gorkov space and inverse propagator}

It is important to note that all effects of interaction are now completely absorbed into $\Phi^{\pm}$, and our new action $S''$ becomes quadratic in the field variables, and therefore, we can compute the functional integral using the standard Gaussian integral result. In order to do that, let us go to the momentum space by introducing the Fourier transforms of the fields
\begin{equation}
\begin{split}
\psi(x) = \frac{1}{\sqrt{V}}\sum_{k}e^{-ik.x}\psi(k), & \ \bar{\psi}(x) = \frac{1}{\sqrt{V}}\sum_{k}e^{ik.x}\bar{\psi}(k)\\
\psi_{C}(x) = \frac{1}{\sqrt{V}}\sum_{k}e^{-ik.x}\psi_{C}(k), & \ \bar{\psi}_{C}(x) = \frac{1}{\sqrt{V}}\sum_{k}e^{ik.x}\bar{\psi}_{C}(k),
\end{split}
\end{equation} 
where $\frac{1}{\sqrt{V}}$ is considered to make the Fourier transformed field variables dimensionless. On the other hand, the temporal component of the four momentum $k = (k_{0}, \vec{k})$ is given by the Matsubara frequencies $k_{0} = - i\omega_{n}$ where $\omega_{n} = (2n + 1)\pi T$. By applying charge conjugation operation to the first relation in the above and comparing it with the third relation, we obtain the relation $\psi_{C}(k) = C\bar{\psi}^{T}(-k)$, and $\bar{\psi}_{C}(k) = \bar{\psi}^{T}(-k)C$. Therefore, in momentum space space, charge conjugation operation flips the sign of 4-momentum.

In order to define the Fourier transformation of $\Phi^{\pm}$, we assume translational invariance $\Phi^{\pm}(x, y) = \Phi^{\pm}(x - y)$ to write
\begin{equation}
\Phi^{\pm}(x - y) = \frac{T}{V}\sum_{k}e^{-ik.(x - y)}\Phi^{\pm}(k),
\end{equation} 
and with this we have the following relation $\Phi^{-}(k) = \gamma^{0}[\Phi^{+}(k)]^{\dagger}\gamma^{0}$.  Now inserting the above Fourier decompositions into the interaction term of the action, we obtain
\begin{equation}
\int_{x, y}\bar{\psi}_{C}(x)\Phi^{+}(x - y)\psi(y) = \frac{1}{T}\sum_{k}\bar{\psi}_{C}(k)\Phi^{+}(k)\psi(k),
\end{equation}
where we have used 
\begin{equation}
\int_{x}e^{-ik.x} = \frac{V}{T}\delta_{k, 0}.
\end{equation}
Now we are going to treat $\bar{\psi}_{C}(k), \ \psi_{C}(k)$ as independent variables in addition to the field variables $\bar{\psi}(k), \ \psi(k)$. As a result of which we may now write the function integral in terms of integral over all four variables by restricting ourselves to 4-momenta in one-half of the full momentum space
\begin{equation}
\begin{split}
\mathcal{D}\bar{\psi}\mathcal{D}\psi & = \prod_{k}d\bar{\psi}(k)d\psi(k)\\
 & = \prod_{k > 0}d\bar{\psi}(k)d\bar{\psi}(- k)d\psi(k)d\psi(-k)\\
 & = \mathcal{N}\prod_{k > 0}d\bar{\psi}(k)d\psi_{C}(k)d\psi(k)d\bar{\psi}_{C}(k),
\end{split}
\end{equation}
with an irrelevant constant $\mathcal{N}$ arises due to change of integration variables. Now we may notice the following relationship between $k < 0$ and $k > 0$ degrees of freedom
\begin{equation}
\begin{split}
\sum_{k < 0}\bar{\psi}_{C}(k)\Phi^{+}(k)\psi(k) & = \sum_{k > 0}\bar{\psi}_{C}( - k)\Phi^{+}( - k)
\psi( - k)\\
 & = \sum_{k > 0}\psi^{T}(k)C\Phi^{+}(- k)C\bar{\psi}_{C}^{T}(k)\\
 & = \sum_{k > 0}[\psi^{T}(k)C\Phi^{+}(- k)C\bar{\psi}_{C}^{T}(k)]^{T}\\
 & = - \sum_{k > 0}\bar{\psi}_{C}(k)C[\Phi^{+}(- k)]^{T}C\psi(k)\\
 & = \sum_{k > 0}\bar{\psi}_{C}(k)\Phi^{+}(k)\psi(k),
\end{split}
\end{equation}
where we have used $C[\Phi^{+}(- k)]^{T}C = - \Phi^{+}(k)$ which can be seen in the following manner
\begin{equation}
\begin{split}
\int_{x, y}\bar{\psi}_{C}(x)\Phi^{+}(x - y)\psi(y) & = - \int_{x, y}\psi^{T}(y)[\Phi^{+}(x - y)]^{T}\bar{\psi}_{C}^{T}(x)\\
 & = - \int_{x, y}\bar{\psi}_{C}(x)C[\Phi^{+}(y - x)]^{T}C\psi(y),
\end{split}
\end{equation} 
which shows $C[\Phi^{+}(y - x)]^{T}C = - \Phi^{+}(x - y)$. As a result, we obtain the result $C[\Phi^{+}(- k)]^{T}C = - \Phi^{+}(k)$ in momentum space. In analogous way, we find the expression for $\bar{\psi}(x)\Phi^{-}(x, y)\psi_{C}(y)$ which yields the following expression of interaction part of the action in momentum space
\begin{equation}
\begin{split}
\frac{1}{2}\int_{x, y} & [\bar{\psi}_{c}(x)\Phi^{+}(x, y)\psi(y) + \bar{\psi}(x)\Phi^{-}(x, y)\psi_{C}(y)]\\
= \frac{1}{T}\sum_{k > 0} & [\bar{\psi}_{C}(k)\Phi^{+}(k)\psi(k) + \bar{\psi}(k)\Phi^{-}(k)\psi_{C}(k)].
\end{split}
\end{equation} 
In order to develop the formalism, we restrict ourselves to a one-component Dirac spinor. With the definition of the Fourier transformation of field variables, the tree-level action can be expressed as
\begin{equation}
\int_{x, y}\bar{\psi}(x)G_{0}^{- 1}(x, y)\psi(y) = \frac{1}{T}\sum_{k}\bar{\psi}(k)(\gamma^{\mu}k_{\mu} + \mu\gamma^{0})\psi(k).
\end{equation} 
Again dividing the sum over $k$ into two parts, we can rewrite the above as
\begin{equation}
\begin{split}
\sum_{k < 0} & \bar{\psi}(k)(\gamma^{\mu}k_{\mu} + \mu\gamma^{0})\psi(k) = \sum_{k > 0}\bar{\psi}(-k)
(- \gamma^{\mu}k_{\mu} + \mu\gamma^{0})\psi(-k)\\
= \sum_{k > 0} & \psi_{C}^{T}(k)C (- \gamma^{\mu}k_{\mu} + \mu\gamma^{0})C\bar{\psi}_{C}^{T}(k)\\
= \sum_{k > 0} & [\psi_{C}^{T}C(- \gamma^{\mu}k_{\mu} + \mu\gamma^{0})C\bar{\psi}_{C}^{T}(k)]^{T}\\
= - \sum_{k > 0} & \bar{\psi}_{C}(k)C(-\gamma_{\mu}^{T}k^{\mu} + \mu\gamma^{0})C\psi_{C}(k)\\
= \sum_{k > 0} & \bar{\psi}_{C}(k)(\gamma^{\mu}k_{\mu} - \mu\gamma^{0})\psi_{C}(k),
\end{split}
\end{equation}
where we have used the fact $C\gamma_{\mu}^{T}C = \gamma_{\mu}$. As a consequence of this, we can now write
\begin{equation}
\begin{split}
\int_{x, y} & \bar{\psi}(x)G_{0}^{-1}(x, y)\psi(y) = \frac{1}{T}\sum_{k > 0}[\bar{\psi}(k)[G_{0}^{+}(k)]^{-1}\psi(k) + \bar{\psi}_{c}(k)[G_{0}^{-}(k)]^{-1}\psi_{C}(k)], 
\end{split}
\end{equation}
where
\begin{equation}
[G_{0}^{\pm}(k)]^{- 1} = \gamma^{\mu}k_{\mu} \pm \gamma^{0}\mu.
\end{equation}
With the new integration measure, tree-level action and interaction part, the partition function can be expressed in the following compact way
\begin{equation}
\mathcal{Z} = \mathcal{N}\mathcal{Z}_{bosons}\mathcal{Z}_{0}\int\mathcal{D}\bar{\Psi}\mathcal{D}\Psi \ 
e^{\sum_{k > 0}\bar{\Psi}(k)\frac{\mathcal{G}^{- 1}}{T}\Psi(k)},
\end{equation} 
where we have abbreviated the integration measure as 
\begin{equation}
\mathcal{D}\bar{\Psi}\mathcal{D}\Psi = \prod_{k > 0}d\bar{\psi}(k)d\psi_{C}(k)d\psi(k)s\bar{\psi}_{C}(k).
\end{equation}
We introduced new spinors
\begin{equation}
\Psi \equiv \begin{bmatrix}
\psi\\
\psi_{C}
\end{bmatrix}, \ \bar{\Psi} \equiv (\bar{\psi}, \bar{\psi}_{C}),
\end{equation}
\textit{w.r.t} which the inverse propagator can be expressed as
\begin{equation}\label{Propagator}
\mathcal{G}^{- 1}(k) = \begin{bmatrix}
[G_{0}^{+}(k)]^{- 1} & \Phi^{-}(k)\\
\Phi^{+}(k) & [G_{0}^{-}(k)]^{-1}
\end{bmatrix}. 
\end{equation}
The two-dimensional space that we introduced by introducing charge-conjugate spinors is called Nambu-Gorkov space. Together with the $4 \times 4$ structure of Dirac space, $\mathcal{G}^{-1}$ is an $8 \times 8$ matrix.

\subsection{Gap equation derivation}

Now we can write Eq. (\ref{Propagator}) in the form of a Dyson-Schwinger equation, in which the inverse propagator can be decomposed into a non-interacting part $\mathcal{G}_{0}^{-1}$ and a self-energy $\Sigma$,
\begin{equation}
\mathcal{G}^{-1} = \mathcal{G}_{0}^{-1} + \Sigma,
\end{equation} 
with
\begin{equation}
\mathcal{G}_{0}^{-1} = \begin{bmatrix}
[G_{0}^{+}]^{-1} & 0\\
0 & [G_{0}^{-}]^{-1}
\end{bmatrix}, \ \Sigma = \begin{bmatrix}
0 & \Phi^{-}\\
\Phi^{+} & 0
\end{bmatrix}.
\end{equation}
The propagator $\mathcal{G}$ is computed by inverting the (\ref{Propagator})
\begin{equation}
\mathcal{G} = \begin{bmatrix}
G^{+} & F^{-}\\
F^{+} & G^{-}
\end{bmatrix},
\end{equation}
with
\begin{equation}
G^{\pm} \equiv ([G_{0}^{\pm}]^{-1} - \Phi^{\mp}G_{0}^{\mp}\Phi^{\pm})^{-1}, \ F^{\pm} \equiv - G_{0}^{\mp}\Phi^{\pm}G^{\pm}.
\end{equation}
The above form of the propagator follows from the identity $\mathcal{G}^{- 1}\mathcal{G} = \mathbf{1}$. The off-diagonal elements $F^{\pm}$ are known as the anomalous propagators. These components describe the propagation of a fermion that is converted into a charge-conjugate fermion or vice-versa which is possible due to the Cooper pair condensate which can be thought of as a reservoir of fermions and fermion-holes. This in a way implies that the symmetry $U(1)$ associated to charge conservation is sponatneously broken. In other words, $\Phi^{\pm}$ is not invariant under the action of $U(1)$ group, and under the transformation $\psi \rightarrow e^{-i\alpha}\psi$, it transforms as
\begin{equation}
\Phi^{\pm} \rightarrow e^{\pm 2i\alpha}\Phi^{\pm}.
\end{equation} 
As a result, $\Phi^{\pm}$ transform in a non-trivial manner under the symmetry transformations of the Lagrangian. However, the order parameter $\Phi^{\pm}$ is invariant under $\mathbb{Z}_{2}$ symmetry.

In order to find the gap equation for the order parameter, we notice that the fermionic propagator must have the following form in Nambu-Gorkov space
\begin{equation}
\mathcal{G}(x, y) = - \langle\Psi(x)\bar{\Psi}(y)\rangle = - \begin{bmatrix}
\langle\psi(x)\bar{\psi}(y)\rangle & \langle\psi(x)\bar{\psi}_{C}(y)\rangle\\
\langle\psi_{C}(x)\bar{\psi}(y)\rangle & \langle\psi_{C}(x)\bar{\psi}_{C}(y)\rangle
\end{bmatrix}.
\end{equation}
Inserting the above relations into (\ref{Gap parameters}), we obtain the relation $\Phi^{+}(x, y) = - g^{2}D(x, y)F^{+}(x, y)$, which in Fourier space becomes
\begin{equation}
\begin{split}
\frac{T}{V}\sum_{p}e^{-p.(x - y)}\Phi^{+}(p) & = - g^{2}\frac{T^{2}}{V^{2}}\sum_{q, k}e^{-i(q + k).(x- y)}D(q)F^{+}(k)\\
 & = - g^{2}\frac{T^{2}}{V^{2}}\sum_{p, k}e^{- ip.(x - y)}D(p - k)F^{+}(k),
\end{split}
\end{equation}
where we have assumed translational invariance of the theory. Comparing the coefficients of the Fourier series in $p$ on both sides, we obtain the following gap equation
\begin{equation}
\Phi^{+}(p) = - g^{2}\frac{T}{V}\sum_{k}D(p - k)F^{+}(k),
\end{equation}
which is an integral equation.

\subsection{Quasiparticle excitations and Bogoliubov coefficients}

Now we are going to compute the various components of the propagator in Nambu-Gorkov space explicitly in order to solve the gap equation. However, before that we look at the fermionic excitations. It would be convenient if we express the inverse tree-level propagators for massless fermions in terms of energy projectors
\begin{equation}
[G_{0}^{\pm}]^{-1} = \gamma^{\mu}K_{\mu} \pm \gamma^{0}\mu = \sum_{e = \pm}[k_{0} \pm (\mu - ek)]
\gamma^{0}\Lambda_{k}^{\pm e},
\end{equation} 
where
\begin{equation}
\Lambda_{k}^{e} = \frac{1}{2}(\mathbf{1} + e\gamma^{0}\vec{\gamma}.\hat{k}).
\end{equation}
It can be checked quite easily that the above operators satisfy the properties of projection operators
\begin{equation}
\Lambda_{k}^{+} + \Lambda_{k}^{-} = \mathbf{1}, \ \Lambda_{k}^{+}\Lambda_{k}^{-} = 0, \ 
\Lambda_{k}^{e}\Lambda_{k}^{e} = \Lambda_{k}^{e}.
\end{equation}
The benefit of having complete set of projection operators is that the inversion of any operator of the form $A = \sum_{i}a_{i}\mathcal{P}_{i}$ can be expressed as $A^{-1} = \sum_{i}a_{i}^{-1}\mathcal{P}_{i}$. Although the only difference in our case is the presence of an addition $\gamma^{0}$ matrix, the matrix $\gamma^{0}$ and $\Lambda_{k}^{e}$ obey the commutation rule $\gamma^{0}\Lambda_{k}^{e} = \Lambda_{k}^{-e}\gamma^{0}$. Using this relation, we find
\begin{equation}
G_{0}^{\pm} = \sum_{e}\frac{\gamma^{0}\Lambda_{k}^{\mp e}}{k_{0} \pm (\mu - ek)}.
\end{equation}
Now we assume the following ansatz for the gap matrix
\begin{equation}
\Phi^{\pm}(K) = \pm \Delta(K)\gamma_{5},
\end{equation} 
where $\Delta(K)$ is assumed to be real. It is important here to note that since $\Phi^{+}$ and $\Phi^{-}$ are related via $\Phi^{-} = \gamma^{0}(\Phi^{+})^{\dagger}\gamma^{0}$, we obtain the expression for $\Phi^{-}$ once we make an ansatz for $\Phi^{+}$. The above ansatz respects the overall anti-symmetry of Cooper pair \textit{w.r.t} exchange of the fermions and it corresponds to even-parity, spin-singlet pairing. 

Inserting the expressions of $[G_{0}^{\pm}]^{-1}, \ G_{0}^{\pm}$, and $\Phi^{\pm}$ into the expressions of propagator and charge-conjugate propagator and using $\{\gamma^{\mu}, \gamma_{5}\} = 0$, as well as $\gamma_{5}^{2} = \mathbf{1}$, we obtain the following expression
\begin{equation}
\begin{split}
G^{\pm}(K) & = \Bigg[\sum_{e}\Big[k_{0} \pm (\mu - ek) - \frac{\Delta^{2}}{k_{0} \mp (\mu - ek)}\Big]
\gamma^{0}\Lambda_{k}^{\pm e}\Bigg]^{-1}\\
 & = \sum_{e}\frac{k_{0} \mp (\mu - ek)}{k_{0}^{2} - (\varepsilon_{k}^{e})^{2}}\gamma^{0}\Lambda_{k}
 ^{\mp e},
\end{split}
\end{equation} 
where
\begin{equation}
\varepsilon_{k}^{e} \equiv \sqrt{(\mu - ek)^{2} + \Delta^{2}}.
\end{equation}
On the other hand, the anomalous propagator reduces to
\begin{equation}
F^{\pm}(K) = \pm \sum_{e}\frac{\Delta(K)\gamma_{5}\Lambda_{k}^{\mp e}}{k_{0}^{2}
 - (\varepsilon_{k}^{e})^{2}}.
\end{equation}
Note that all the components of the Nambu-Gorkov propagator have the same poles, $k_{0} = \pm\varepsilon_{k}^{e}$, and these correspond to the excitation energies for quasi-particles $(e = +)$
and quasi-antiparticles $(e = -)$ (both for the upper sign) and quasi-holes $(e = +)$ and quasi-anti-holes $(e = -)$ (both for the lower sign). We also note that $\Delta$ is indeed the gap in the quasiparticle energy spectrum which is the reason behind frictionless charge transport in a fermionic superfluid.

Further insights into the nature of Cooper pairs can be obtained from the computation of charge density and the occupation numbers from the partition function. To start with, pressure is defined as
\begin{equation}
P = \frac{T}{V}\log\mathcal{Z},
\end{equation} 
which can also be expressed as
\begin{equation}
P = \frac{1}{2}\frac{T}{V}\sum_{K}\text{Tr}\log\mathcal{G}^{-1} + \frac{T}{V}\log\mathcal{Z}_{0} + \frac{T}{V}\log\mathcal{Z}_{bosons},
\end{equation}
where trace is taken over both Dirac space and Nambu-Gorkov space. In the above expression, we summed over all $K$, but taken care of overcounting by multiplying by $\frac{1}{2}$. Since we are interested in charge density $n$ which is nothing but the derivative of $P$ \textit{w.r.t} the chemical potential 
$\mu$, we obtain the following expression for $n$
\begin{equation}
\begin{split}
n & = \frac{1}{2}\frac{T}{V}\sum_{K}\text{Tr}\left(\mathcal{G}\frac{\partial\mathcal{G}^{-1}}{\partial\mu}\right)\\
 & = \frac{1}{2}\frac{T}{V}\sum_{K}\text{Tr}[\gamma^{0}(G^{+} - G^{-})],
\end{split}
\end{equation}
where in the last step, we have used the explicit form of $\mathcal{G}^{-1}$ and the trace over Nambu-Gorkov space. Inserting the expression of propagators and using $\text{Tr}[\Lambda_{k}^{e}] = 2$, we obtain the following expression for number density
\begin{equation}
\begin{split}
n & = - 2\frac{T}{V}\sum_{K}\sum_{e}\frac{\mu - ek}{k_{0}^{2} - (\varepsilon_{k}^{e})^{2}}\\
 & = 2\sum_{e}\int\frac{d^{3}k}{(2\pi)^{3}}\frac{\mu - ek}{2\varepsilon_{k}^{e}}\tanh\left(
 \frac{\varepsilon_{k}^{e}}{2T}\right).
\end{split}
\end{equation}
In obtaining the above result, we used the summation over Matsubara frequencies $k_{0} = -i\omega_{n}$ where $\omega_{n} = (2n + 1)\pi T$
\begin{equation}
T\sum_{k_{0}}\frac{1}{k_{0}^{2} - a^{2}} = - \frac{1}{2a}\tanh\left(\frac{a}{2T}\right),
\end{equation}
and we have also considered thermodynamic limit in which 
\begin{equation}
\frac{1}{V}\sum_{\vec{k}} \rightarrow \int\frac{d^{3}k}{(2\pi)^{3}}.
\end{equation}
Defining Fermi-Dirac distribution function $f(x) = \frac{1}{e^{\frac{x}{T}} + 1}$, we may write the following expression for number density
\begin{equation}
\begin{split}
n & = 2\sum_{e}e\int\frac{d^{3}k}{(2\pi)^{3}}\Bigg[\frac{1}{2}\left(1 - \frac{k - e\mu}{\varepsilon_{k}^{e}}\right) + \frac{k - e\mu}{\varepsilon_{k}^{e}}f(\varepsilon_{k}^{e})\Bigg]\\
 & = 2\sum_{e}e\int\frac{d^{3}k}{(2\pi)^{3}}\Big[|u_{k}^{e}|^{2}f(\varepsilon_{k}^{e}) + |v_{k}^{e}|^{2}
 [1 - f(\varepsilon_{k}^{e})]\Big],
\end{split}
\end{equation}
with
\begin{equation}
|u_{k}^{e}|^{2} = \frac{1}{2}\left(1 + \frac{k - e\mu}{\varepsilon_{k}^{e}}\right), \ 
|v_{k}^{e}|^{2} = \frac{1}{2}\left(1 - \frac{k - e\mu}{\varepsilon_{k}^{e}}\right).
\end{equation}
The above result shows that the quasiparticles are mixtures of fermions with occupation $f$ and fermion-holes with occupation $1 - f$, where the mixing coefficients are known as the Bogoliubov coefficients $|u_{k}^{e}|^{2}$ and $|v_{k}^{e}|^{2}$ with $|u_{k}^{e}|^{2} + |v_{k}^{e}|^{2} = 1$. One may check in the gapless limit, the above expression of number density reduces to the standard result
\begin{equation}
\begin{split}
n_{\Delta = 0} & = 2\sum_{e}e\int\frac{d^{3}k}{(2\pi)^{3}}\Big[\Theta(e\mu - k) + \text{sgn}(k - e\mu)
f(|k - e\mu|)\Big]\\
 & = 2\sum_{e}e\int\frac{d^{3}k}{(2\pi)^{3}}f(k - e\mu).
\end{split}
\end{equation}
Now let us consider the zero temperature limit in the expression of number density. Since $\varepsilon_{k}^{e} > 0$, $f(\varepsilon_{k}^{e}) \rightarrow \Theta(- \varepsilon_{k}^{e}) = 0$ at zero temperature, and thus neglecting the contribution coming from the antiparticles, we obtain the following final expression
\begin{equation}
n_{T = 0} \simeq 2\int\frac{d^{3}k}{(2\pi)^{3}}\frac{1}{2}\left(1 - \frac{k - \mu}{\sqrt{(k - \mu)^{2} + \Delta^{2}}}\right).
\end{equation}

\subsection{Gap equation solution and critical temperature}

Inserting the ansatz for gap matrix mentioned earlier and the expression of anomalous propagator into the gap equation, we obtain the following relation
\begin{equation}
\Delta(P)\gamma_{5} = - g^{2}\frac{T}{V}\sum_{K}D(P - K)\frac{\Delta(K)\gamma_{5}\Lambda_{k}^{-}}{k_{0}^{2} - \varepsilon_{k}^{2}},
\end{equation}
where we have neglected the contribution coming from the antiparticles and abbreviated $\varepsilon_{k} = \varepsilon_{k}^{+}$. In order to avoid the matrix structure of the above equation, we multiply both sides of the above equation with $\gamma_{5}$ and take trace over Dirac space to get
\begin{equation}
\Delta(P) = - \frac{g^{2}}{2}\frac{T}{V}\sum_{K}D(P - K)\frac{\Delta(K)}{k_{0}^{2} - \varepsilon_{k}^{2}},
\end{equation}
where we have used $\text{Tr}[\Lambda_{k}^{e}] = 2$. Now we assume that the four-fermion interaction is point-like, i.e., the inverse boson propagator can be approximated by the boson mass squared $D^{-1}(Q) = - Q^{2} + M^{2} \simeq M^{2}$, and as a result $\Delta(P)$ becomes independent of momenta $P$. After the Matsubara summation, we obtain the following result
\begin{equation}\label{gap equation 2}
\Delta = G\int\frac{d^{3}k}{(2\pi)^{3}}\frac{\Delta}{\varepsilon_{k}}\tanh\left(\frac{\varepsilon_{k}}{2T}\right),
\end{equation}
where the effective coupling constant $G$ is expressed as
\begin{equation}
G = \frac{g^{2}}{2M^{2}}.
\end{equation}
It is important to note that while $g$ is dimensionless, the ffective coupling constant $G$ has mass dimensions $-2$.

Now we look at the solution of the gap equation near zero temperature where we can make the approximation $\tanh\left(\frac{\varepsilon_{k}}{2T}\right) = 1$. Further, we assume that the interaction is non-zero for fermions in the vicinity of the Fermi surface $[\mu - \delta, \mu + \delta]$ with
\begin{equation}
\Delta_{0} \ll \delta \ll \mu ,
\end{equation}  
where $\Delta_{0} = \Delta(T = 0)$. The above assumption corresponds to weak coupling limit, and as a result of it, the gap equation becomes
\begin{equation}
\Delta_{0} \simeq \frac{\mu^{2}G}{2\pi^{2}}\int_{0}^{\delta}d\xi \ \frac{\Delta_{0}}{\sqrt{\xi^{2} + \Delta_{0}^{2}}},
\end{equation}
where we have made an approximation $dk \ k^{2} \simeq dk \ \mu^{2}$, and introduced the new integration variable $\xi = k - \mu$ \textit{w.r.t} which the integrand has a symmetry of $\xi \rightarrow - \xi$ such that one can restrict integration from $\xi \in [0, \delta]$. It is important to note $\Delta_{0} = 0$ is an obvious solution of the above equation, however, we are interested in a non-trivial solution of $\Delta_{0}$. In order to get that we compute the integral first which results in
\begin{equation}
\int\frac{d\xi}{\sqrt{\xi^{2} + \Delta_{0}^{2}}} = \log\Big[2\left(\xi + \sqrt{\xi^{2} + \Delta_{0}^{2}}\right)\Big],
\end{equation}
from which we find
\begin{equation}
\Delta_{0} \simeq 2\delta \ e^{ - \frac{2\pi^{2}}{G\mu^{2}}}.
\end{equation}
This is the famous result for the BCS gap which corresponds to instability of Fermi surface \textit{w.r.t} the formation of a Cooper pair condensate at weak coupling. Now we use the gap equation to determine the critical temperature $T_{c}$ for the superconducting phase transition. In BCS theory, this phase transition is second order in nature as the gap vanishes continuously at the critical point. In order to obtain the expression of critical temperature, we assume that we are at slightly below the critical temperature, the gap is non-zero, and we may divide both side of the equation (\ref{gap equation 2}), and then take the limit $\Delta \rightarrow 0$ limit which results in
\begin{equation}
1 \simeq \frac{G\mu^{2}}{2\pi^{2}}\int_{0}^{\delta}\frac{d\xi}{\xi}\tanh\left(\frac{\xi}{2T_{c}}\right).
\end{equation}
With the new integration variable $z = \frac{\xi}{2T_{c}}$ and after integration by parts, we obtain
\begin{equation}
\begin{split}
\frac{2\pi^{2}}{G\mu^{2}} & = \log(z)\tanh(z)\Big|_{0}^{\frac{\delta}{2T_{c}}} - \int_{0}^{\frac{\delta}{2T_{c}}}dz \ \frac{\log(z)}{\cosh^{2}z}\\
 & \simeq \log\left(\frac{\delta}{2T_{c}}\right) - \int_{0}^{\infty}dz \ \frac{\log(z)}{\cosh^{2}z}\\
 & = \log\left(\frac{\delta}{2T_{c}}\right) + \gamma - \log\left(\frac{\pi}{4}\right), 
\end{split}
\end{equation}
where $\gamma$ is the Euler-Mascheroni constant. Solving the above equation for $T_{c}$, we obtain
\begin{equation}
T_{c} = \frac{e^{\gamma}}{\pi}\Delta_{0} \simeq 0.57\Delta_{0}.
\end{equation}

\section{Two-Component Theory with Fermi Surface Mismatch}
\label{sec:S4}

\subsection{16$\times$16 propagator structure}

In this section, we consider our original with two massless Dirac field theory unified in a fermionic doublet formalism. As a consequence, now the fermion propagator introduced in previous section will now become a $16 \times 16$ matrix: 2 degrees of freedom from the two fermions species, 2 from fermion/charge-conjugate fermions (Nambu-Gorkov basis), and 4 from spin-$\frac{1}{2}$ and particle/antiparticle degrees of freedom (Dirac space).

In this case, we want the fermions of different flavors to form Cooper pairs which we demand by considering the following ansatz for the gap matrix
\begin{equation}
\Phi^{\pm} = \pm \Delta\sigma_{1}\gamma_{5},
\end{equation}  

with the off-diagonal and symmetric Pauli matrix $\sigma_{1}$ since the above ansatz lead to an overall antisymmetric Cooper pairs. We now promote the inverse tree-level propagator to a matrix in flavor space which is of the following form
\begin{equation}
\begin{split}
[G_{0}^{\pm}]^{- 1} & = \begin{bmatrix}
\gamma^{\mu}K_{\mu} \pm \mu_{1}\gamma^{0} & 0\\
0 & \gamma^{\mu}K_{\mu} \pm \mu_{2}\gamma^{0}
\end{bmatrix}\\
 & = \sum_{e = \pm}\gamma^{0}\Lambda_{k}^{\pm e}\begin{bmatrix}
 k_{0} \pm (\mu_{1} - ek) & 0\\
 0 & k_{0} \pm (\mu_{2} - ek)
 \end{bmatrix},
\end{split}
\end{equation}
where we introduced the effective chemical potentials of two flavors $\mu_{1} = \mu + \frac{g}{2}\bar{b}$ and $\mu_{2} = \mu - \frac{g}{2}\bar{b}$. Since we are working in the massless Dirac field theory, the chemical potentials are the same as their Fermi-momenta $\mu_{i} = k_{F, i}$. Below we choose to work in the following parametrization
\begin{equation}
\bar{\mu} = \frac{\mu_{1} + \mu_{2}}{2} = \mu, \ \delta\mu = \frac{\mu_{1} - \mu_{2}}{2} = \frac{g}{2}\bar{b},
\end{equation}
and without loss of generality that under SSB $g, \bar{b} > 0 $ which makes $\mu_{1} > \mu_{2}$. With this, the tree-level propagator can be expressed as
\begin{equation}
G_{0}^{\pm} = \sum_{e = \pm}\gamma^{0}\Lambda_{k}^{\mp e}\begin{bmatrix}
\frac{1}{k_{0} \pm (\mu_{1} - ek)} & 0\\
0 & \frac{1}{k_{0} \pm (\mu_{2} - ek)}
\end{bmatrix}.
\end{equation}

\subsection{Eight quasiparticle poles}

In order to compute the quasiparticle excitations, we need to compute the full propagator which is of the following form
\begin{equation}
G^{\pm} = ([G_{0}^{\pm}]^{-1} - \Phi^{\mp}G_{0}^{\mp}\Phi^{\pm})^{-1},
\end{equation}
where
\begin{equation}
\Phi^{\mp}G_{0}^{\mp}\Phi^{\pm} = \sum_{e}\gamma^{0}\Lambda_{k}^{\pm e}\begin{bmatrix}
\frac{\Delta^{2}}{k_{0} \pm (\mu_{1} - ek)} & 0\\
0 & \frac{\Delta^{2}}{k_{0} \mp (\mu_{1} - ek)}
\end{bmatrix}.
\end{equation}
As a result, we find the following form of the propagator
\begin{equation}
G^{\pm} = \sum_{e}\gamma^{0}\Lambda_{k}^{\mp e}\begin{bmatrix}
\frac{k_{0} \mp (\mu_{2} - ek)}{(k_{0} \pm \delta\mu)^{2} - (\varepsilon_{k}^{e})^{2}} & 0\\
0 & \frac{k_{0} \mp (\mu_{1} - ek)}{(k_{0} \mp \delta\mu)^{2} - (\varepsilon_{k}^{e})^{2}}
\end{bmatrix},
\end{equation}
where we have denoted
\begin{equation}
\varepsilon_{k}^{e} = \sqrt{(\bar{\mu} - ek)^{2} + \Delta^{2}}.
\end{equation}
In the case of absence of mismatch, this would be single-particle fermionic excitations in the superfluid. In the above, we have used the following identity
\begin{equation}
[k_{0} \pm (\mu_{1} - ek)][k_{0} \mp (\mu_{2} - ek)] = (k_{0} \pm \delta\mu)^{2} - (\bar{\mu} - ek)^{2}.
\end{equation}
From the above expression of propagator, we may notice the existence of 4 poles which are 
\begin{equation}
\varepsilon_{k}^{e} + \delta\mu, \ \varepsilon_{k}^{e} - \delta\mu, \ - \varepsilon_{k}^{e} + \delta\mu, \ - \varepsilon_{k}^{e} - \delta\mu, 
\end{equation}
which includes anti-particles ($e = -$) for which there are 8 poles. In the absence of pairing, the 8 poles correspond to fermions of species 1 and 2, and fermion-holes of species 1 and 2, and the same for the anti-fermions. On the other hand, in the presence of mismatch or in other words pairing, the new quasiparticles are mixtures of the original ones, however, the number branches remains the same. One may note here that at the low momenta, two of the excitation branches are fermions (fermion holes) of species 1 while, at large momenta, they become fermion holes (fermions) of species 2, and in between, they are mixture thereof. For the other two quasiparticles, the above statements remain the same except and the labels 1 and 2 must be interchanged. We will soon see that the mismatch leads to a reduction of the effective energy gap, and for $\delta\mu > \Delta$, the effective gap vanishes. However, later we discuss the fact that the gapless scenario corresponds to an unstable state.

\subsection{Free-energy density --- CJT/2PI derivation}

In the last section, we have discussed Cooper pairing without showing the stability by comparing the free-energy densities of superfluid and normal state of matter. In the case of pairing with mismatch, the free-energy comparison becomes important. Hence, in this section, we compute the free-energy density and show as a side result, that in the absence of mismatch, the superfluid state is preferred over the normal state of matter.

We may start with the expression of free-energy density which is of the following form (derived from the previous section) 
\begin{equation}\label{free-energy density 2}
\Omega = - \frac{1}{2}\frac{T}{V}\sum_{K}\text{Tr}\log\left(\frac{\mathcal{G}^{-1}}{T}\right) + \frac{\Delta^{2}}{G},
\end{equation}
where $\mathcal{G}$ is nothing but the propagator in Nambu-Gorkov space where we have neglected the contributions from the fluctuations. In the above expression, trace if taken over Nambu-Gorkov space, Dirac space, and internal flavor space.

In order to provide a mathematically consistent derivation, we follow CJT or two-particle irreducible (2PI) formalism \cite{CJT1974,Schmitt2015} in which, we use a more general form of the free-energy density
\begin{equation}\label{CJT}
\Omega = - \frac{1}{2}\frac{T}{V}\sum_{K}\text{Tr}\log\left(\frac{\mathcal{G}^{-1}}{T}\right) + 
\frac{1}{4}\frac{T}{V}\sum_{K}\text{Tr}[1 - \mathcal{G}_{0}^{- 1}\mathcal{G}],
\end{equation}
$\mathcal{G}_{0}^{- 1}$ is the inverse tree-level propagator in Nambu-Gorkov space. The CJT 
effective action is a functional of the Nambu-Gorkov propagator $\mathcal{G}$
\begin{equation}
\Gamma[\mathcal{G}] = \frac{1}{2}\frac{T}{V}\sum_{K}\text{Tr}\log\left(\frac{\mathcal{G}^{-1}}
{T}\right) - \frac{1}{2}\frac{T}{V}\sum_{K}\text{Tr}[1 - \mathcal{G}_{0}^{- 1}\mathcal{G}] + 
\Gamma_{2}[\mathcal{G}],
\end{equation}
where $\Gamma_{2}[\mathcal{G}]$ is the contribution coming from the set of all 2PI diagrams. Extremizing the above effective action \textit{w.r.t} the propagator yields the Dyson equation $\mathcal{G}^{ - 1} = \mathcal{G}_{0}^{- 1} + \Sigma$ which in this context is nothing but the gap equation for the superfluid energy gap. For computational purpose, we must truncate $\Gamma_{2}[\mathcal{G}]$ at a certain number of loops. For example, in order to derive the gap equation at one-loop level,the effective action must be computed at two-loop level. It is important here to note that the free-energy density is the same as effective action density at the stationary point up to a negative sign. This follows from the expression $\Gamma_{2}[\mathcal{G}] = \frac{1}{4}\frac{T}{V}\text{Tr}[\Sigma\mathcal{G}]$ and expressiong $\Sigma$ in terms of $\mathcal{G}_{0}$ and $\mathcal{G}$ via the Dyson-Schwinger equation. The free-energy density (\ref{CJT}) is identical to the one given in (\ref{free-energy density 2}) if the latter is evaluated at the stationary point, or in other words, if we replace $\frac{\Delta^{2}}{2G}$ by a momentum integral with the help of the gap equation.  

We now start computing the two terms in the free-energy density (\ref{CJT}) separately. For the first term, we use the fact $\text{Tr}\log = \log\text{det}$ and
\begin{equation}
\text{det}\begin{bmatrix}
A & B\\
C & D
\end{bmatrix} = \text{det}(AD - BD^{-1}CD).
\end{equation}
Given the full propagator of the following form
\begin{equation}
\mathcal{G}^{ - 1} = \begin{bmatrix}
[G_{0}^{+}]^{- 1} & \Phi^{-}\\
\Phi^{+} & [G_{0}^{-}]^{- 1}
\end{bmatrix},
\end{equation}
we obtain the following result
\begin{equation}
\text{Tr}\log\left(\frac{\mathcal{G}^{-1}}{T}\right) = \text{Tr}\log\left(\frac{[G_{0}^{+}]^{-1}[G_{0}^{-}]^{-1} - \Phi^{-}G_{0}^{-}\Phi^{+}[G_{0}^{-}]^{-1}}{T^{2}}\right).
\end{equation}
With the help of the expression of tree-level propagators, we compute the following quantities first
\begin{equation}
[G_{0}^{+}]^{-1}[G_{0}^{-}]^{-1} = \sum_{e}\Lambda_{k}^{e}\begin{bmatrix}
k_{0}^{2} - (\mu_{1} - ek)^{2} & 0\\
0 & k_{0}^{2} - (\mu_{2} - ek)^{2}
\end{bmatrix},
\end{equation}
whereas
\begin{equation}
\Phi^{-}G_{0}^{-}\Phi^{+}[G_{0}^{-}]^{-1} = \Delta^{2}\sum_{e}\Lambda_{k}^{-e}\begin{bmatrix}
\frac{k_{0} - (\mu_{1} - ek)}{k_{0} - (\mu_{2} - ek)} & 0\\
0 & \frac{k_{0} - (\mu_{2} - ek)}{k_{0} - (\mu_{1} - ek)}
\end{bmatrix}.
\end{equation}
Therefore, we finally obtain the following expression
\begin{equation}
\begin{split}
[G_{0}^{+}]^{-1} & [G_{0}^{-}]^{-1} - \Phi^{-}G_{0}^{-}\Phi^{+}[G_{0}^{-}]^{-1}\\
= \sum_{e}\Lambda_{k}^{-e} & \begin{bmatrix}
\frac{k_{0} - (\mu_{1} - ek)}{k_{0} - (\mu_{2} - ek)}[(k_{0} + \delta\mu)^{2} - (\varepsilon_{k}^{e})^{2}] & 0\\
0 & \frac{k_{0} - (\mu_{2} - ek)}{k_{0} - (\mu_{1} - ek)}[(k_{0} - \delta\mu)^{2} - (\varepsilon_{k}^{e})^{2}]
\end{bmatrix}.
\end{split}
\end{equation}
In Dirac space, the above matrix has the form $a_{+}\mathcal{P}_{+} + a_{-}\mathcal{P}_{-}$ where $\mathcal{P}_{\pm}$ are the complete, orthogonal projection operators. For such a matrix, $a_{\pm}$ are the degenerate eigenvalues with degeneracy being $\text{Tr}\mathcal{P}_{\pm}$ which leads to the result $\text{Tr}\log(a_{+}\mathcal{P}_{+} + a_{-}\mathcal{P}_{-}) = \text{Tr}[\mathcal{P}_{+}]\log a_{+} + \text{Tr}[\mathcal{P}_{-}]\log a_{-}$. In our case, the above mentioned degeneracy of eigenvalues are $\text{Tr}[\Lambda_{k}^{-e}] = 2$. Therefore, we obtain
\begin{equation}
\begin{split}
- \frac{1}{2} & \frac{T}{V}\sum_{K}\text{Tr}\log\mathcal{G}^{-1} = - \frac{T}{V}\sum_{K}\sum_{e}\Bigg[
\log\Big[\frac{k_{0} - (\mu_{1} - ek)}{k_{0} - (\mu_{2} - ek)} \frac{(k_{0} + \delta\mu)^{2} - (\varepsilon_{k}^{e})^{2}}{T^{2}}\Big]\\
 & + \log\Big[\frac{k_{0} - (\mu_{2} - ek)}{k_{0} - (\mu_{1} - ek)} \frac{(k_{0} - \delta\mu)^{2} - (\varepsilon_{k}^{e})^{2}}{T^{2}}\Big]\Bigg]\\
 & = - \frac{T}{V}\sum_{K}\sum_{e}\Bigg[\log\left(\frac{(\varepsilon_{k}^{e} + \delta\mu)^{2}
 - k_{0}^{2}}{T^{2}}\right) + \log\left(\frac{(\varepsilon_{k}^{e} - \delta\mu)^{2} - k_{0}^{2}}{T^{2}}\right)\Bigg]\\
 & = - 2\sum_{e}\int\frac{d^{3}k}{(2\pi)^{3}}\Bigg[\varepsilon_{k}^{e} + T\log\left(1 + 
 e^{- \frac{\varepsilon_{k}^{e} + \delta\mu}{T}}\right) + T\log\left(1 + e^{- \frac{\varepsilon_{k}^{e} - \delta\mu}{T}}\right)\Big],
\end{split}
\end{equation}
where in order to get the last expression, we performed summation over fermionic Matsubara frequencies defined earlier.

Now, we look at the second term of the free-energy density. After doing the trace operation over Nambu-Gorkov basis, we obtain the following result
\begin{equation}
\text{Tr}[\mathbf{1} - \mathcal{G}_{0}^{- 1}\mathcal{G}] = \text{Tr}[2\mathbf{1} - [G_{0}^{+}]^{- 1}G^{+} - [G_{0}^{-}]^{- 1}G^{-}],
\end{equation}
where the anomalous propagators do not appear. Now, the terms inside the square bracket can be expressed as
\begin{equation}
[G_{0}^{\pm}]^{-1}G^{\pm} = \sum_{e}\Lambda_{k}^{\mp e}\begin{bmatrix}
1 + \frac{\Delta^{2}}{(k_{0} \pm \delta\mu)^{2} - (\varepsilon_{k}^{e})^{2}} & 0\\
0 & 1 + \frac{\Delta^{2}}{(k_{0} \mp \delta\mu)^{2} - (\varepsilon_{k}^{e})^{2}}
\end{bmatrix},
\end{equation} 
using which we obtain the following result after doing the trace operation and summation over Matsubara frequencies
\begin{equation}
\begin{split}
\frac{1}{4}\frac{T}{V} & \sum_{K}\text{Tr}[\mathbf{1} - \mathcal{G}_{0}^{- 1}\mathcal{G}]\\
 & = - \frac{T}{V}\sum_{K}\sum_{e}\Big[\frac{\Delta^{2}}{(k_{0} + \delta\mu)^{2} - (\varepsilon_{k}^{e})^{2}} + \frac{\Delta^{2}}{(k_{0} - \delta\mu)^{2} - (\varepsilon_{k}^{e})^{2}}\Big]\\
 & = \sum_{e}\int\frac{d^{3}k}{(2\pi)^{3}}\frac{\Delta^{2}}{\varepsilon_{k}^{e}}[1 - f(\varepsilon_{k}
 + \delta\mu) - f(\varepsilon_{k}^{e} - \delta\mu)].
\end{split}
\end{equation}
Inserting the above results into the expression of free-energy density, we obtain the following result
\begin{equation}
\begin{split}
\Omega & = - 2\sum_{e}\int\frac{d^{3}k}{(2\pi)^{3}}\Bigg[\varepsilon_{k}^{e} + T\log\left(1 + 
e^{ - \frac{\varepsilon_{k}^{e} - \delta\mu}{T}}\right) + T\log\left(1 + e^{ - \frac{\varepsilon_{k}^{e} + \delta\mu}{T}}\right)\\
 & - \frac{\Delta^{2}}{2\varepsilon_{k}^{e}}[1 - f(\varepsilon_{k}^{e} - \delta\mu) 
 - f(\varepsilon_{k}^{e} + \delta\mu)]\Bigg].
\end{split}
\end{equation}

\subsection{Zero-temperature free energy and integral evaluation}

Now we evaluate the free-energy density at zero temperature for which we are going to use the following relations
\begin{equation}
\lim_{T\rightarrow 0}T\log(1 + e^{- x/T}) = - x\Theta(-x), \ \lim_{T\rightarrow 0}f(x) = \Theta(-x),
\end{equation}
which yield
\begin{equation}
\Omega = - 2\sum_{e}\int\frac{d^{3}k}{(2\pi)^{3}}\Big[\varepsilon_{k}^{e} + (\delta\mu - \varepsilon_{k}^{e})\Theta(\delta\mu - \varepsilon_{k}^{e}) - \frac{\Delta^{2}}{2\varepsilon_{k}^{e}}\Theta(\varepsilon_{k}^{e} - \delta\mu)\Big].
\end{equation}
We are allowing $\varepsilon_{k}^{e} > \delta\mu$ in order to allow $\Delta < \delta\mu$ which is the case when there is no energy gap left in the excitation spectrum.

Now note that by setting $\Delta = 0$, we get non-superfluid free-energy density $\Omega_{0}$ in which $\varepsilon_{k}^{e}$ reduces to $|k - e\bar{\mu}|$ which leads to the following integral expression
\begin{equation}
\Omega_{0} = 2\int\frac{d^{3}k}{(2\pi)^{3}}[(k - \mu_{1})\Theta(\mu_{1} - k) + (k - \mu_{2})\Theta(\mu_{2} - k) - 2k].
\end{equation}
In this form, we essentially recover the zero-temperature expression for the free energy $\Omega_{0} = \varepsilon - \mu_{1}n_{1} - \mu_{2}n_{2}$ with the energy density $\varepsilon$ and the charge densities of the two fermion species $n_{1}, \ n_{2}$, and after subtracting the vacuum contribution we obtain the following expected result,
\begin{equation}
\Omega_{0} = - \frac{\mu_{1}^{4}}{12\pi^{2}} - \frac{\mu_{2}^{4}}{12\pi^{2}}.
\end{equation}
Now, we are going to compute the free-energy density for the superfluid state. In order to compute the integral expression, we abbreviate the integral as
\begin{equation}
\Omega = \sum_{e}\int_{0}^{\infty}I_{\Delta}^{e},
\end{equation} 
where the integral stands for the integral over the modulus of the three-momentum $k$, and the gap $\Delta$ is assumed to be vanishing everywhere in momentum space except for a small vicinity around the Fermi surface, in this case around the average Fermi surface $k \in [\bar{\mu} - \delta, \bar{\mu} + \delta]$. Moreover, it is also assumed that $\Delta \ll \delta \ll \bar{\mu}$, and we are also going to assume that $\delta\mu$ is of the order of gap $\Delta$ such that $\delta\mu \ll \delta$. Then we may write
\begin{equation}
\begin{split}
\Omega & = \Omega_{0} + \sum_{e}\int_{\bar{\mu} - \delta}^{\bar{\mu} + \delta}I_{\Delta}^{e}
 - \sum_{e}\int_{\bar{\mu} - \delta}^{\bar{\mu} + \delta}I_{0}^{e}\\
 & \simeq \Omega_{0} + \int_{\bar{\mu} - \delta}^{\bar{\mu} + \delta}(I_{\Delta}^{+} - I_{0}^{+}), 
\end{split}
\end{equation}
where we have set the antiparticle gap to zero, and this is possible since at zero temperature and positive chemical potentials the anti-particles play no role. As a result, we have the following expression
\begin{equation}
\Omega = \Omega_{0} + \Delta\Omega,
\end{equation}
where
\begin{equation}
\begin{split}
\Delta\Omega & = - \frac{1}{\pi^{2}}\int_{\bar{\mu} - \delta}^{\bar{\mu} + \delta}dk \ k^{2}\Bigg[
\varepsilon_{k}^{+} - \frac{\Delta^{2}}{2\varepsilon_{k}^{+}} + \left(\delta\mu - \varepsilon_{k}^{+}
+ \frac{\Delta^{2}}{2\varepsilon_{k}^{+}}\right)\\
 & \times\Theta(\delta\mu - \varepsilon_{k}^{+}) - [|k - \bar{\mu}| + (\delta\mu - |k - \bar{\mu}|)
 \Theta(\delta\mu - |k - \bar{\mu}|)]\Bigg],
\end{split}
\end{equation}
is nothing but the free energy difference between the superfluid and non-superfluid phases. Now we are going to compute the above three integrals separately. From the first integral, we obtain
\begin{equation}
\begin{split}
\int_{\bar{\mu} - \delta}^{\bar{\mu} + \delta} & dk \ k^{2}\left(\varepsilon_{k}^{+}
 - \frac{\Delta^{2}}{2\varepsilon_{k}^{+}}\right) = \delta\sqrt{\delta^{2} + \Delta^{2}}\left(\bar{\mu}^{2} + \frac{\delta^{2}}{2} - \frac{\Delta^{2}}{4}\right)\\
 & + \frac{\Delta^{4}}{8}\log\left(\frac{\sqrt{\delta^{2} + \Delta^{2}} + \delta}{\sqrt{\delta^{2} + \Delta^{2}} - \delta}\right)\\
 & = \bar{\mu}^{2}\delta^{2} + \frac{\bar{\mu}^{2}\Delta^{2}}{2} + \frac{\delta^{4}}{2}
 + \mathcal{O}(\Delta^{4}).
\end{split}
\end{equation}
Using the fact $\delta \gg \delta\mu$, we may write the second integral as
\begin{equation}
\begin{split}
\int_{\bar{\mu} - \delta}^{\bar{\mu} + \delta} & dk \ k^{2}\left(\delta\mu - \varepsilon_{k}^{+}
+ \frac{\Delta^{2}}{2\varepsilon_{k}^{+}}\right)\Theta(\delta\mu - \varepsilon_{k}^{+})\\
= \Theta(\delta\mu - \Delta) & \int_{\bar{\mu} - \sqrt{\bar{\mu}^{2} - \Delta^{2}}}^{\bar{\mu} 
+ \sqrt{\bar{\mu}^{2} - \Delta^{2}}}dk \ k^{2}\left(\delta\mu - \varepsilon_{k}^{+}
+ \frac{\Delta^{2}}{2\varepsilon_{k}^{+}}\right)\\
= \Theta(\delta\mu - \Delta) & \Big[\delta\mu\sqrt{\delta\mu^{2} - \Delta^{2}}\left(\bar{\mu}^{2}
+ \frac{\delta\mu^{2}}{6} + \frac{\Delta^{2}}{12}\right)\\
 & - \frac{\Delta^{4}}{8}\log\left(\frac{\delta\mu + \sqrt{\delta\mu^{2} - \Delta^{2}}}{\delta\mu 
 - \sqrt{\delta\mu^{2} - \Delta^{2}}}\right)\Big]\\
= \Theta(\delta\mu - \Delta) & \delta\mu\bar{\mu}^{2}\sqrt{\delta\mu^{2} - \Delta^{2}}
 + \mathcal{O}(\Delta^{4}). 
\end{split}
\end{equation}
In the above, we have neglected terms of the form $\Delta^{4}, \delta\mu^{4}, \delta\mu^{2}\Delta^{2}$, and higher order terms. Lastly, the third contribution is the following
\begin{equation}
\begin{split}
\int_{\bar{\mu} - \delta}^{\bar{\mu} + \delta} & dk \ k^{2}[|k - \bar{\mu}| + (\delta\mu - |k - \bar{\mu}|)\Theta(\delta\mu - |k - \bar{\mu}|)]\\
 & = \bar{\mu}^{2}\delta^{2} + \frac{\delta^{4}}{2} + \bar{\mu}^{2}\delta\mu^{2}
 + \frac{\delta\mu^{4}}{6}.
\end{split}
\end{equation}
In order to be consistent with the previously omitted terms, we neglect the term $\delta\mu^{4}$ from the above contribution. Therefore, summing up all the contributions, we finally obtain the following expression
\begin{equation}
\Delta\Omega \simeq \frac{\bar{\mu}^{2}\delta\mu^{2}}{\pi^{2}} - \frac{\bar{\mu}^{2}\Delta^{2}}{2\pi^{2}} - \Theta(\delta\mu - \Delta)\frac{\bar{\mu}^{2}\delta\mu\sqrt{\delta\mu^{2} - \Delta^{2}}}{\pi^{2}}.
\end{equation}

\subsection{Chandrasekhar--Clogston limit}

We may note that when $\delta\mu \propto \bar{b} = 0$ (the case without mismatch), then $\mu_{1} = \mu_{2}$, $\mu \equiv \bar{\mu} = \mu_{1} = \mu_{2}$, and as a result, the free-energy density difference reduces to 
\begin{equation}
\Delta\Omega \simeq - \frac{\mu^{2}\Delta^{2}}{2\pi^{2}},
\end{equation}
This contribution to the free energy density is known as the condensation energy density and it essentially shows that the free energy density is lowered by the gap, and therefore, the superfluid state wins over the non-superfluid state for all nonzero values of $\Delta$.

Now we turn our discussion to the case of mismatch, but keep it smaller than the gap $0 < \delta\mu < \Delta$. In this case, we find that
\begin{equation}
\Delta\Omega \simeq \frac{\bar{\mu}^{2}\delta\mu^{2}}{\pi^{2}} - \frac{\bar{\mu}^{2}\Delta^{2}}{2\pi^{2}}.
\end{equation}
The above expression shows that the mismatch essentially induces an additional, positive, contribution to the free energy: now the system not only gain free energy from pairing but also have to pay a price in free energy. The superfluid state is now only preferred over the non-superfluid state for
\begin{equation}
\delta\mu < \frac{\Delta}{\sqrt{2}}.
\end{equation}
This is called the Chandrasekhar-Clogston limit. For $\delta\mu$ beyond this limit, the superfluid state breaks down, and it depends on the specific system under consideration whether $\Delta$ or $\delta\mu$ is larger.

\section{Application to the Dynamical Mismatch Model}
\label{sec:S5}

\subsection{Full free-energy difference with $\zeta$-dependent terms}

In order to apply the above formalism in our model, we consider a generic situation in which we only assume $\delta\mu \ll \delta \ll \bar{\mu}$ which leads to the following contribution to difference in free-energy density
\begin{equation}
\begin{split}
\Delta\Omega & = - \frac{1}{\pi^{2}}\Bigg[\delta\sqrt{\delta^{2} + \Delta^{2}}\left(\bar{\mu}^{2} + \frac{\delta^{2}}{2} - \frac{\Delta^{2}}{4}\right)\\
 & + \frac{\Delta^{4}}{8}\log\left(\frac{\sqrt{\delta^{2} + \Delta^{2}} + \delta}{\sqrt{\delta^{2} + \Delta^{2}} - \delta}\right)\Bigg]\\
 & - \frac{1}{\pi^{2}}\Theta(\delta\mu - \Delta)\Bigg[\delta\mu\sqrt{\delta\mu^{2} - \Delta^{2}}
 \left(\bar{\mu}^{2} + \frac{\delta\mu^{2}}{6} + \frac{\Delta^{2}}{12}\right)\\
 & - \frac{\Delta^{4}}{8}\log\left(\frac{\delta\mu + \sqrt{\delta\mu^{2} - \Delta^{2}}}{\delta\mu 
 - \sqrt{\delta\mu^{2} - \Delta^{2}}}\right)\Bigg]\\
 & - \frac{1}{\pi^{2}}\left(\bar{\mu}^{2}\delta^{2} + \frac{\delta^{4}}{2} + \bar{\mu}^{2}\delta\mu^{2}
 + \frac{\delta\mu^{4}}{6}\right) + \frac{2\zeta}{3}\delta\mu^{4},
\end{split}
\end{equation}
where the last term is essentially comes from our model where we express $\bar{b}$ in terms of $\delta\mu$. Assuming the weak coupling criterion for the gap $\Delta \ll \delta \ll \bar{\mu}$, we may write the above expression as
\begin{equation}
\begin{split}
\Delta\Omega & = - \frac{1}{\pi^{2}}\Bigg[\left(\delta^{2} + \frac{\Delta^{2}}{2}\right)\left(\bar{\mu}^{2} + \frac{\delta^{2}}{2} - \frac{\Delta^{2}}{4}\right) + \frac{\Delta^{4}}{4}\log\left(\frac{2\delta}{\Delta}\right)\\
 & + \Theta(\delta\mu - \Delta)\Big[\delta\mu\sqrt{\delta\mu^{2} - \Delta^{2}}\left(\bar{\mu}^{2} 
 + \frac{\delta\mu^{2}}{6} + \frac{\Delta^{2}}{12}\right)\\
 & - \frac{\Delta^{4}}{8}\log\left(\frac{\delta\mu + \sqrt{\delta\mu^{2} - \Delta^{2}}}{\delta\mu 
 - \sqrt{\delta\mu^{2} - \Delta^{2}}}\right)\Big]\\
 & + \left(\bar{\mu}^{2}\delta^{2} + \frac{\delta^{4}}{2} + \bar{\mu}^{2}\delta\mu^{2}
 + \frac{\delta\mu^{4}}{6}\right) - \frac{2\pi^{2}\zeta}{3}\delta\mu^{4}\Bigg].
\end{split}
\end{equation}
Now if we assume $\delta\mu < \Delta$, then the step function vanishes and the above expression reduces to the following form
\begin{equation}
\begin{split}
\Delta\Omega & = - \frac{1}{\pi^{2}}\Bigg[\left(\delta^{2} + \frac{\Delta^{2}}{2}\right)\left(\bar{\mu}^{2} + \frac{\delta^{2}}{2} - \frac{\Delta^{2}}{4}\right) + \frac{\Delta^{4}}{4}\log\left(\frac{2\delta}{\Delta}\right)\\
 & + \left(\bar{\mu}^{2}\delta^{2} + \frac{\delta^{4}}{2} + \bar{\mu}^{2}\delta\mu^{2}
 + \frac{\delta\mu^{4}}{6}\right) - \frac{2\pi^{2}\zeta}{3}\delta\mu^{4}\Bigg]\\
 & = - \frac{1}{\pi^{2}}\Bigg[2\bar{\mu}^{2}\delta^{2} + \delta^{4} + \bar{\mu}^{2}
 \left(\frac{\Delta^{2}}{2} + \delta\mu^{2}\right) + \frac{\Delta^{4}}{4}\left(\log\left(
 \frac{2\delta}{\Delta}\right) - \frac{1}{2}\right)\\
 & - \frac{2\pi^{2}}{3}\delta\mu^{4}\left(\zeta - \frac{1}{4\pi^{2}}\right)\Bigg].
\end{split}
\end{equation}

\subsection{Decomposition into $(\DOmega)_0$ and $(\DOmega)_1$}

Since we are interested in $\zeta > \frac{1}{4\pi^{2}}$ as in that condition only $\delta\mu \neq 0$, the above expression clearly shows that there is a competition. In order to see under what circumstances, the mismatch would be preferred, we decompose $\Delta\Omega$ in two parts --- (i) independent of $\delta\mu$ which we call $(\Delta\Omega)_{0}$ and (ii) the part which depends on $\delta\mu$ which denote by $(\Delta\Omega)_{1}$, and they are expressed as follows
\begin{equation}
\begin{split}
(\Delta\Omega)_{0} & = - \frac{1}{\pi^{2}}\Bigg[2\bar{\mu}^{2}\delta^{2} + \delta^{4} + \bar{\mu}^{2}
 \frac{\Delta^{2}}{2} +  + \frac{\Delta^{4}}{4}\left(\log\left(\frac{2\delta}{\Delta}\right)
 - \frac{1}{2}\right)\Bigg]\\
(\Delta\Omega)_{1} & = \frac{2}{3}\delta\mu^{4}\left(\zeta - \frac{1}{4\pi^{2}}\right) - \frac{1}{\pi^{2}}\bar{\mu}^{2}\delta\mu^{2}\\
 & = \bar{\mu}^{4}\Bigg[\frac{2}{3}\left(\frac{\delta\mu}{\bar{\mu}}\right)^{4}\left(\zeta - \frac{1}{4\pi^{2}}\right) - \frac{1}{\pi^{2}}\left(\frac{\delta\mu}{\bar{\mu}}\right)^{2}\Bigg],
\end{split}
\end{equation}
which clearly shows that $(\Delta\Omega)_{1} < 0$ as long as the following condition is satisfied
\begin{equation}
\Big|\frac{\delta\mu}{\bar{\mu}}\Big| < \sqrt{\frac{3}{2\pi^{2}\left(\zeta - \frac{1}{4\pi^{2}}\right)}}.
\end{equation}
On the other hand, we assumed that $\delta\mu \ll \bar{\mu}$. We may note that, these two conditions can be satisfied simultaneously. This shows clearly that mismatch would be preferred when $\delta\mu < \Delta$. Now we consider the other scenario, namely, when $\delta\mu > \Delta$ in which 
\begin{equation}
\begin{split}
\Delta\Omega & = - \frac{1}{\pi^{2}}\Bigg[\left(\delta^{2} + \frac{\Delta^{2}}{2}\right)\left(\bar{\mu}^{2} + \frac{\delta^{2}}{2} - \frac{\Delta^{2}}{4}\right) + \frac{\Delta^{4}}{4}\log\left(\frac{2\delta}{\Delta}\right)\\
 & + \Big[\delta\mu\sqrt{\delta\mu^{2} - \Delta^{2}}\left(\bar{\mu}^{2} + \frac{\delta\mu^{2}}{6} 
 + \frac{\Delta^{2}}{12}\right)\\
 & - \frac{\Delta^{4}}{8}\log\left(\frac{\delta\mu + \sqrt{\delta\mu^{2} - \Delta^{2}}}{\delta\mu 
 - \sqrt{\delta\mu^{2} - \Delta^{2}}}\right)\Big]\\
 & + \left(\bar{\mu}^{2}\delta^{2} + \frac{\delta^{4}}{2} + \bar{\mu}^{2}\delta\mu^{2}
 + \frac{\delta\mu^{4}}{6}\right) - \frac{2\pi^{2}\zeta}{3}\delta\mu^{4}\Bigg]\\
 & = - \frac{1}{\pi^{2}}\Bigg[2\bar{\mu}^{2}\delta^{2} + \delta^{4} + \bar{\mu}^{2}\frac{\Delta^{2}}{2}
 + \frac{\Delta^{4}}{2}\left(\log\left(\frac{2\delta}{\Delta}\right) - \frac{1}{2}\right)\Bigg]\\
 & + \frac{2}{3}\delta\mu^{4}\left(\zeta - \frac{1}{4\pi^{2}}\right) - \frac{1}{\pi^{2}}\bar{\mu}^{2}
 \delta\mu^{2}\\
 & - \frac{1}{\pi^{2}}\Big[\delta\mu\sqrt{\delta\mu^{2} - \Delta^{2}}\left(\bar{\mu}^{2} + \frac{\delta
 \mu^{2}}{6} + \frac{\Delta^{2}}{12}\right)\\
 & - \frac{\Delta^{4}}{8}\log\left(\frac{\delta\mu + \sqrt{\delta\mu^{2} - \Delta^{2}}}{\delta\mu 
 - \sqrt{\delta\mu^{2} - \Delta^{2}}}\right)\Big].
\end{split}
\end{equation}
As earlier, in the last equality, we have separated the $\delta\mu$ independent and dependent terms, and now we are going to look at $\delta\mu$ dependent terms by rewriting them in the following manner by first defining $\delta_{1} = \frac{\delta\mu}{\bar{\mu}}, \ \delta_{2} = \frac{\Delta}{\bar{\mu}}$
\begin{equation}
\begin{split}
(\Delta\Omega)_{1}(\delta_{1}, \delta_{2}) & = \bar{\mu}^{4}\Bigg[\frac{2}{3}\delta_{1}^{4}\left(\zeta - \frac{1}{4\pi^{2}}\right) - \frac{1}{\pi^{2}}\delta_{1}^{2}\\
 & - \frac{1}{\pi^{2}}\Big[\delta_{1}\sqrt{\delta_{1}^{2} - \delta_{2}^{2}}\left(1 + \frac{\delta_{1}
 ^{2}}{6} + \frac{\delta_{2}^{2}}{12}\right)\\
 & - \frac{\delta_{2}^{4}}{8}\log\left(\frac{\delta_{1} + \sqrt{\delta_{1}^{2} - \delta_{2}^{2}}}
 {\delta_{1}  - \sqrt{\delta_{1}^{2} - \delta_{2}^{2}}}\right)\Big]\Bigg].
\end{split}
\end{equation}
The color plot of the function $\frac{(\Delta\Omega)_{1}(\delta_{1}, \delta_{2})}{\bar{\mu}^{4}}$ is shown below in figure \ref{Figure 1}.
\begin{figure}[b]
\includegraphics[height = 12cm, width = 15cm]{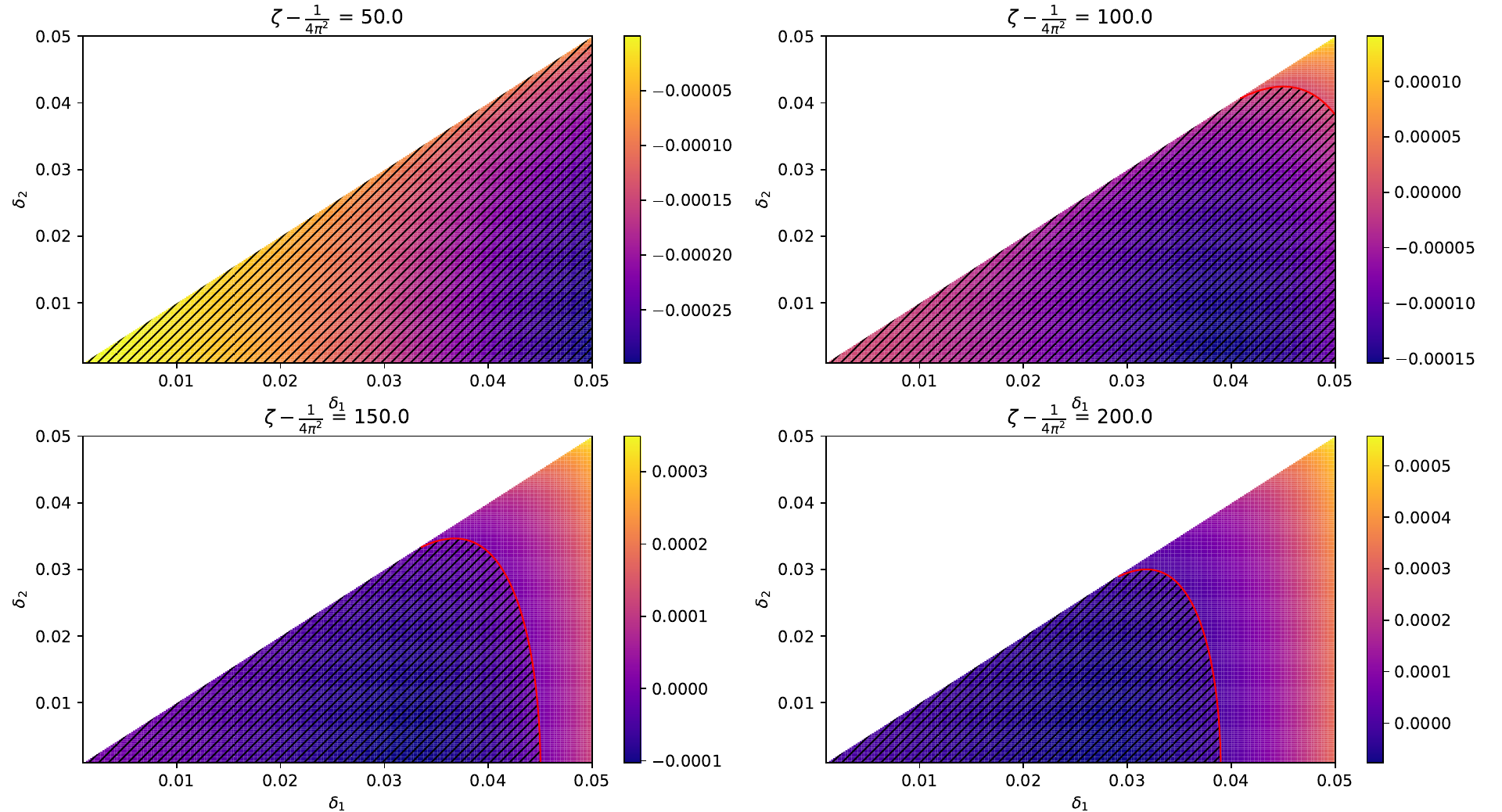}
\caption{Color plot of the function $\frac{(\Delta\Omega)_{1}(\delta_{1}, \delta_{2})}{\bar{\mu}^{4}}$ with a marked region in which the function takes negative value}
\label{Figure 1}
\end{figure}

\section{Fluctuations around the mean field in the two-component Dirac field}
In the $SU(2)$ symmetry breaking phase, we may write the mean-field Lagrangian density as
\begin{equation}
\mathcal{L}_{MF} = \bar{\psi}\begin{bmatrix}
[G_{0}^{+}]^{-1} & \Phi^{-}\\
\Phi^{+} & [G_{0}^{-}]^{-1}
\end{bmatrix}\psi
+ \frac{\text{Tr}[\Phi^{+}\Phi^{-}]}{4G},
\end{equation}
where $\Phi^{\pm}$ and $[G_{0}^{\pm}]^{-1}$ were defined in previous section.

In order to introduce fluctuations around the mean-field, it is essential that we allow for the gap $\Delta$ to become complex. Denoting the bosonic fluctuation field by $\eta(X)\in\mathbb{C}$, we can 
write
\begin{equation}
\Delta(X) = \Delta e^{2i\vec{q}.\vec{x}} + \eta(X),
\end{equation}
where $\Delta$ is the constant value of the gap parameter, determined from the gap equation. In addition to the fluctuations, we have also introduced a topological mode, determined by the externally given three-vector $\vec{q}$, and it is necessary to introduce a superflow, and thus required to describe the hydrodynamics of the superfluid.

The fluctuations $\eta(X)$ describe the Goldstone mode and a massive mode, and they must be considered as a dynamical field \textit{w.r.t} which the partition function becomes
\begin{equation}
\mathcal{Z} = \int\mathcal{D}\bar{\psi}\mathcal{D}\psi\mathcal{D}\eta^{*}\mathcal{D}\eta \ e^{-S_{E}}.
\end{equation}
Due to the superflow, the inverse fermionic propagator now depends on $\bf{x}$ in a nontrivial way. With a simple transformation of the fields, however, we can get rid of this dependence, and these transformations are given by
\begin{equation}
\psi'(X) = e^{i\vec{q}.\vec{x}}\psi(X), \ \eta'(X) = e^{-2i\vec{q}.\vec{x}}\eta(X).
\end{equation}
In terms of the above transformed fields, the mean-field Lagrangian with fluctuations can be expressed as
\begin{equation}
\mathcal{L}_{MF + fl} = \bar{\psi}'(\mathcal{G}^{-1} + h)\psi' - \frac{\Delta^{2}}{G} - \frac{\Delta}{G}(\eta' + \eta'^{*}) - \frac{|\eta'|^{2}}{G}
\end{equation}
where
\begin{equation}
h = \begin{bmatrix}
0 & -\eta'^{*}\gamma_{5}\tau_{1}\\
\eta'\gamma_{5}\tau_{1} & 0
\end{bmatrix}
\end{equation}
and
\begin{equation}
\mathcal{G}^{-1} = \begin{bmatrix}
i\slashed{\partial} + \gamma^{0}\mu_{1} + \vec{\gamma}.\vec{q} & 0 & 0 & - \Delta\gamma_{5}\\
0 & i\slashed{\partial} - \gamma^{0}\mu_{1} - \vec{\gamma}.\vec{q} & - \Delta\gamma_{5} & 0\\
0 & \Delta\gamma_{5} & i\slashed{\partial} + \gamma^{0}\mu_{2} + \vec{\gamma}.\vec{q} & 0\\
\Delta\gamma_{5} & 0 & 0 & i\slashed{\partial} - \gamma^{0}\mu_{2} - \vec{\gamma}.\vec{q}
\end{bmatrix}. 
\end{equation}
In the basis of the field $\psi'$ the $x$ dependence in the off-diagonal components of $\mathcal{G}
^{-1}$ is gone, and the superflow appears in the form of the term $\pm\vec{\gamma}.\vec{q}$ in the diagonal components.

As the fermionic fields only appear quadratically in the action, we can integrate them out which leads to the partition function
\begin{equation}
\mathcal{Z} = \int\mathcal{D}\eta'^{*}\mathcal{D}\eta' \ e^{-S_{E}[\eta'^{*},\eta']}
\end{equation}
where
\begin{equation}
\begin{split}
S_{E}[\eta'^{*},\eta'] & = - \frac{1}{2}\int_{X}\text{Tr}\log\left(\mathcal{G}^{-1} + h\right)\\
 & + \int_{X}\left(\frac{\Delta^{2}}{G} + \frac{\Delta}{G}(\eta' + \eta'^{*}) + \frac{|\eta'|^{2}}
{G}\right),
\end{split} 
\end{equation}
where the trace is taken over Nambu-Gorkov space and Dirac space. Expanding the trace of logarithm 
term for small fluctuations, we may write 
\begin{equation}
\text{Tr}\log\left(\mathcal{G}^{-1} + h\right) = \text{Tr}\log\mathcal{G}^{-1} + \text{Tr}\log
(1 + \mathcal{G}h),
\end{equation} 
and then we use the expansion of the logarithm $\log(1 + x) = x - \frac{x^{2}}{2} + \frac{x^{3}}{3}
+ \ldots$. Further, keeping terms up to second order in $h$ and writing the space-time arguments explicitly, we obtain
\begin{equation}
\begin{split}
\frac{1}{2}\int_{X} & \text{Tr}\log(\mathcal{G}^{-1} + h) = \frac{1}{2}\int_{X}\text{Tr}\log
\mathcal{G}^{-1} + \frac{1}{2}\int_{X}\text{Tr}[\mathcal{G}(X, X)h(X)]\\
 & - \frac{1}{4}\int_{X, Y}\text{Tr}[\mathcal{G}(X, Y)h(Y)\mathcal{G}(Y, X)h(X)] + \mathcal{O}
 (\eta'^{3}).
\end{split}
\end{equation}
Combining all of the above information, we may write the total action as
\begin{equation}
S_{E}[\eta'^{*}, \eta'] = S^{(0)} + S^{(1)} + S^{(2)},
\end{equation}
where
\begin{equation}
\begin{split}
S^{(0)} & = - \frac{1}{2}\int_{X}\text{Tr}\log\mathcal{G}^{-1} + \frac{V}{T}\frac{\Delta^{2}}{G}\\
S^{(1)} & = - \frac{1}{2}\int_{X}\text{Tr}[\mathcal{G}(X, X)h(X)] + \frac{\Delta}{G}\int_{X}
(\eta'^{*} + \eta')\\
S^{(2)} & = \frac{1}{4}\int_{X, Y}\text{Tr}[\mathcal{G}(X, Y)h(Y)\mathcal{G}(Y, X)h(X)] + 
\int_{X}\frac{\eta'^{*}\eta'}{G},
\end{split}
\end{equation}
where we replaced the trivial space-time integral in the imaginary time formalism of thermal 
field theory by $\frac{V}{T}$. Moreover, note that $S^{(0)}$ is the purely fermionic mean-field
effective action that does not contain any fluctuations. Hence, the partition function reduces
to the following form
\begin{equation}
\mathcal{Z} = e^{-S^{(0)}}\int\mathcal{D}\eta'^{*}\mathcal{D}\eta' \ e^{- S^{(1)} - S^{(2)}}.
\end{equation}
In order to evaluate the contributions coming from $S^{(1)}$ and $S^{(2)}$, we first write the 
propagator in Nambu-Gorkov space in terms of both normal and anomalous propagators, 
\begin{equation}
\mathcal{G}(X, Y) = \begin{bmatrix}
G^{+}(X, Y) & F^{-}(X, Y)\\
F^{+}(X, Y) & G^{-}(X, Y)
\end{bmatrix}.
\end{equation}
Now, introducing the Fourier transforms for the propagators as
\begin{equation}
\begin{split}
G^{\pm}(X - Y) & = \frac{T}{V}\sum_{K}e^{-iK.(X - Y)}G^{\pm}(K)\\
F^{\pm}(X - Y) & = \frac{T}{V}\sum_{K}e^{-iK.(X - Y)}F^{\pm}(K),
\end{split}
\end{equation}
assuming translational invariance, we may now write
\begin{equation}
\begin{split}
S^{(1)} & = - \frac{1}{2}\text{Tr}\int_{X}\Big[\frac{T}{V}\sum_{K}F^{-}(K) + \frac{\Phi^{-}}{2G}\Big]
\tau_{1}\gamma_{5}\eta'(X)\\
 & + \frac{1}{2}\text{Tr}\int_{X}\Big[\frac{T}{V}\sum_{K}F^{+}(K) + \frac{\Phi^{+}}{2G}\Big]
\tau_{1}\gamma_{5}\eta'^{*}(X)
\end{split}
\end{equation}
where the trace is taken over both Dirac space and flavor space, and we have introduced $\Phi^{\pm} = \pm\Delta\gamma_{5}\tau_{1}$. The reason for this particular way of writing the result is that we recover the mean-field gap equation, and therefore, at the point where the mean-field gap equation is fulfilled, we have $S^{(1)} = 0$. The reason is that the gap equation is obtained by minimizing the free energy with
respect to $\Delta$, and this is nothing but looking for the point where the variation of the gap vanishes to linear order. Therefore, it is clear that $S^{(1)}$ must vanish at the mean-field solution.
Therefore, we are left with $S^{(2)}$ which we evaluate next. For the fluctuation fields, we introduce real and imaginary parts in the following manner
\begin{equation}
\eta'(X) = \frac{1}{\sqrt{2}}[\eta_{1}'(X) + i\eta_{2}'(X)],
\end{equation}
and their Fourier transforms,
\begin{equation}
\eta_{i}'(X) = \frac{1}{\sqrt{TV}}\sum_{K}e^{-iK.X}\eta_{i}'(K), \ i = 1, 2.
\end{equation}
For the second term on the \textit{r.h.s} of $S^{(2)}$, we obtain
\begin{equation}
\int_{X}\frac{\eta'^{*}\eta'}{G} = \frac{1}{2T^{2}}\sum_{K}\left(\frac{\eta_{1}'(K)\eta_{1}'(-K)}{G}
+ \frac{\eta_{2}'(K)\eta_{2}'(-K)}{G}\right).
\end{equation}
The trace over Nambu-Gorkov space in the first term on the \textit{r.h.s} leads to
\begin{equation}
\begin{split}
\text{Tr}[\mathcal{G}h\mathcal{G}h] & = - \text{Tr}[G^{+}\eta'^{*}\tau_{1}\gamma_{5}G^{-}\eta'\tau_{1}\gamma_{5}] - \text{Tr}[G^{-}\eta'\tau_{1}\gamma_{5}G^{+}\eta'^{*}\tau_{1}\gamma_{5}]\\
 & + \text{Tr}[F^{+}\eta'^{*}\tau_{1}\gamma_{5}F^{+}\eta'^{*}\tau_{1}\gamma_{5}] + \text{Tr}[F^{-}\eta'\tau_{1}\gamma_{5}F^{-}\eta'\tau_{1}\gamma_{5}],
\end{split}
\end{equation}
where we have omitted all space-time arguments for the sake of simplicity of notation. Going to 
momentum space and to the basis of $\eta_{1}', \eta_{2}'$ yields for the first of these terms
\begin{equation}
\begin{split}
\text{Tr} & [G^{+}\eta'^{*}\tau_{1}\gamma_{5}G^{-}\eta'\tau_{1}\gamma_{5}]\\
 & = \frac{1}{2TV}\sum_{K, P}[\eta_{1}'(K)\eta_{1}'(-K) + \eta_{2}'(K)\eta_{2}'(-K) - i\eta_{1}'(K)\eta_{2}'(-K) + i\eta_{2}'(K)\eta_{1}'(-K)]\\
 & \times \text{Tr}[G^{+}(P)\tau_{1}\gamma_{5}G^{-}(P + K)\tau_{1}\gamma_{5}],
\end{split}
\end{equation}
and analogously for the three other terms. In order to write the result in a compact manner, we 
introduce the following abbreviations,
\begin{equation}
\begin{split}
\Pi^{\pm}(K) & \equiv \frac{1}{2}\frac{T}{V}\sum_{P}\text{Tr}[G^{\pm}(P)\gamma_{5}\tau_{1}
G^{\mp}(P + K)\gamma_{5}\tau_{1}]\\
\Sigma^{\pm}(K) & \equiv \frac{1}{2}\frac{T}{V}\sum_{P}\text{Tr}[F^{\pm}(P)\gamma^{5}\tau_{1}
F^{\pm}(P + K)\gamma_{5}\tau_{1}]. 
\end{split}
\end{equation}
Putting everything together, the quadratic contribution to the action $S^{(2)}$ can be expressed 
as
\begin{equation}
S^{(2)} = \frac{1}{2}\sum_{K}\eta'(K)\frac{D^{-1}(K)}{T^{2}}\eta'^{\dagger}(K), \ \eta'^{\dagger}
(K) = \eta'(-K),
\end{equation}
where
\begin{equation}
D^{-1}(K) = \begin{bmatrix}
\frac{1}{G} - \frac{\Pi^{+} + \Pi^{-}}{2} + \frac{\Sigma^{+} + \Sigma^{-}}{2} & i\frac{\Pi^{+} - \Pi^{-}}{2} - i\frac{\Sigma^{+} - \Sigma^{-}}{2}\\
- i\frac{\Pi^{+} - \Pi^{-}}{2} - i\frac{\Sigma^{+} - \Sigma^{-}}{2} & \frac{1}{G} - \frac{\Pi^{+} + \Pi^{-}}{2} - \frac{\Sigma^{+} + \Sigma^{-}}{2}
\end{bmatrix},
\end{equation}
and the two component field
\begin{equation}
\eta'(K) \equiv \begin{bmatrix}
\eta_{1}'(K) & \eta'_{2}(K)
\end{bmatrix}.
\end{equation}
Now using the following properties
\begin{equation}
G^{+}(-K) = - G^{-}(K), \ F^{+}(-K) = - F^{-}(K),
\end{equation}
which can be checked easily from the mathematical form of the propagators given in previous section. Therefore, we also find the relations $\Pi^{+}(K) = \Pi^{-}(-K)$ and $\Sigma^{+}(K) = \Sigma^{-}(K)$.
The above relations make it possible to express everything in terms of $\Sigma^{+}$ and $\Pi^{+}$, and
therefore, for convenience, we drop the plus sign from the superscript. Imposing these relations, we find
\begin{equation}
D^{-1}(K) = \begin{bmatrix}
\frac{1}{G} - \bar{\Pi}(K) + \Sigma(K) & i\delta\Pi(K)\\
-i\delta\Pi(K) & \frac{1}{G} - \bar{\Pi}(K) - \Sigma(K)
\end{bmatrix}
\end{equation}
where
\begin{equation}
\bar{\Pi}(K) = \frac{\Pi(K) + \Pi(-K)}{2}, \ \delta\Pi(K) = \frac{\Pi(K) - \Pi(-K)}{2}. 
\end{equation}
Before going into evaluation of the boson propagator explicitly for small momenta, we may write down an expression formally for the free energy density of the system which can be obtained by performing the integration over the fluctuations in the partition function 
\begin{equation}
\Omega = \frac{\Delta^{2}}{G} - \frac{1}{2}\frac{T}{V}\sum_{K}\log\left(\frac{\mathcal{G}^{-1}}{T}\right)
+ \frac{1}{2}\frac{T}{V}\sum_{K}\text{Tr}\log\left(\frac{D^{-1}(K)}{T^{2}}\right).
\end{equation}
In the following, we set the superflow to zero for the sake of mathematical simplicity, $\vec{q} = 0$, such that we can work with the fermionic propagators in momentum space that we already derived
\begin{equation}
\begin{split}
G^{\pm}(K) & = \sum_{e}\gamma^{0}\Lambda_{k}^{\mp e}\begin{bmatrix}
\frac{k_{0} \mp (\mu_{2} - ek)}{(k_{0} \pm \delta\mu)^{2} - (\varepsilon_{k}^{e})^{2}} & 0\\
0 & \frac{k_{0} \mp (\mu_{1} - ek)}{(k_{0} \mp \delta\mu)^{2} - (\varepsilon_{k}^{e})^{2}}
\end{bmatrix}\\
F^{\pm}(K) & = \pm\sum_{e}\Delta\gamma_{5}\Lambda_{k}^{\mp e}\begin{bmatrix}
0 & \frac{1}{(k_{0} \mp \delta\mu)^{2} - (\varepsilon_{k}^{e})^{2}}\\
\frac{1}{(k_{0} \pm \delta\mu)^{2} - (\varepsilon_{k}^{e})^{2}} & 0
\end{bmatrix}
\end{split}
\end{equation}
Defining $\xi_{k}^{e} = k - e\bar{\mu}$ we may write $\varepsilon_{k}^{e} = \sqrt{(\xi_{k}^{e})
^{2} + \Delta^{2}}$ where
\begin{equation}
\frac{\Delta}{G} = \sum_{e = \pm}\int\frac{d^{3}p}{(2\pi)^{3}}\frac{\Delta}{2\varepsilon_{p}^{e}}
[1 - 2f(\varepsilon_{p}^{e})].
\end{equation}
In order to compute $\Pi(K)$ and $\Sigma(K)$, we define $Q \equiv P + K$ and
\begin{equation}
\begin{split}
\varepsilon_{1} \equiv \varepsilon_{p}^{e_{1}}, & \ \varepsilon_{2} \equiv \varepsilon_{q}^{e_{2}}\\
\xi_{1} \equiv p - e_{1}\bar{\mu}, & \ \xi_{2} \equiv q - e_{2}\bar{\mu}. 
\end{split}
\end{equation}
With the above abbreviations, and the following result
\begin{equation}
-\text{Tr}[\gamma^{0}\Lambda_{p}^{-e_{1}}\gamma_{5}\gamma^{0}\Lambda_{q}^{e_{2}}\gamma_{5}]
= 1 + e_{1}e_{2}\hat{p}.\hat{q},
\end{equation}
we obtain
\begin{equation}
\begin{split}
\Pi(K) & = - \frac{1}{2}\frac{T}{V}\sum_{P}\sum_{e_{1}, e_{2}}(1 + e_{1}e_{2}\hat{p}.\hat{q})\Bigg[
\frac{(p_{0} + \delta\mu) + e_{1}\xi_{1}}{(p_{0} + \delta\mu)^{2} - \varepsilon_{1}^{2}}\\
 & \times \frac{(q_{0} - \delta\mu) - e_{2}\xi_{2}}{(q_{0} - \delta\mu)^{2} - \varepsilon_{2}^{2}}
 + \frac{(p_{0} - \delta\mu) + e_{1}\xi_{1}}{(p_{0} - \delta\mu)^{2} - \varepsilon_{1}^{2}}\\
 & \times \frac{(q_{0} + \delta\mu) - e_{2}\xi_{2}}{(q_{0} + \delta\mu)^{2} - \varepsilon_{2}^{2}}
 \Bigg]\\
\implies\Pi(-K) & = - \frac{1}{2}\frac{T}{V}\sum_{P}\sum_{e_{1}, e_{2}}(1 + e_{1}e_{2}\hat{p}.\hat{q})\Bigg[\frac{(-p_{0} + \delta\mu) + e_{1}\xi_{1}}{(p_{0} - \delta\mu)^{2} - \varepsilon_{1}^{2}}\\
 & \times \frac{(-q_{0} - \delta\mu) - e_{2}\xi_{2}}{(q_{0} + \delta\mu)^{2} - \varepsilon_{2}^{2}}
 + \frac{(- p_{0} - \delta\mu) + e_{1}\xi_{1}}{(p_{0} + \delta\mu)^{2} - \varepsilon_{1}^{2}}\\
 & \times \frac{(- q_{0} + \delta\mu) - e_{2}\xi_{2}}{(q_{0} - \delta\mu)^{2} - \varepsilon_{2}^{2}}
 \Bigg]\\
 & = - \frac{1}{2}\frac{T}{V}\sum_{P}\sum_{e_{1}, e_{2}}(1 + e_{1}e_{2}\hat{p}.\hat{q})\Bigg[\frac{
 (p_{0} - \delta\mu) - e_{1}\xi_{1}}{(p_{0} - \delta\mu)^{2} - \varepsilon_{1}^{2}}\\
 & \times \frac{(q_{0} + \delta\mu) + e_{2}\xi_{2}}{(q_{0} + \delta\mu)^{2} - \varepsilon_{2}^{2}}
 + \frac{(p_{0} + \delta\mu) - e_{1}\xi_{1}}{(p_{0} + \delta\mu)^{2} - \varepsilon_{1}^{2}}\\
 & \times \frac{(q_{0} - \delta\mu) + e_{2}\xi_{2}}{(q_{0} - \delta\mu)^{2} - \varepsilon_{2}^{2}}
 \Bigg]. 
\end{split}
\end{equation}
The above relations lead to the following expression of $\bar{\Pi}(K)$
\begin{equation}
\begin{split}
\bar{\Pi}(K) & = - \frac{1}{2}\frac{T}{V}\sum_{e_{1}, e_{2}}\sum_{P}(1 + e_{1}e_{2}\hat{p}.\hat{q})\\ 
 & \times\Bigg[
\frac{(p_{0} + \delta\mu)(q_{0} - \delta\mu) - e_{1}e_{2}\xi_{1}\xi_{2}}{[(p_{0} + \delta\mu)^{2} 
- \varepsilon_{1}^{2}][(q_{0} - \delta\mu)^{2} - \varepsilon_{2}^{2}]}\\
 & + \frac{(p_{0} - \delta\mu)(q_{0} + \delta\mu) - e_{1}e_{2}\xi_{1}\xi_{2}}{[(p_{0} - \delta\mu)^{2} 
- \varepsilon_{1}^{2}][(q_{0} + \delta\mu)^{2} - \varepsilon_{2}^{2}]}\Bigg]\\
 & = - \frac{1}{2}\frac{T}{V}\sum_{e_{1}, e_{2}}\sum_{P}(1 + e_{1}e_{2}\hat{p}.\hat{q})\frac{1}{4}\Bigg[\Big[\frac{1}{p_{0} + \delta\mu - e_{1}\varepsilon_{1}}\\
 & + \frac{1}{p_{0} + \delta\mu + e_{1}\varepsilon_{1}}\Big]\Big[\frac{1}{q_{0} - \delta\mu + e_{2}
 \varepsilon_{2}} + \frac{1}{q_{0} - \delta\mu - e_{2}\varepsilon_{2}}\Big]\\
 & - \frac{e_{1}e_{2}\xi_{1}\xi_{2}}{e_{1}e_{2}\varepsilon_{1}\varepsilon_{2}}\Big[\frac{1}{p_{0} 
 + \delta\mu - e_{1}\varepsilon_{1}}\\
 & - \frac{1}{p_{0} + \delta\mu + e_{1}\varepsilon_{1}}\Big]\Big[\frac{1}{q_{0} - \delta\mu - e_{2}
 \varepsilon_{2}} - \frac{1}{q_{0} - \delta\mu + e_{2}\varepsilon_{2}}\Big]\Bigg]
\end{split}
\end{equation}
On the other hand, $\delta\Pi(K)$ can be expressed as
\begin{equation}
\begin{split}
\delta\Pi(K) & = \frac{1}{2}\frac{T}{V}\sum_{e_{1}, e_{2}}\sum_{P}(1 + e_{1}e_{2}\hat{p}.\hat{q})\\
 & \times\Bigg[\frac{- e_{1}\xi_{1}(q_{0} - \delta\mu) + e_{2}\xi_{2}(p_{0} + \delta\mu)}{
 [(p_{0} + \delta\mu)^{2} - \varepsilon_{1}^{2}][(q_{0} - \delta\mu)^{2} - \varepsilon_{2}^{2}]}\\
 & - \frac{e_{1}\xi_{1}(q_{0} + \delta\mu) - e_{2}\xi_{2}(p_{0} - \delta\mu)}{[(p_{0} - \delta\mu)^{2} 
 - \varepsilon_{1}^{2}][(q_{0} + \delta\mu)^{2} - \varepsilon_{2}^{2}]}\Bigg]\\
 & = \frac{1}{2}\frac{T}{V}\sum_{e_{1}, e_{2}}\sum_{P}(1 + e_{1}e_{2}\hat{p}.\hat{q})\\
 & \times\Bigg[- \frac{\xi_{1}}{4\varepsilon_{1}}\left(\frac{1}{q_{0} - \delta\mu + e_{2}\varepsilon_{2}}
 + \frac{1}{q_{0} - \delta\mu - e_{2}\varepsilon_{2}}\right)\\
 & \times\left(\frac{1}{p_{0} + \delta\mu - e_{1}\varepsilon_{1}} - \frac{1}{p_{0} + \delta\mu + e_{1}
 \varepsilon_{1}}\right)\\
 & + \frac{\xi_{2}}{4\varepsilon_{2}}\left(\frac{1}{p_{0} + \delta\mu - e_{1}\varepsilon_{1}} + 
 \frac{1}{p_{0} + \delta\mu + e_{1}\varepsilon_{1}}\right)\\
 & \times\left(\frac{1}{q_{0} - \delta\mu - e_{2}\varepsilon_{2}}
 - \frac{1}{q_{0} - \delta\mu + e_{2}\varepsilon_{2}}\right)\\
 & - \frac{\xi_{1}}{4\varepsilon_{1}}\left(\frac{1}{p_{0} - \delta\mu - e_{1}\varepsilon_{1}} - \frac{1}
 {p_{0} - \delta\mu + e_{1}\varepsilon_{1}}\right)\\
 & \times\left(\frac{1}{q_{0} + \delta\mu - e_{2}\varepsilon_{2}}
 + \frac{1}{q_{0} + \delta\mu + e_{2}\varepsilon_{2}}\right)\\
 & + \frac{\xi_{2}}{4\varepsilon_{2}}\left(\frac{1}{p_{0} - \delta\mu - e_{1}\varepsilon_{1}} + \frac{1}
 {p_{0} - \delta\mu + e_{1}\varepsilon_{1}}\right)\\
 & \times\left(\frac{1}{q_{0} + \delta\mu - e_{2}\varepsilon_{2}}
 - \frac{1}{q_{0} + \delta\mu + e_{2}\varepsilon_{2}}\right)\Bigg].
\end{split}
\end{equation}
Finally, the expression of $\Sigma(K)$ is given by
\begin{equation}
\begin{split}
\Sigma(K) & = \frac{1}{2}\frac{T}{V}\sum_{e_{1}, e_{2}}\sum_{P}(1 + e_{1}e_{2}\hat{p}.\hat{q})
\Delta^{2}\Bigg[\frac{1}{[(p_{0} - \delta\mu)^{2} - \varepsilon_{1}^{2}]}\\
 & \times \frac{1}{[(q_{0} - \delta\mu)^{2} - \varepsilon_{2}^{2}]} + \frac{1}{[(p_{0} + \delta\mu)^{2} 
 - \varepsilon_{1}^{2}]}\\
 & \times\frac{1}{[(q_{0} + \delta\mu)^{2} - \varepsilon_{2}^{2}]}\Bigg]\\
 & = \frac{1}{2}\frac{T}{V}\sum_{e_{1}, e_{2}}\sum_{P}(1 + e_{1}e_{2}\hat{p}.\hat{q})
\Delta^{2}\Bigg[\frac{1}{4e_{1}e_{2}\varepsilon_{1}\varepsilon_{2}}\\
 & \left(\frac{1}{p_{0} - \delta\mu - e_{1}\varepsilon_{1}} - \frac{1}{p_{0} - \delta\mu + e_{1}\varepsilon_{1}}\right)\\
 & \times \left(\frac{1}{q_{0} - \delta\mu - e_{2}\varepsilon_{2}} - \frac{1}{q_{0} - \delta\mu + e_{2}\varepsilon_{2}}\right)\\
 & + \frac{1}{4e_{1}e_{2}\varepsilon_{1}\varepsilon_{2}}\left(\frac{1}{p_{0} + \delta\mu - e_{1}\varepsilon_{1}} - \frac{1}{p_{0} + \delta\mu + e_{1}\varepsilon_{1}}\right)\\
 & \times \left(\frac{1}{q_{0} + \delta\mu - e_{2}\varepsilon_{2}} - \frac{1}{q_{0} + \delta\mu + e_{2}\varepsilon_{2}}\right)\Bigg].
\end{split}
\end{equation}
After doing the Matsubara summation, we obtain the following results 
\begin{equation}
\begin{split}
\Sigma(K) & = \frac{\Delta^{2}}{8V}\sum_{e_{1},e_{2}=\pm1}\sum_{\bm{p}}
\big(1 + e_{1}e_{2}\,\hat{p}\!\cdot\!\hat{q}\big)\\
 & \times \Bigg\{
\frac{\,n_{F}(e_{1}\varepsilon_{1}+\delta\mu)
     - n_{F}(e_{2}\varepsilon_{2}+\delta\mu)\,}
     {e_{1}\varepsilon_{1}e_{2}\varepsilon_{2}\,[k_{0} - (e_{2}\varepsilon_{2}-e_{1}\varepsilon_{1})]}
\\
 & +\frac{\,n_{F}(e_{1}\varepsilon_{1}-\delta\mu)
       - n_{F}(e_{2}\varepsilon_{2}-\delta\mu)\,}
       {e_{1}\varepsilon_{1}e_{2}\varepsilon_{2}\,[k_{0} - (e_{2}\varepsilon_{2}-e_{1}\varepsilon_{1})]}
\Bigg\}\\
\bar{\Pi}(K) & = -\frac{1}{8V}\sum_{e_{1},e_{2}=\pm1}\sum_{\bm{p}}\big(1 + e_{1}e_{2}\,\hat{p}\!\cdot\!\hat{q}\big)\\
 & \times \Bigg\{(1+\Xi)\,
\frac{\,n_{F}(E_{1}^{-}) - n_{F}(E_{2}^{+})\,}
     {\,k_{0} - \big(E_{2}^{+} - E_{1}^{-}\big)\,}
\\
 & +(1-\Xi)\,
\frac{\,n_{F}(E_{1}^{-}) - n_{F}(E_{2}^{-})\,}
     {\,k_{0} - \big(E_{2}^{-} - E_{1}^{-}\big)\,}
\\
 & +(1-\Xi)\,
\frac{\,n_{F}(E_{1}^{+}) - n_{F}(E_{2}^{+})\,}
     {\,k_{0} - \big(E_{2}^{+} - E_{1}^{+}\big)\,}
\\
 & +(1+\Xi)\,
\frac{\,n_{F}(E_{1}^{+}) - n_{F}(E_{2}^{-})\,}
     {\,k_{0} - \big(E_{2}^{-} - E_{1}^{+}\big)\,}
\Bigg\}\\
\delta\Pi(K) & = \frac{1}{8V}\sum_{e_{1},e_{2}=\pm1}\sum_{\bm{p}}
\big(1 + e_{1}e_{2}\,\hat{p}\!\cdot\!\hat{q}\big)\\
 & \times\Bigg\{
\Big(-\frac{\xi_{1}}{\varepsilon_{1}} - \frac{\xi_{2}}{\varepsilon_{2}}\Big)
\frac{\,n_{F}(E_{1}^{-}) - n_{F}(E_{2}^{+})\,}
     {\,k_{0} - \big(E_{2}^{+} - E_{1}^{-}\big)\,}
\\
 & +\Big(-\frac{\xi_{1}}{\varepsilon_{1}} + \frac{\xi_{2}}{\varepsilon_{2}}\Big)
\frac{\,n_{F}(E_{1}^{-}) - n_{F}(E_{2}^{-})\,}
     {\,k_{0} - \big(E_{2}^{-} - E_{1}^{-}\big)\,}
\\
 & +\Big(\frac{\xi_{1}}{\varepsilon_{1}} - \frac{\xi_{2}}{\varepsilon_{2}}\Big)
\frac{\,n_{F}(E_{1}^{+}) - n_{F}(E_{2}^{+})\,}
     {\,k_{0} - \big(E_{2}^{+} - E_{1}^{+}\big)\,}
\\
 & +\Big(\frac{\xi_{1}}{\varepsilon_{1}} + \frac{\xi_{2}}{\varepsilon_{2}}\Big)
\frac{\,n_{F}(E_{1}^{+}) - n_{F}(E_{2}^{-})\,}
     {\,k_{0} - \big(E_{2}^{-} - E_{1}^{+}\big)\,}
\Bigg\},
\end{split}
\end{equation}
where the pole energies are
\[
\begin{aligned}
E_{1}^{-} &= e_{1}\varepsilon_{1} - \delta\mu, &
E_{1}^{+} &= -\,e_{1}\varepsilon_{1} - \delta\mu,\\
E_{2}^{+} &= \delta\mu - e_{2}\varepsilon_{2}, &
E_{2}^{-} &= \delta\mu + e_{2}\varepsilon_{2}.
\end{aligned}
\]
\[
\begin{aligned}
E_{2}^{+} - E_{1}^{-} &= 2\delta\mu - e_{2}\varepsilon_{2} - e_{1}\varepsilon_{1},\\
E_{2}^{-} - E_{1}^{-} &= 2\delta\mu + e_{2}\varepsilon_{2} - e_{1}\varepsilon_{1},\\
E_{2}^{+} - E_{1}^{+} &= 2\delta\mu + e_{1}\varepsilon_{1} - e_{2}\varepsilon_{2},\\
E_{2}^{-} - E_{1}^{+} &= 2\delta\mu + e_{1}\varepsilon_{1} + e_{2}\varepsilon_{2}.
\end{aligned}
\]
and
\[
\Xi = \frac{\xi_{1}\xi_{2}}{\varepsilon_{1}\varepsilon_{2}},
\]
We start from the definition:
\[
\delta \Pi(K) = \frac{1}{8V} \sum_{e_1,e_2 = \pm 1} \sum_{\mathbf p} \big(1 + e_1 e_2 \, \hat p \cdot \hat q \big) \big(T_1 + T_2 + T_3 + T_4\big),
\]
where
\[
\begin{aligned}
T_1 &= \Big(-\frac{\xi_1}{\varepsilon_1} - \frac{\xi_2}{\varepsilon_2}\Big) \frac{n_F(E_1^-) - n_F(E_2^+)}{k_0 - (E_2^+ - E_1^-)},\\
T_2 &= \Big(-\frac{\xi_1}{\varepsilon_1} + \frac{\xi_2}{\varepsilon_2}\Big) \frac{n_F(E_1^-) - n_F(E_2^-)}{k_0 - (E_2^- - E_1^-)},\\
T_3 &= \Big(\frac{\xi_1}{\varepsilon_1} - \frac{\xi_2}{\varepsilon_2}\Big) \frac{n_F(E_1^+) - n_F(E_2^+)}{k_0 - (E_2^+ - E_1^+)},\\
T_4 &= \Big(\frac{\xi_1}{\varepsilon_1} + \frac{\xi_2}{\varepsilon_2}\Big) \frac{n_F(E_1^+) - n_F(E_2^-)}{k_0 - (E_2^- - E_1^+)}.
\end{aligned}
\]
The energies and other quantities are defined as
\[
\begin{aligned}
E_1^- &= e_1 \varepsilon_1 - \delta \mu, & E_1^+ &= - e_1 \varepsilon_1 - \delta \mu,\\
E_2^+ &= \delta \mu - e_2 \varepsilon_2, & E_2^- &= \delta \mu + e_2 \varepsilon_2,\\
\varepsilon_1 &= \varepsilon_p^{e_1}, \quad \varepsilon_2 = \varepsilon_q^{e_2},\\
\xi_1 &= p - e_1 \bar \mu, \quad \xi_2 = q - e_2 \bar \mu.
\end{aligned}
\]
Setting \(k_0 = 0\):
\[
\begin{aligned}
D_1 &\equiv k_0 - (E_2^+ - E_1^-) = -2 \delta\mu + e_1 \varepsilon_1 + e_2 \varepsilon_2,\\
D_2 &\equiv k_0 - (E_2^- - E_1^-) = -2 \delta\mu - e_2 \varepsilon_2 + e_1 \varepsilon_1,\\
D_3 &\equiv k_0 - (E_2^+ - E_1^+) = -e_{1}\varepsilon_{1} + e_{2}\varepsilon_{2} -2\delta\mu ,\\
D_4 &\equiv k_0 - (E_2^- - E_1^+) = -2 \delta\mu - e_1 \varepsilon_1 - e_2 \varepsilon_2.
\end{aligned}
\]
Note for $\delta\mu = 0$, $D_1 = - D_4, D_2 = - D_3$.
Define
\[
\begin{aligned}
N_1 &= -\frac{\xi_1}{\varepsilon_1} - \frac{\xi_2}{\varepsilon_2}, \quad
N_2 = -\frac{\xi_1}{\varepsilon_1} + \frac{\xi_2}{\varepsilon_2},\\
N_3 &= \frac{\xi_1}{\varepsilon_1} - \frac{\xi_2}{\varepsilon_2} = - N_2, \quad
N_4 = \frac{\xi_1}{\varepsilon_1} + \frac{\xi_2}{\varepsilon_2} = - N_1.
\end{aligned}
\]
Consider the pairs \(T_1+T_4\) and \(T_2+T_3\). At \(k_0 = 0\) and $\delta\mu = 0$:
\[
\begin{aligned}
T_1 + T_4 &= \frac{N_1}{D_1} (A_1 + A_4),\\ 
\quad A_1 &= n_F(E_1^-) - n_F(E_2^+), \quad A_4 = n_F(E_1^+) - n_F(E_2^-),\\
T_2 + T_3 &= \frac{N_2}{D_2} (A_2 + A_3),\\
\quad A_2 &= n_F(E_1^-) - n_F(E_2^-), \quad A_3 = n_F(E_1^+) - n_F(E_2^+).
\end{aligned}
\]
Using the definitions of \(E_1^\pm, E_2^\pm\) and the Fermi function symmetry, we have
\[
A_1 + A_4 = 0, \quad A_2 + A_3 = 0,
\]
which implies
\[
T_1 + T_4 = 0, \quad T_2 + T_3 = 0.
\]
Therefore, the full sum vanishes:
\[
\delta \Pi(K) \big|_{k_0 = 0} = \frac{1}{8V} \sum_{e_1,e_2} \sum_{\mathbf p} (1 + e_1 e_2 \hat p \cdot \hat q)(T_1+T_2+T_3+T_4) = 0.
\]
This cancellation occurs for each \((e_1,e_2,\mathbf p)\) individually, so the vanishing of \(\delta\Pi(K)\) at \(k_0 = 0\) is exact only for $\delta\mu = 0$.

In continuum limit, we can replace $\frac{1}{V}\sum_{\vec{p}} = \int\frac{d^{3}p}{(2\pi)^{3}}$. Consider small $K$ compared to $\bar{\mu}$, and $\bar{\mu} \gg \delta\mu, \Delta$, the above expression can be expressed as
\begin{equation}
\Sigma(K) = \Sigma_{0} + \alpha_{1}k_{0}^{2} + \alpha_{2}k^{2} + \alpha_{3}k_{0}k.
\end{equation}
In a similar manner, we may write
\begin{equation}
\bar{\Pi}(K) = \bar{\Pi}_{0} + \bar{\alpha}_{1}k_{0}^{2} + \bar{\alpha}_{2}\vec{k}^{2} + \bar{\alpha}_{3}k_{0}k,
\end{equation}
and in small $K$ limit, $\delta\Pi$ can also be expanded as
\begin{equation}
\delta\Pi(K) = \alpha k_{0} + \beta |\vec{k}|,
\end{equation}
since it is a odd function. With the above information, we may express $D^{-1}(K)$ as
\begin{equation}
D^{-1}(K) = \begin{bmatrix}
a_{0} + a_{1}k_{0}^{2} + a_{2}k^{2} + a_{3}k_{0}k & i(\alpha k_{0} + \beta k)\\
- i(\alpha k_{0} + \beta k) & b_{0} + b_{1}k_{0}^{2} + b_{2}k^{2} + b_{3}k_{0}k
\end{bmatrix},
\end{equation}
where
\begin{equation}
\begin{split}
a_{0} & = \frac{1}{G} + (\Sigma_{0} - \bar{\Pi}_{0}), \ a_{1} = \alpha_{1} - \bar{\alpha}_{1}, \ a_{2} = 
\alpha_{2} - \bar{\alpha}_{2}\\
b_{0} & = \frac{1}{G} - (\Sigma_{0} + \bar{\Pi}_{0}), \ b_{1} = - (\alpha_{1} + \bar{\alpha}_{1}), \ b_{2} = - (\alpha_{2} + \bar{\alpha}_{2})\\
a_{3} & = \alpha_{3} - \bar{\alpha}_{3}, \ b_{3} = - (\alpha_{3} + \bar{\alpha}_{3}).
\end{split}
\end{equation}
From the vanishing determinant condition of $D^{-1}(K)$ with small $K$ approximation, we obtain low-energy excitation spectrum of Goldstone modes, given by
\begin{equation}
\begin{split}
a_{0}[b_{1}k_{0}^{2} + b_{2}k^{2} + b_{3}k_{0}k] & + b_{0}[a_{1}k_{0}^{2} + a_{2}k^{2} + a_{3}k_{0}k]\\
 & = \alpha^{2}k_{0}^{2} + \beta^{2} k^{2} + 2\alpha\beta k_{0}k,
\end{split}  
\end{equation}
which can also be expressed as $k_{0} = uk$ where
\begin{equation}
\begin{split}
- u^{2} & (\alpha^{2} - a_{0}b_{1} - b_{0}a_{1}) + (a_{0}b_{2} + b_{0}a_{2} - \beta^{2})\\
 & - (2\alpha\beta - a_{0}b_{3} - b_{0}a_{3}) u = 0.
\end{split}
\end{equation}
However, we also know that
\begin{equation}
0 = b_{0} = \frac{1}{G} - \bar{\Pi}_{0} - \Sigma_{0}.
\end{equation}
With the above information, we may now write
\begin{equation}
u^{2}(\alpha^{2} - a_{0}b_{1}) + \beta^{2} - a_{0}b_{2} + (2\alpha\beta - a_{0}b_{3}) u = 0,
\end{equation}
which has the following two roots
\begin{equation}
u_{\pm} = \frac{- \alpha\beta + \frac{1}{2} a_{0}b_{3} \pm \sqrt{(\alpha\beta - a_{0}b_{3}/2 )^{2} 
+ (a_{0}b_{2} - \beta^{2})(\alpha^{2} - a_{0}b_{1})}}{(\alpha^{2} - a_{0}b_{1})}.
\end{equation}
%
%
\begin{figure}[b]
\includegraphics[height = 7cm, width = 12.5cm]{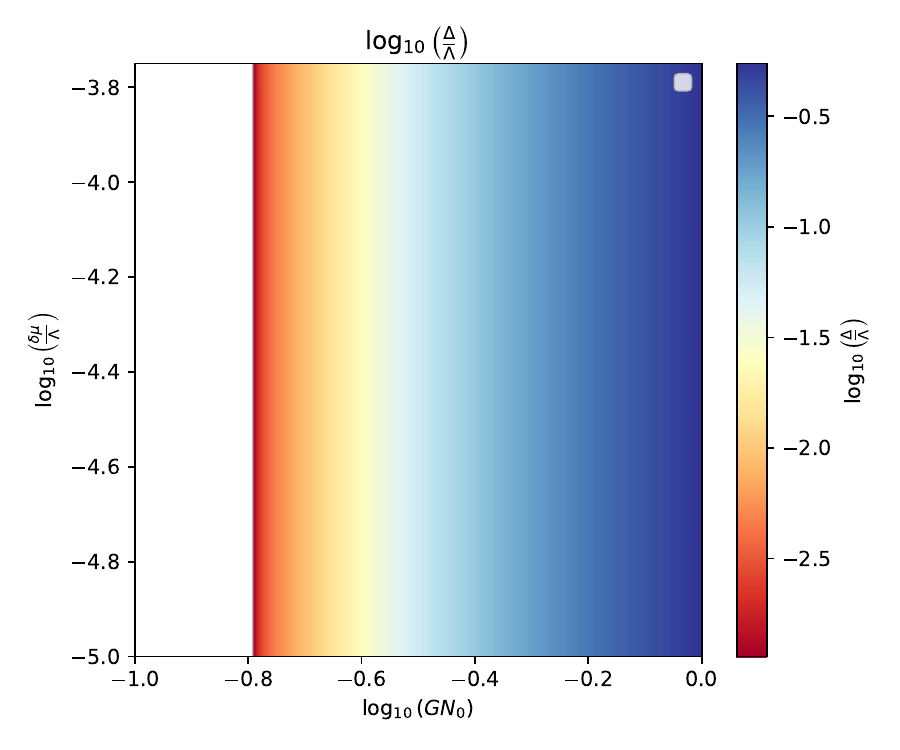}
\caption{Solution of gap parameter with non-zero $\delta\mu$.}
\label{Figure 5}
\end{figure}

\section{Heat capacity}

As we have seen previously, considering two fermionic species pairing
\begin{equation}
\mu_{\pm} = \bar{\mu} \pm \delta\mu, \ \delta\mu = \frac{g}{2}\bar{b},
\end{equation}
where $\delta\mu$ measures Fermi surface mismatch. We defined 
\begin{equation}
\varepsilon_{\vec{k}}^{e} = \sqrt{(k - e\bar{\mu})^{2} + \Delta^{2}},
\end{equation}
and after diagonalization using Nambu-Gorkov basis, we obtain the following two branches
of quasi-particle spectrum
\begin{equation}
E_{\vec{k}}^{\pm} = E_{\vec{k}} \pm \delta\mu, \ E_{\vec{k}} = \sqrt{(k - \bar{\mu})^{2}} 
+ \Delta^{2}.
\end{equation}

The grand potential at temperature $T$ is expressed as
\begin{equation}
\begin{split}
\Omega & = \frac{\Delta^{2}}{G} - 2\sum_{\vec{k}}\left(E_{\vec{k}} + T\log\left(1 
+ e^{-\frac{E_{\vec{k}}^{+}}{T}}\right) + T\log\left(1 + e^{-\frac{E_{\vec{k}}^{-}}{T}}
\right)\right),
\end{split}
\end{equation}
where we may convert summation into integral by
\begin{equation}
\sum_{\vec{k}} \rightarrow \mathcal{N}(0)\int d\xi.
\end{equation}
The heat capacity is expressed as
\begin{equation}
C_{V} = - T\frac{\partial^{2}\Omega}{\partial T^{2}},
\end{equation}
and here we also consider
\begin{equation}
\Delta \simeq 2\delta e^{-\frac{2\pi^{2}}{G\bar{\mu}^{2}}}, \ \mathcal{N}(0) = \frac{\bar{\mu}^{2}}{\pi^{2}}.
\end{equation}

\section{Different mean field approximation}
In this section, we consider the same model but with the consideration of a different mean-field ansatz which is also spatially uniform. Like earlier we consider a system consisting of two massless fermion species denoted by Dirac fields $\psi_{1}$ and $\psi_{2}$. The model that we start is described by the following action
\begin{equation}\label{action 1}
\begin{split}
S & = \int d^{4}x \Big[\bar{\psi}i\slashed{\partial}\psi - \frac{1}{4}\mathbf{b}_{\mu\nu}.\mathbf{b}^
{\mu\nu} + \frac{1}{2}m_{b}^{2}\mathbf{b}_{\mu}.\mathbf{b}^{\mu} - \frac{g}{2}\bar{\psi}\gamma^{\mu}
\boldsymbol{\tau}.\mathbf{b}_{\mu}\psi - \frac{\zeta g^{4}}{24}(\mathbf{b}_{\mu}.\mathbf{b}^{\mu})^{2}\Big],
\end{split} 
\end{equation} 
where
\begin{equation}\label{two component Dirac field}
\psi = \begin{bmatrix}
\psi_{1}\\
\psi_{2}
\end{bmatrix}, \ \bar{\psi} = \begin{bmatrix}
\bar{\psi}_{1} & \bar{\psi}_{2}
\end{bmatrix}
\end{equation}
and $\mathbf{b}_{\mu} = (b_{\mu}^{1}, b_{\mu}^{2}, b_{\mu}^{3})$ is a triplet vector field with the field strength tensor being
\begin{equation}\label{field strength tensor}
\mathbf{b}_{\mu\nu} = \partial_{\mu}\mathbf{b}_{\nu} - \partial_{\nu}\mathbf{b}_{\mu}.
\end{equation}
The above action (\ref{action 1}) is invariant under $SU(2)_{V}$ group defined by
\begin{equation}\label{symmetry}
\psi \rightarrow U\psi, \ \bar{\psi} \rightarrow \bar{\psi}U^{\dagger}, \ \boldsymbol{\tau}.\mathbf{b}_{\mu} \rightarrow U\boldsymbol{\tau}.\mathbf{b}_{\mu}U^{\dagger},
\end{equation}
where $U\in SU(2)$. In the action (\ref{action 1}), $\boldsymbol{\tau} = (\tau^{1}, \tau^{2}, \tau^{3})$ are the Pauli matrices. The action (\ref{action 1}) is invariant under $SU(2)_{V} \times U(1)$.

Under the relativistic mean field approximation, we assume $\langle b_{\mu}^{a}\rangle = \bar{b}_{1}\delta_{\mu}^{0}\delta^{a1} + \bar{b}_{2}\delta_{\mu}^{0}\delta^{a2}$ which reduces the action (\ref{action 1}) to the following mean-field approximation
\begin{equation}\label{RMF action 1}
\begin{split}
S & = \int d^{4}x \Big[\bar{\psi}i\slashed{\partial}\psi + \frac{1}{2}m_{b}^{2}(\bar{b}_{1}^{2} 
 + \bar{b}_{2}^{2}) - \frac{g}{2}\bar{\psi}\gamma^{0}(\tau^{1}\bar{b}_{1} + \tau^{2}\bar{b}_{2})\psi 
 - \frac{\zeta g^{4}}{24}(\bar{b}_{1}^{2} + \bar{b}_{2}^{2})^{2}\Big].
\end{split}
\end{equation}
Now we define the quantities $\Delta_{0} = \bar{b}_{1} - i\bar{b}_{2}, \ \Delta_{0}^{*} = \bar{b}_{1}
 + i\bar{b}_{2}$. At finite temperature and chemical potential, the Euclidean action is given by
\begin{equation}\label{Euclidean action 1}
\begin{split}
S^{\beta} & = \int_{0}^{\beta}d\tau\int d^{3}x \ \Bigg[\bar{\psi}\Big[\gamma^{0}(\partial_{\tau} - \mu)
 + i\gamma^{k}\partial_{k} + \frac{g}{2}\gamma^{0}[\Delta_{0}\tau^{+} + \Delta_{0}^{*}\tau^{-}]\Big]\psi
 - \frac{1}{2}m_{b}^{2}|\Delta_{0}|^{2} + \frac{\zeta g^{4}}{24}|\Delta_{0}|^{4}\Big],
\end{split}
\end{equation}
and as a result, the Euclidean Lagrangian density for the two-component spinor field can be expressed as
\begin{equation}\label{Euclidean Lagrangian 1}
\mathcal{L}_{\psi}^{\beta} = \bar{\psi}\mathcal{M}\psi,
\end{equation}
where
\begin{equation}\label{expression of the quadration form}
\mathcal{M} = \begin{bmatrix}
\gamma^{0}(\partial_{\tau} - \mu) + i\gamma^{k}\partial_{k} & \Delta_{0}\gamma^{0}\\
\Delta_{0}^{*}\gamma^{0} & \gamma^{0}(\partial_{\tau} - \mu) + i\gamma^{k}\partial_{k}
\end{bmatrix}.
\end{equation}
Under the degeneracy limit (or small temperature limit) such that $\beta\mu \gg 1$, the partition function defined by
\begin{equation}\label{partition function}
\mathcal{Z} = \int\mathcal{D}\bar{\psi}\mathcal{D}\psi \ e^{-S^{\beta}},
\end{equation}
can be computed exactly and it is related to the product of determinant of the quadratic form $\mathcal{M}$ over each momentum $\vec{k}$ and Matsubara frequencies $\omega_{n} = \frac{(2n + 1)
\pi}{\beta}$. We may note that
\begin{equation}
\mathcal{M}^{2} = \begin{bmatrix}
K^{2} + |\Delta_{0}|^{2} & 2K_{0}\Delta_{0}^{*}\\
2K_{0}\Delta_{0} & K^{2} + |\Delta_{0}|^{2}
\end{bmatrix},
\end{equation}
where $K^{\mu} = ((i\omega_{n} - \mu), \vec{k})$, and the determinant of the above matrix can be 
expressed as
\begin{equation}
\begin{split}
\text{det}\mathcal{M}^{2} & = (K^{2} + |\Delta_{0}|^{2})^{2} - 4|\Delta_{0}|^{2}K_{0}^{2}\\
 & = (K_{0}^{2} - \vec{k}^{2} + |\Delta_{0}|^{2})^{2} - 4|\Delta_{0}|^{2}K_{0}^{2}\\
 & = \{(K_{0} - |\Delta_{0}|)^{2} - \vec{k}^{2}\}\{(K_{0} + |\Delta_{0}|)^{2} - \vec{k}^{2}\}\\
 & = \{(i\omega_{n} - \mu_{+})^{2} - |\vec{k}|^{2}\}\{(i\omega_{n} - \mu_{-})^{2} - |\vec{k}|^{2}\}, 
\end{split}
\end{equation}
where $\mu_{\pm} = \mu \pm \frac{g}{2}|\Delta_{0}|$. Using the above expression, we may now express 
the logarithm of partition function as
\begin{equation}\label{logarithm of partition function}
\begin{split}
\frac{1}{\beta V} & \log\mathcal{Z} = \frac{\mu_{+}^{4}}{12\pi^{2}} + \frac{\mu_{-}^{4}}{12\pi^{2}} 
 + \frac{1}{2}m_{b}^{2}|\Delta_{0}|^{2} - \frac{\zeta g^{4}}{24}|\Delta_{0}|^{4}\\
 & = \frac{\mu^{4}}{6\pi^{2}} + |\Delta_{0}|^{2}\left(\frac{m_{b}^{2}}{2} + \frac{g^{2}\mu^{2}}{4\pi^{2}}
 \right) + \left(\frac{g^{4}}{96\pi^{2}} - \frac{\zeta g^{4}}{24}\right)|\Delta_{0}|^{4}.
\end{split}
\end{equation}
The associated free-energy density can be expressed as
\begin{equation}\label{free-energy density}
\begin{split}
f = - \frac{1}{\beta V}\log\mathcal{Z} & = \frac{g^{4}}{24}\left(\zeta - \frac{1}{4\pi^{2}}\right)
|\Delta_{0}|^{4} - |\Delta_{0}|^{2}\left(\frac{m_{b}^{2}}{2} + \frac{g^{2}\mu^{2}}{4\pi^{2}}\right)\\ 
 & - \frac{\mu^{4}}{6\pi^{2}}.
\end{split}
\end{equation}
We may note that the above expression of free-energy density is invariant under the action of $U(1)$
group $\Delta_{0} \rightarrow \Delta_{0}e^{i\varphi}$. If we consider for time-being $\zeta = 0$, then the from the field equation, we get
\begin{equation}\label{zeta = 0 equation of motion}
m_{b}^{2}|\Delta_{0}| = \frac{g}{2}\langle\bar{\psi}\gamma^{0}\tau^{3}\psi\rangle = \frac{g}{2}
(n_{1} - n_{2}).
\end{equation}
On the other hand, in that case, the total number density can be expressed as
\begin{equation}\label{number density}
\begin{split}
n & = \frac{\partial}{\partial\mu}\left(\frac{1}{\beta V}\log\mathcal{Z}\right) = 
\frac{\partial}{\partial\mu_{+}}\left(\frac{1}{\beta V}\log\mathcal{Z}\right)
 + \frac{\partial}{\partial\mu_{-}}\left(\frac{1}{\beta V}\log\mathcal{Z}\right) 
 = \frac{\mu_{+}^{3}}{3\pi^{2}} + \frac{\mu_{-}^{3}}{3\pi^{2}}. 
\end{split}
\end{equation}
The above expression can also be expressed as
\begin{equation}\label{number density relations}
n = n_{1} + n_{2}, \ n_{1} = \frac{\mu_{-}^{3}}{3\pi^{2}}, \ n_{2} = \frac{\mu_{+}^{3}}{3\pi^{2}}.
\end{equation} 
From the above expression, we also find the following relation
\begin{equation}\label{relation 1}
- g|\Delta_{0}| = (3\pi^{2}n_{1})^{1/3} - (3\pi^{2}n_{2})^{1/3},
\end{equation}
which combines with (\ref{zeta = 0 equation of motion}) leads to the following relation
\begin{equation}\label{relation 2}
\begin{split}
(bn_{1})^{1/3} - (bn_{2})^{1/3} & = - \frac{g^{2}}{6\pi^{2}}(bn_{1} - bn_{2})\\
\implies (bn_{1})^{2/3} + (bn_{1})^{1/3}(bn_{2})^{1/3} & + (bn_{2})^{2/3} = - \frac{6\pi^{2}}{g^{2}},
\end{split}
\end{equation}
where $b = \frac{3\pi^{2}}{m_{b}^{3}}$. Defining $S = (bn_{1})^{1/3} + (bn_{2})^{1/3}$ and $P = (bn_{1})^{1/3}(bn_{2})^{1/3}$, we will find the following relations
\begin{equation}
S^{2} - P = - \frac{6\pi^{2}}{g^{2}}, \ S^{3} - 3PS = bn,
\end{equation}
and combining the above set of equations, we get the following relation
\begin{equation}
2S^{3} + \frac{18\pi^{2}}{g^{2}}S + bn = 0,
\end{equation}
whose real root is given by
\begin{equation}
\begin{split}
S & = - \Bigg[\left(\frac{bn}{4} + \sqrt{\left(\frac{bn}{4}\right)^{2} + \left(\frac{3\pi^{2}}{g^{2}}\right)^{3}}\right)^{1/3}\\
 & - \left( - \frac{bn}{4} + \sqrt{\left(\frac{bn}{4}\right)^{2} + \left(\frac{3\pi^{2}}{g^{2}}\right)^{3}}\right)^{1/3}\Bigg],
\end{split}
\end{equation}
which shows $S < 0$, however, $S$ by definition should be positive definite. Therefore, the only solution for $\zeta = 0$ case is $n_{1} = n_{2}$.

On the other hand, for $\zeta > \frac{1}{4\pi^{2}}$, $\bar{b}$ chooses a non-zero expectation value which minimizes the free-energy density, and its expectation value is given by
\begin{equation}
|\Delta_{0}| = \sqrt{\frac{6}{g^{4}}\frac{\left(m_{b}^{2} + \frac{g^{2}\mu^{2}}{2\pi^{2}}\right)}{\left(\zeta - \frac{1}{4\pi^{2}}\right)}}.
\end{equation}
Note that the dispersion relation for the fermions can be obtained vanishing of the determinant 
$\mathcal{M}$ in momentum space which shows the energy dispersion relations are
\begin{equation}
\varepsilon_{\pm}(\mathbf{k}) = |\mathbf{k}| \pm \frac{g}{2}|\Delta_{0}|.
\end{equation}
This shows that for a given chemical potential $\mu$, the Fermi radius of two particle species become different 
\begin{equation}
k_{F}^{\pm} = \mu_{\mp} = \mu \mp \frac{g}{2}|\Delta_{0}| = \mu \mp \sqrt{\frac{3}{2g^{2}}\frac{\left(m_{b}^{2} + \frac{g^{2}\mu^{2}}{2\pi^{2}}\right)}{\left(\zeta - \frac{1}{4\pi^{2}}
\right)}}.
\end{equation}
In order for $k_{F}^{+} > 0$, the following condition must be satisfied
\begin{equation}
\begin{split}
\frac{2g^{2}}{3}\mu^{2} & > \frac{\left(m_{b}^{2} + \frac{g^{2}\mu^{2}}{2\pi^{2}}\right)}{\left(\zeta - \frac{1}{4\pi^{2}}\right)}\\
\implies \mu^{2} &  > \frac{3m_{b}^{2}}{2g^{2}\left(\zeta - \frac{1}{\pi^{2}}\right)}.
\end{split}
\end{equation}
The above equation shows that for $\zeta > \frac{1}{\pi^{2}}$, if the chemical potential satisfies the above inequality, two different Fermi surfaces of different radius form. On the other hand, for $\frac{1}{\pi^{2}} > \zeta > \frac{1}{4\pi^{2}}$, there will be a single Fermi sphere whereas for $0 \leq \zeta < \frac{1}{4\pi^{2}}$, $|\Delta_{0}|$ satisfy the following equation of motion
\begin{equation}
\begin{split}
m_{b}^{2}|\Delta_{0}| & = \frac{\zeta g^{4}}{6}|\Delta_{0}|^{3} + \frac{g}{2}(n_{1} - n_{2}) = \frac{\zeta g^{4}}{6}|\Delta_{0}|^{3}\\
 & - \frac{g}{6\pi^{2}}(\mu_{+}^{3} - \mu_{-}^{3})\\
\implies m_{b}^{2}|\Delta_{0}| & = \frac{\zeta g^{4}}{6}|\Delta_{0}|^{3} - \frac{\mu^{2}}{2\pi^{2}}
 g^{2}|\Delta_{0}| - \frac{g^{4}}{24\pi^{2}}|\Delta_{0}|^{3}\\
\implies |\Delta_{0}| & \Big[\frac{g^{4}}{6}\left(\zeta - \frac{1}{4\pi^{2}}\right)|\Delta_{0}|^{2} 
- \left(m_{b}^{2} + \frac{g^{2}\mu^{2}}{2\pi^{2}}\right)\Big] = 0,
\end{split}
\end{equation}
which shows that $|\Delta_{0}| = 0$ for $0 \leq \zeta < \frac{1}{4\pi^{2}}$. It is important to note 
here that non-zero value of $|\Delta_{0}|$ breaks $SU(2)_{V} \times U(1)$ symmetry of the system to
$U(1)_{V} \times U(1)$ spontaneously.

\section{Mean-field order parameter with spatially varying phase}

Unlike the previous section, here we consider the mean field parameters $\bar{b}_{1}$ and $\bar{b}_{2}$ are spatially varying. We specifically consider the complex-valued mean field parameters to be of the following form
\begin{equation}
\Delta_{0} = \bar{\Delta}e^{i\mathbf{q}.\mathbf{x}}, \ \Delta_{0}^{*} = \bar{\Delta}e^{- i\mathbf{q}.\mathbf{x}}.
\end{equation}
We may note that in this case, the Maxwell term reduces to the following form
\begin{equation}
\begin{split}
- \frac{1}{4} & \mathbf{b}_{\mu\nu}.\mathbf{b}^{\mu\nu} = - \frac{1}{4}\Big[(\partial_{\mu}\bar{b}_{1}
 \eta_{\nu 0} - \partial_{\nu}\bar{b}_{1}\eta_{\mu 0})\delta^{i1}\\
 & + (\partial_{\mu}\bar{b}_{2}\eta_{\nu 0} - \partial_{\nu}\bar{b}_{2}\eta_{\mu 0})\delta^{i2}\Big]
 \Big[(\partial^{\mu}\bar{b}_{1}\eta^{\nu 0} - \partial^{\nu}\bar{b}_{1}\eta^{\mu 0})\delta^{i1}\\
 & + (\partial^{\mu}\bar{b}_{2}\eta^{\nu 0} - \partial^{\nu}\bar{b}_{2}\eta^{\mu 0})\delta^{i2}\Big]\\
 & = - \frac{1}{4}\Big[(\partial_{\mu}\bar{b}_{1}\eta_{\nu 0} - \partial_{\nu}\bar{b}_{1}\eta_{\mu 0})
 (\partial^{\mu}\bar{b}_{1}\eta^{\nu 0} - \partial^{\nu}\bar{b}_{1}\eta^{\mu 0})\\
 & + (\partial_{\mu}\bar{b}_{2}\eta_{\nu 0} - \partial_{\nu}\bar{b}_{2}\eta_{\mu 0})(\partial^{\mu}
 \bar{b}_{2}\eta^{\nu 0} - \partial^{\nu}\bar{b}_{2}\eta^{\mu 0})\Big]\\
 & = - \frac{1}{2}\left(\partial_{\mu}\bar{b}_{1}\partial^{\mu}\bar{b}_{1} - (\partial_{0}
 \bar{b}_{1})^{2} + \partial_{\mu}\bar{b}_{2}\partial^{\mu}\bar{b}_{2} - (\partial_{0}\bar{b}_{2})^{2}
 \right)\\
 & = \frac{1}{2}[(\mathbf{\nabla}\bar{b}_{1})^{2} + (\mathbf{\nabla}\bar{b}_{2})^{2}] = \frac{1}{2}
 \mathbf{\nabla}\Delta_{0}.\mathbf{\nabla}\Delta_{0}^{*} = \frac{1}{2}\mathbf{q}^{2}\bar{\Delta}^{2},
\end{split}
\end{equation}
which makes the RMF action as
\begin{equation}
\begin{split}
S & = \int d^{4}x \Big[\bar{\psi}i\slashed{\partial}\psi + \frac{1}{2}(m_{b}^{2} + \mathbf{q}^{2})\bar{\Delta}^{2}\\ 
 & - \frac{g\bar{\Delta}}{2}\bar{\psi}\gamma^{0}(\tau^{1}\cos(\mathbf{q}.\mathbf{x}) - \tau^{2} 
 \sin(\mathbf{q}.\mathbf{x}))\psi - \frac{\zeta g^{4}}{24}\bar{\Delta}^{4}\Big].
\end{split}
\end{equation}
At a finite temperature and chemical potential, the Euclidean action is now given by
\begin{equation}
\begin{split}
S^{\beta} & = \int_{0}^{\beta}d\tau\int d^{3}x \ \Bigg[\bar{\psi}\Big[\gamma^{0}(\partial_{\tau} - \mu)
 + i\gamma^{k}\partial_{k}\\
 & + \frac{g}{2}\bar{\Delta}\gamma^{0}[e^{i\mathbf{q}.\mathbf{x}}\tau^{+} + e^{-i\mathbf{q}.\mathbf{x}}
 \tau^{-}]\Big]\psi\\
 & - \frac{1}{2}(m_{b}^{2} + \mathbf{q}^{2})\bar{\Delta}^{2} + \frac{\zeta g^{4}}{24}
 \bar{\Delta}^{4}\Big],
\end{split}
\end{equation}
and therefore, the quadratic form of Dirac fermions in this case is given by
\begin{equation}
\mathcal{M}_{\mathbf{q}} = \begin{bmatrix}
\gamma^{0}(\partial_{\tau} - \mu) + i\gamma^{k}\partial_{k} & \bar{\Delta}\gamma^{0}e^{i\mathbf{q}.\mathbf{x}}\\
\bar{\Delta}\gamma^{0}e^{-i\mathbf{q}.\mathbf{x}} & \gamma^{0}(\partial_{\tau} - \mu) + i\gamma^{k}\partial_{k}
\end{bmatrix}.
\end{equation}
We may note further that the square of the above matrix can be expressed as
\begin{equation}
\mathcal{M}_{\mathbf{q}}^{2} = \begin{bmatrix}
(\partial_{\tau} - \mu)^{2} + \nabla^{2} & g\bar{\Delta}e^{i\mathbf{q}.\mathbf{x}}(\partial_{\tau} - \mu) \\
 + \frac{g^{2}}{4}\bar{\Delta}^{2} & - \frac{g}{2}\vec{\gamma}.\vec{q}\gamma^{0}\bar{\Delta}
 e^{i\vec{q}.\vec{x}}\\
 & \\
g\bar{\Delta}e^{-i\mathbf{q}.\mathbf{x}}(\partial_{\tau} - \mu) & (\partial_{\tau} - \mu)^{2} + \nabla^{2}\\
+ \frac{g}{2}\vec{\gamma}.\vec{q}\gamma^{0}\bar{\Delta}e^{-i\vec{q}.\vec{x}} &  + \frac{g^{2}}{4}\bar{\Delta}^{2},
\end{bmatrix} 
\end{equation}
and the determinant of the above matrix in Fourier domain is
\begin{equation}
[(i\omega_{n} - \mu_{+})^{2} - \mathbf{k}^{2}][(i\omega_{n} - \mu_{-})^{2} - \mathbf{k}^{2}] + \frac{g^{2}}{4}\mathbf{q}^{2}\bar{\Delta}^{2}.
\end{equation}
Therefore, the logarithm of partition function can be evaluated as
\begin{equation}
\begin{split}
\log\mathcal{Z} & = \log\mathcal{Z}_{1} + \log\mathcal{Z}_{2}\\
\log\mathcal{Z}_{1} & = V\sum_{n}\int\frac{d^{3}k}{(2\pi)^{3}}\log\left([(i\omega_{n} - \mu_{+})^{2} - \mathbf{k}^{2}][(i\omega_{n} - \mu_{-})^{2} - \mathbf{k}^{2}]\right)\\
\log\mathcal{Z}_{2} & = \frac{g^{2}}{4}\mathbf{q}^{2}\bar{\Delta}^{2}V\sum_{n}\int\frac{d^{3}k}{(2\pi)^{3}}\\
 & \times \frac{1}{[(i\omega_{n} - \mu_{+})^{2} - \mathbf{k}^{2}][(i\omega_{n} - \mu_{-})^{2} - \mathbf{k}^{2}]} + \mathcal{O}(g^{4}|\mathbf{q}|^{4}\bar{\Delta}^{4}),
\end{split}
\end{equation}
where $\log\mathcal{Z}_{1}$ is already calculated earlier. On the other hand, the leading part of $\log\mathcal{Z}_{2}$ can be expressed as
\begin{equation}
\begin{split}
\log\mathcal{Z}_{2} & = - \frac{\beta}{2\pi i}\frac{g^{2}}{4}\mathbf{q}^{2}\bar{\Delta}^{2}V\int\frac{d^{3}k}{(2\pi)^{3}}\oint_{\mathcal{C}_{F}}dz \ n_{F}(z)\\
 & \times \frac{1}{[(z - \mu_{+})^{2} - \mathbf{k}^{2}][(z - \mu_{-})^{2} - \mathbf{k}^{2}]},
\end{split}
\end{equation}
where the poles of the above integrand lies on real axis and they are located at $z = \mu_{+} \pm k$ and $z = \mu_{-} \pm k$. Computing the residues, we finally obtain
\begin{equation}
\begin{split}
\log\mathcal{Z}_{2} & = \beta V \frac{g^{2}}{4}\mathbf{q}^{2}\bar{\Delta}^{2}\int\frac{d^{3}k}{(2\pi)^{3}}\frac{1}{2k(\mu_{+} - \mu_{-})}\Bigg[\frac{n_{F}(\mu_{+} + k) + n_{F}(\mu_{-} - k)}{2k 
+ \mu_{+} - \mu_{-}}\\
 & + \frac{n_{F}(\mu_{+} - k) + n_{F}(\mu_{-} + k)}{2k - (\mu_{+} - \mu_{-})}\Bigg],
\end{split}
\end{equation}
where $n_{F}$ is the Fermi-Dirac distribution function. At zero temperature limit, the above expression reduces to the following
\begin{equation}
\begin{split}
\log\mathcal{Z}_{2} & = \beta V\frac{g^{2}}{16\pi^{2}}\mathbf{q}^{2}\bar{\Delta}^{2}\int_{0}^{\infty}\frac{k \ dk}{(\mu_{+} - \mu_{-})}\\
 & \times \left(\frac{\Theta(k - \mu_{-})}{2k + \mu_{+} - \mu_{-}} + \frac{\Theta(k - \mu_{+})}{2k - (\mu_{+} - \mu_{-})}\right).
\end{split}
\end{equation}
Since the above expression is diverging, we remove it from the free-energy density, and as a result, the total free-energy density can be expressed as
\begin{equation}
f_{\text{tot}} = \frac{g^{4}}{24}\left(\zeta - \frac{1}{4\pi^{2}}\right)\bar{\Delta}^{4} - \bar{\Delta}^{2}\left(\frac{m_{b}^{2}}{2} + \frac{g^{2}\mu^{2}}{4\pi^{2}} - \frac{\mathbf{q}^{2}}{2}\right) 
- \frac{\mu^{4}}{6\pi^{2}}.
\end{equation}
The above expression clearly shows that spatial variation of order parameters $\Delta_{0}, \ \Delta_{0}^{*}$ leads to positive contribution to the free-energy as the spatial variation of order parameter 
cost energy.
%



\begin{thebibliography}{0}%
\makeatletter
\providecommand \@ifxundefined [1]{%
 \@ifx{#1\undefined}
}%
\providecommand \@ifnum [1]{%
 \ifnum #1\expandafter \@firstoftwo
 \else \expandafter \@secondoftwo
 \fi
}%
\providecommand \@ifx [1]{%
 \ifx #1\expandafter \@firstoftwo
 \else \expandafter \@secondoftwo
 \fi
}%
\providecommand \natexlab [1]{#1}%
\providecommand \enquote  [1]{``#1''}%
\providecommand \bibnamefont  [1]{#1}%
\providecommand \bibfnamefont [1]{#1}%
\providecommand \citenamefont [1]{#1}%
\providecommand \href@noop [0]{\@secondoftwo}%
\providecommand \href [0]{\begingroup \@sanitize@url \@href}%
\providecommand \@href[1]{\@@startlink{#1}\@@href}%
\providecommand \@@href[1]{\endgroup#1\@@endlink}%
\providecommand \@sanitize@url [0]{\catcode `\\12\catcode `\$12\catcode `\&12\catcode `\#12\catcode `\^12\catcode `\_12\catcode `\%12\relax}%
\providecommand \@@startlink[1]{}%
\providecommand \@@endlink[0]{}%
\providecommand \url  [0]{\begingroup\@sanitize@url \@url }%
\providecommand \@url [1]{\endgroup\@href {#1}{\urlprefix }}%
\providecommand \urlprefix  [0]{URL }%
\providecommand \Eprint [0]{\href }%
\providecommand \doibase [0]{https://doi.org/}%
\providecommand \selectlanguage [0]{\@gobble}%
\providecommand \bibinfo  [0]{\@secondoftwo}%
\providecommand \bibfield  [0]{\@secondoftwo}%
\providecommand \translation [1]{[#1]}%
\providecommand \BibitemOpen [0]{}%
\providecommand \bibitemStop [0]{}%
\providecommand \bibitemNoStop [0]{.\EOS\space}%
\providecommand \EOS [0]{\spacefactor3000\relax}%
\providecommand \BibitemShut  [1]{\csname bibitem#1\endcsname}%
\let\auto@bib@innerbib\@empty
\end{thebibliography}%


\begin{thebibliography}{99}

\bibitem{Sarma1963}
G.~Sarma, J.\ Phys.\ Chem.\ Solids \textbf{24}, 1029 (1963).

\bibitem{LiuWilczek2003}
W.~V.~Liu and F.~Wilczek, Phys.\ Rev.\ Lett.\ \textbf{90}, 047002 (2003).

\bibitem{Gubankova2005}
E.~Gubankova, E.~G.~Mishchenko, and F.~Wilczek, Phys.\ Rev.\ Lett.\ \textbf{94}, 110402 (2005).

\bibitem{FuldeFerrell1964}
P.~Fulde and R.~A.~Ferrell, Phys.\ Rev.\ \textbf{135}, A550 (1964).

\bibitem{LarkinOvchinnikov1965}
A.~I.~Larkin and Yu.~N.~Ovchinnikov, Zh.\ Eksp.\ Teor.\ Fiz.\ \textbf{47}, 1136 (1964) [Sov.\ Phys.\ JETP \textbf{20}, 762 (1965)].

\bibitem{Matsuda2007}
Y.~Matsuda and H.~Shimahara, J.\ Phys.\ Soc.\ Jpn.\ \textbf{76}, 051005 (2007).

\bibitem{Kinnunen2018}
J.~J.~Kinnunen \textit{et al.}, Rep.\ Prog.\ Phys.\ \textbf{81}, 046401 (2018).

\bibitem{Zheng2014}
Z.~Zheng \textit{et al.}, Sci.\ Rep.\ \textbf{4}, 6535 (2014).

\bibitem{Chandrasekhar1962}
B.~S.~Chandrasekhar, Appl.\ Phys.\ Lett.\ \textbf{1}, 7 (1962).

\bibitem{Clogston1962}
A.~M.~Clogston, Phys.\ Rev.\ Lett.\ \textbf{9}, 266 (1962).

\bibitem{Gubankova2003}
E.~Gubankova, W.~V.~Liu, and F.~Wilczek, Phys.\ Rev.\ Lett.\ \textbf{91}, 032001 (2003).

\bibitem{Reddy2005}
S.~Reddy and G.~Rupak, Phys.\ Rev.\ C \textbf{71}, 025201 (2005).

\bibitem{Gubankova2006}
E.~Gubankova, A.~Schmitt, and F.~Wilczek, Phys.\ Rev.\ B \textbf{74}, 064505 (2006).

\bibitem{Zwierlein2006}
M.~W.~Zwierlein, A.~Schirotzek, C.~H.~Schunck, and W.~Ketterle, Science \textbf{311}, 492 (2006).

\bibitem{Partridge2006}
G.~B.~Partridge \textit{et al.}, Science \textbf{311}, 503 (2006).

\bibitem{Schunck2007}
C.~H.~Schunck \textit{et al.}, Science \textbf{316}, 867 (2007).

\bibitem{Wehling2014}
T.~O.~Wehling, A.~M.~Black-Schaffer, and A.~V.~Balatsky, Adv.\ Phys.\ \textbf{63}, 1 (2014).

\bibitem{Nandkishore2016}
R.~Nandkishore, Phys.\ Rev.\ B \textbf{93}, 020506 (2016).

\bibitem{Bednik2015}
G.~Bednik, A.~A.~Zyuzin, and A.~A.~Burkov, Phys.\ Rev.\ B \textbf{92}, 035153 (2015).

\bibitem{Yang2014}
S.~A.~Yang, H.~Pan, and F.~Zhang, Phys.\ Rev.\ Lett.\ \textbf{113}, 046401 (2014).

\bibitem{Meng2012}
T.~Meng and L.~Balents, Phys.\ Rev.\ B \textbf{86}, 054504 (2012).

\bibitem{Bai2025}
X.~Bai, W.~LiMing, and T.~Zhou, New J.\ Phys.\ \textbf{27}, 013003 (2025).

\bibitem{Gubbels2012}
K.~B.~Gubbels and H.~T.~C.~Stoof, arXiv:1205.0568 (2012).

\bibitem{Boettcher2015}
I.~Boettcher, T.~K.~Herbst, J.~M.~Pawlowski, N.~Strodthoff, L.~von Smekal, and C.~Wetterich, Phys.\ Lett.\ B \textbf{742}, 86 (2015).

\bibitem{He2006}
L.~He, M.~Jin, and P.~Zhuang, Phys.\ Rev.\ D \textbf{74}, 036005 (2006).

\bibitem{He2006b}
L.~He, M.~Jin, and P.~Zhuang, Phys.\ Rev.\ B \textbf{73}, 214527 (2006).

\bibitem{He2009}
L.~He and P.~Zhuang, Phys.\ Rev.\ B \textbf{79}, 024511 (2009).

\bibitem{Liao2003}
J.~Liao and P.~Zhuang, Phys.\ Rev.\ D \textbf{68}, 114016 (2003).

\bibitem{Huang2007}
X.~Huang, X.~Hao, and P.~Zhuang, Int.\ J.\ Mod.\ Phys.\ E \textbf{16}, 2307 (2007).

\bibitem{ShovkovyHuang2003}
I.~Shovkovy and M.~Huang, Phys.\ Lett.\ B \textbf{564}, 205 (2003).

\bibitem{Schmitt2010}
A.~Schmitt, in \textit{Dense Matter in Compact Stars} (Springer, 2010), pp.~29--59.

\bibitem{CJT1974}
J.~M.~Cornwall, R.~Jackiw, and E.~Tomboulis, Phys.\ Rev.\ D \textbf{10}, 2428 (1974).

\bibitem{Schmitt2015}
A.~Schmitt, \textit{Introduction to Superfluidity}, Lect.\ Notes Phys.\ Vol.~888 (Springer, 2015).

\bibitem{Walecka1974}
J.~D.~Walecka, Ann.\ Phys.\ (N.Y.) \textbf{83}, 491 (1974).

\bibitem{Bodmer1991}
A.~R.~Bodmer, Nucl.\ Phys.\ A \textbf{526}, 703 (1991).

\bibitem{MuellerSerot1996}
H.~M\"uller and B.~D.~Serot, Nucl.\ Phys.\ A \textbf{606}, 508 (1996).

\bibitem{Ohsaku2004}
T.~Ohsaku, Int.\ J.\ Mod.\ Phys.\ B \textbf{18}, 1771 (2004).

\bibitem{BCS1957}
J.~Bardeen, L.~N.~Cooper, and J.~R.~Schrieffer, Phys.\ Rev.\ \textbf{106}, 162 (1957).

\bibitem{SM}
See Supplemental Material for complete derivations.

\bibitem{Forbes2005}
M.~M.~Forbes, E.~Gubankova, W.~V.~Liu, and F.~Wilczek, Phys.\ Rev.\ Lett.\ \textbf{94}, 017001 (2005).

\bibitem{Jo2009}
G.-B.~Jo, Y.-R.~Lee, J.-H.~Choi, C.~A.~Christensen, T.~H.~Kim, J.~H.~Thywissen, D.~E.~Pritchard, and W.~Ketterle, Science \textbf{325}, 1521 (2009).

\bibitem{ValleyPol2015}
H.~Pan, S.~A.~Yang, and F.~Zhang, Phys.\ Rev.\ B \textbf{91}, 155423 (2015).

\bibitem{TBG2021}
Y.-H.~Zhang, D.~Mao, and T.~Senthil, Phys.\ Rev.\ Res.\ \textbf{3}, L032035 (2021).

\bibitem{Koponen2007}
T.~K.~Koponen \textit{et al.}, Phys.\ Rev.\ Lett.\ \textbf{99}, 120403 (2007).

\bibitem{Yang2026}
F.~Yang, R.~Li, J.~Liu, and B.~Yan, arXiv:2606.01785.

\bibitem{WuYip2003}
S.-T.~Wu and S.~Yip, Phys.\ Rev.\ A \textbf{67}, 053603 (2003).

\bibitem{HuangShovkovy2004}
M.~Huang and I.~A.~Shovkovy, Phys.\ Rev.\ D \textbf{70}, 051501(R) (2004); Phys.\ Rev.\ D \textbf{70}, 094030 (2004).

\end{thebibliography}
\end{document}